\documentclass[aps,prd,twocolumn,tightenlines,nofootinbib,superscriptaddress,floatfix]{revtex4}

\usepackage[english]{babel}
\usepackage{verbatim}
\usepackage{times}
\usepackage{natbib}
\usepackage{amssymb,amsbsy,amsmath,amsfonts,amstext}
\usepackage{mathrsfs}
\usepackage{graphicx,xcolor}
\usepackage{epsf,epsfig,latexsym,amsthm,fancyhdr,rotating}
\usepackage{slashed}
\usepackage{esint}
\usepackage{nicefrac}
\usepackage{blindtext}
\usepackage{float}

\newcommand{\be}{\begin{eqnarray}}
\newcommand{\ee}{\end{eqnarray}}
\newcommand{\ba}{\begin{array}}
\newcommand{\ea}{\end{array}}
\newcommand{\bi}{\begin{itemize}}
\newcommand{\ei}{\end{itemize}}

\begin{document}
\title{Deeply-virtual Compton process $e^- N \to e^- \gamma \pi N$ to study nucleon to resonance transitions}

\author{Kirill M. Semenov-Tian-Shansky}
\affiliation{Department of Physics, Kyungpook National University, Daegu 41566, Korea}
\author{Marc Vanderhaeghen}
\affiliation{Institut f\"ur Kernphysik and $\text{PRISMA}^+$ Cluster of Excellence, Johannes Gutenberg Universit\"at, D-55099 Mainz, Germany}

\date{\today}

\begin{abstract}

We study the deeply-virtual Compton scattering (DVCS) process $e^- N \to e^- \gamma \pi N$ involving the transition between a nucleon and a nucleon resonance in the $\pi N$ system, within the framework of generalized parton distributions (GPDs). For the four lowest-lying nucleon resonances, $\Delta(1232)$, 
$P_{11}(1440)$, $D_{13}(1520)$, and $S_{11}(1535)$,
we express the DVCS amplitude in the Bjorken limit in terms of corresponding nucleon-to-resonance GPDs. 
Building upon the knowledge of the well studied electromagnetic nucleon-to-resonance transition form factors, which map the quark charge densities in transverse position space, the corresponding GPDs will open the prospect to also access the longitudinal momentum distributions of quarks in the transition. We provide estimates for cross sections and beam-spin asymmetries in the first and second $\pi N$ resonance regions in the kinematics of forthcoming CLAS12 data from Jefferson Lab.

\end{abstract}

\maketitle

\section{Introduction}

The framework of Generalized Parton Distributions (GPDs)
\cite{Mueller:1998fv,Radyushkin:1997ki,Ji:1996nm}
was found to be extremely
instructive and convenient both to extract and interpret the information on the dynamics
of hadron constituents from studies of hard exclusive reactions, such as
the Deeply Virtual Compton Scattering
(DVCS) and Deeply Virtual Meson electro-Production (DVMP). These studies (see
Refs.~\cite{Goeke:2001tz,Diehl:2003ny,Belitsky:2005qn,Boffi:2007yc}
for a review) represent a major
step towards understanding of strong interaction phenomena in the strong-coupling regime
and a description of the internal structure of hadrons
in terms of the fundamental degrees of freedom of Quantum Chromodynamics (QCD).

The hadronic structural information encoded in GPDs include femto-photographic ``images'' of
hadrons (and nuclei)
\cite{Ralston:2001xs,Dupre:2016mai,Dupre:2017hfs}
in the transverse plane that can be revealed by the Fourier transform of
GPDs to the impact parameter space
\cite{Burkardt:2000za}.
Moreover, the first Mellin moments of GPDs provide access to the Gravitational Form Factors (GFFs)
of hadrons defined from the hadronic matrix elements of the QCD Energy-Momentum Tensor (EMT). In
particular, Ji's sum rule
\cite{Ji:1996ek}
permits to quantify the fractions of the total hadron's angular momentum carried by quarks of a given
flavor and by gluons. Another important fundamental quantity, which can be obtained from the
hadronic EMT matrix elements, is the $D$-term form factor. It is related to the stress tensor,
that characterizes the spatial distribution of forces
inside a hadronic system
\cite{Polyakov:2002yz,Polyakov:2018zvc}.
First empirical analyses aimed at accessing the pressure distribution inside the proton were recently reported~\cite{Burkert:2018bqq,Kumericki:2019ddg,Dutrieux:2021nlz,Burkert:2021ith}.

Since the early days of the GPD formalism development it was tempting
\cite{Polyakov:1998sz,Frankfurt:1998jq}
to consider {\it non-diagonal} DVCS and DVMP reactions in which, instead of the final state nucleon,
a nucleon-meson system is produced. This final state meson-nucleon system resonates at invariant
mass values corresponding to the production of nucleon resonances.
Studies of these reactions may give access to
$N \to N^*$
and
$N \to \Delta$
transition GPDs
\cite{Frankfurt:1999xe,Goeke:2001tz}.
Also, a possible description in terms of more general
$N \to \pi N$ GPDs
\cite{Polyakov:2006dd}
was sketched in
\cite{Goeke:2001tz}. In particular, it was shown in \cite{Guichon:2003ah} how the $N \to \pi N$ GPDs in the threshold region are related to the nucleon GPDs through soft-pion theorems, allowing to make parameter-free predictions for the $N \to \pi N$ DVCS process in the $\pi N$ threshold region.

The reason to be particularly interested in studies of non-diagonal DVCS and DVMP
reactions and their description within the GPD framework can be seen manyfold.

Firstly, the $N \to N^*$
and
$N \to \Delta$
GPDs describing nucleon-to-resonance transitions can probe the transition
matrix elements of the QCD EMT
\cite{Ozdem:2019pkg,Polyakov:2020rzq,Azizi:2020jog,Ozdem:2022zig}
complementing the studies of electromagnetic transition form factors
\cite{Pascalutsa:2006up,Aznauryan:2011qj,Burkert:2022hjz}. 
While the Fourier transform of the electromagnetic transition form factors map the spatial transverse charge densities of quarks which are active in the excitation of a nucleon to a $N^\ast$ or $\Delta$ resonance, the $N \to N^\ast$ and $N \to \Delta$ GPDs in addition access the longitudinal momentum distributions of the active quarks in this transition. 
The extraction of this information from the experimental data and its detailed physical
interpretation will significantly enhance our understanding of the QCD dynamics
of resonance formation.

Furthermore, the non-diagonal DVCS and DVMP reactions are methodologically very similar to
the electro-excitation of
$N^\ast$
and
$\Delta$
at {\it low} photon virtuality.  All sophisticated partial wave (PW) analysis methods
\cite{Arndt:2006bf}
developed for that case can be adopted. The key difference is that,
as dictated by the QCD factorization theorem
\cite{Ji:1996nm},
the excitation of a nucleon resonance in this case occurs by the quark (and gluon)
QCD string operators:
\begin{eqnarray}
\label{AtoB} 
&&\langle R| \bar \psi (0) [0; \,z] \psi (z) |N\rangle , \nonumber \\
&&\langle R| G_{\alpha \beta}^a(0)\ [0,z]^{ab}\ G_{\mu \nu}^{b}(z) |N\rangle,
\end{eqnarray}
with $R = N^\ast, \Delta$. 
The operators in Eq.~(\ref{AtoB}) are defined on the light-cone $z^2=0$; and $[0;z]$ stand for the
corresponding Wilson lines. 
The major new information, compared to the usual electro- and
photo-production, are the following.
\begin{itemize}
\item{ The relevant operators are non-local, which leads to the possibility to study
the excitation of the resonances with
quark-gluon probes of {\it arbitrary} spin. Particularly, spin $J=2$ production
(graviproduction) of resonances
\cite{Kobzarev:1962wt}
can be addressed.}
\item  They involve {\it explicitly} gluonic degrees of freedom which can not
be accessed directly with electromagnetic or weak probes.
\end{itemize}

The pioneering studies of non-diagonal
$N\to\Delta$ DVCS relying on the large-$N_c$ limit framework
\cite{Frankfurt:1999xe}
for $N \to \Delta$ GPDs
were reported in
Refs.~\cite{Guidal:2003ji,Guichon:2003ah}.
Since the first analysis of the CLAS6 data from Jefferson Lab (JLab) presented in
\cite{Moreno:2009oga},
the non-diagonal DVCS and DVMP have recently seen both
experimental
\cite{SDiehlTrento}
and theoretical
\cite{Kroll:2022roq}
attention.

In this paper we consider the GPD-based description of the deeply-virtual
$e^- N \to e^- \gamma R \to e^- \gamma \pi N$ process in the first $\pi N$-resonance region, corresponding with the $\Delta(1232)$ resonance excitation, as well as in the second $\pi N$-resonance region, corresponding with the excitation of the  
$P_{11}(1440)$, $D_{13}(1520)$, and $S_{11}(1535)$ resonances.  
We construct phenomenological models for $N \to R$ transition GPDs; discuss the relevant experimental observables and present our
estimates  of cross sections and beam spin asymmetries  (BSA) for the kinematical conditions of the JLab@12 GeV experimental setup.

The outline of this paper is as follows. 
In Section~\ref{sec2} we introduce the $e^- N \to e^- \gamma \pi N$ kinematics and give the expression for the cross section. 
In Section~\ref{sec3} we then discuss the 
$e^- N \to e^- \gamma R$ amplitude for a nucleon resonance $R$ in the first and second nucleon resonance region. 
We subsequently discuss the Bethe-Heitler process, in which the outgoing photon is emitted from the electron line, and the deeply-virtual Compton scattering process which originates from 
the Compton subprocess on a quark line. 
In each case, we discuss the amplitude successively for $R = \Delta(1232)$, $P_{11}(1440)$, $D_{13}(1520)$, and $S_{11}(1535)$ resonances. 
With the knowledge of the $e^- N \to e^- \gamma R$ amplitude, we then discuss in Section~\ref{sec4} the $R \to \pi N$ decay to calculate the full amplitude for the $e^- N \to e^- \gamma R \to e^- \gamma \pi N$  process for the nucleon resonances in first and second $\pi N$ resonance regions. In Section~\ref{sec5}, 
we present our results in kinematics of forthcoming data by the CLAS12 experiment at JLab
and give our conclusions and outlook in Section~\ref{sec6}. 

\section{$e^- N \to e^- \gamma \pi N$ kinematics and cross section}
\label{sec2}

In this work we study the hard exclusive process 
\begin{eqnarray}
    e^-(k)+N(p) &\rightarrow& e^-(k^\prime)+ \gamma(q^\prime) + R(p_R) \nonumber \\
    &\rightarrow&  e^-(k^\prime)+ \gamma(q^\prime) + \pi(p_\pi) + N(p^\prime),
    \nonumber \\
    \label{eq:reaction}
\end{eqnarray}
as a probe to access the GPDs for the transition from a nucleon to a nucleon resonance $R$ which decays into the $\pi N$ system.  

\begin{figure}[h]
    \centering
    \includegraphics[width = 8.5cm]{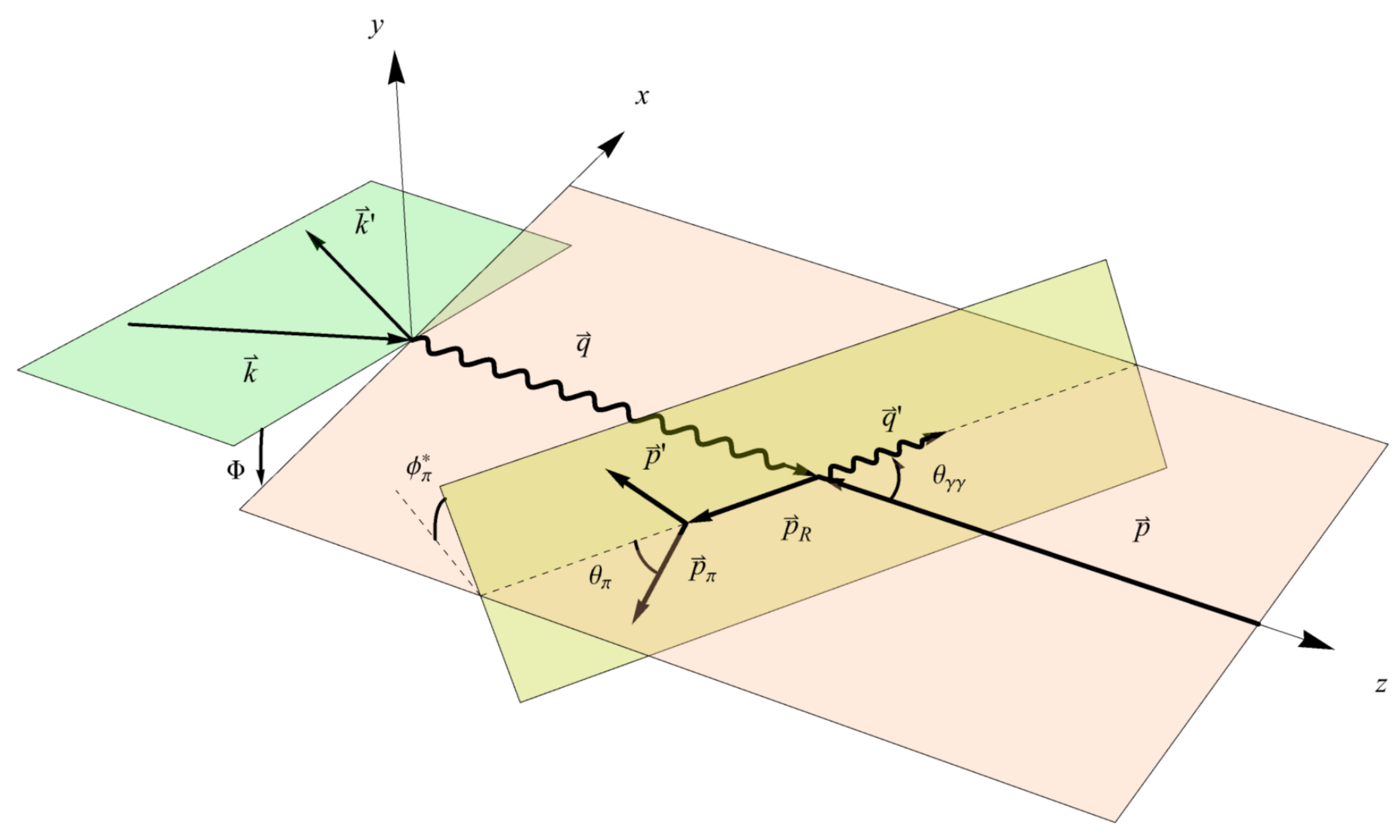}
    \caption{Planes defining the scattering angles which characterize the $ e^- N \to e^- \gamma \pi N$  process. The angles $\Phi$ and $\phi^*_\pi$ are defined with respect to the $xz$-plane, which is the scattering plane of the virtual photons with four-momenta $q$ and $q'$.}
    \label{fig:kin_plane}
\end{figure}

To specify the kinematics, it is useful to introduce the following four-momenta:
\begin{eqnarray}
    q = k - k^\prime, \quad p_R = p_\pi + p^\prime, \quad \Delta = p_R - p.
\end{eqnarray}
The process~(\ref{eq:reaction}) is defined by 
$8$
kinematical invariants, which we choose as:
\begin{eqnarray}
    s&=&(k+p)^2,\quad
    Q^2=-q^2, \quad
    x_B=Q^2/(2 p\cdot q), \nonumber \\
    t&=&\Delta^2, \quad M^2_{\pi N} = p_R^2, \quad \Phi,\quad \theta^*_\pi, \quad \phi^*_\pi,
\end{eqnarray}
where $\Phi$ denotes the angle of the initial electron plane relative to the production plane, spanned by the vectors $\vec q$ and $\vec q^{\, \prime}$, defined in the c.m. of the $\gamma^\ast p$ system ($\vec p + \vec q$ = 0). Furthermore, the angle $\theta^*_\pi$ ($\phi^*_\pi$) denotes the polar (azimuthal) angle respectively of the pion in the rest frame of the $\pi N$ system, relative to the c.m. direction of the momentum $\vec p_R$. 
In Fig.~\ref{fig:kin_plane} we show the three different scattering planes defined by these angles.

In the following analysis it will be convenient to study the decay of the resonance $R$ in the rest frame of the $\pi N$ system,  in which the resonance is at rest.  We will designate those kinematic quantities in the $\pi N$ rest frame by a $^*$.   In that system, the value of the pion three-momentum is given by:
\begin{eqnarray}
| \vec p^{\,*}_\pi | = \frac{1}{2 M_{\pi N}} \lambda^{1/2}(M^2_{\pi N}, M_N^2, m_\pi^2),
\label{eq:pimomstar}
\end{eqnarray}
where $M_N$ ($m_\pi$) denote the nucleon (pion) mass respectively, with $\lambda(x,y,z) \equiv x^2 + y^2 + z^2 - 2 x y - 2 x z - 2 y z$ the K\"all\'en triangle function. 

The 7-fold differential unpolarized cross section for the $ e^- N \to e^- \gamma \pi N$ process is then given  by:
\begin{eqnarray}
\frac{d \sigma}{d Q^2 d x_B d t d \Phi d M^2_{\pi N} d \Omega^*_\pi} 
&&= \frac{1}{(2 \pi)^7} \frac{x_B \, y^2}{32 \, Q^4 \sqrt{ 1 + \frac{4 M_N^2 x_B^2}{Q^2} }}\nonumber \\
&&\hspace{-2.25cm} \times \frac{| \vec p^{\,*}_\pi |}{4 M_{\pi N}} 
 \overline{\sum_i}\sum_f\left|\mathcal{M}(e^- N \to e^- \gamma \pi N) \right|^2, 
    \nonumber \\
\label{eq:cross1}
\end{eqnarray}
with $y \equiv p \cdot q / p \cdot k$, where $\mathcal{M}(e^- N \to e^- \gamma \pi N)$ is the invariant amplitude for the studied process, 
and where the sums correspond with an average (sum) over the initial (final) particle helicities. 

Instead of the pion polar angle  $\theta^*_\pi$, it is equivalent to use the invariant mass of the $\pi \gamma$ system as independent kinematic variable, using the relation:
\begin{eqnarray}
M^2_{\pi \gamma} &\equiv& (p_\pi + q^\prime)^2 \nonumber \\
&=& m_\pi^2 + \frac{(W^2 - M^2_{\pi N})}{2 M^2_{\pi N}} \left[ M^2_{\pi N} + m^2_\pi - M^2_N \right. \nonumber \\
&&\left. \hspace{2cm} + \lambda^{1/2}(M^2_{\pi N}, M_N^2, m_\pi^2) 
\cos \theta^*_\pi \right], \;\;\;
    \nonumber \\
\end{eqnarray}
with squared invariant mass of the $\gamma^\ast N$ system given by:
 \begin{eqnarray}
 W^2 \equiv (q + p)^2 = M_N^2 + Q^2 (1/x_B - 1). 
\end{eqnarray}
The cross section can then equivalently be expressed in terms of $M_{\pi \gamma}$ using the conversion
\begin{eqnarray}
d M^2_{\pi \gamma} = \frac{(W^2 - M^2_{\pi N})}{M_{\pi N}} \, | \vec p^{\,*}_\pi | \, d \cos \theta^*_\pi.
\end{eqnarray}

\section{$e^- N \to e^- \gamma R$ amplitude for $R = \Delta(1232)$, $P_{11}(1440)$, $D_{13}(1520)$, and $S_{11}(1535)$ resonances}
\label{sec3}

In this work we estimate the amplitude of the process of Eq.~(\ref{eq:reaction}) in an isobar model in which the nucleon resonance $R$ is 
produced in a first step, which then decays into the $\pi N$ system. We will consider the transition from a target nucleon to the prominent  
$\Delta(1232)$ resonance, with isospin 3/2 and spin-parity $J^P = \frac{3}{2}^+$, 
as well as the transitions to the nucleon resonances (with isospin 1/2) which are excited in the second resonance region: 
$P_{11}(1440)$ with $J^P = \frac{1}{2}^+$, $D_{13}(1520)$ with $J^P = \frac{3}{2}^-$, 
and $S_{11}(1535)$ with $J^P = \frac{1}{2}^-$. The formalism can be straightforwardly applied to other higher resonances with the same spin-parity assignments. In this Section, we will discuss the $e^- N \to e^- \gamma R$ amplitude, which will then be combined in Section~\ref{sec4} with the $R \to \pi N$ decay to calculate the full amplitude for the $e^- N \to e^- \gamma \pi N$ process. 

\begin{figure}[h]
    \centering
    \includegraphics[width = 3.8cm]{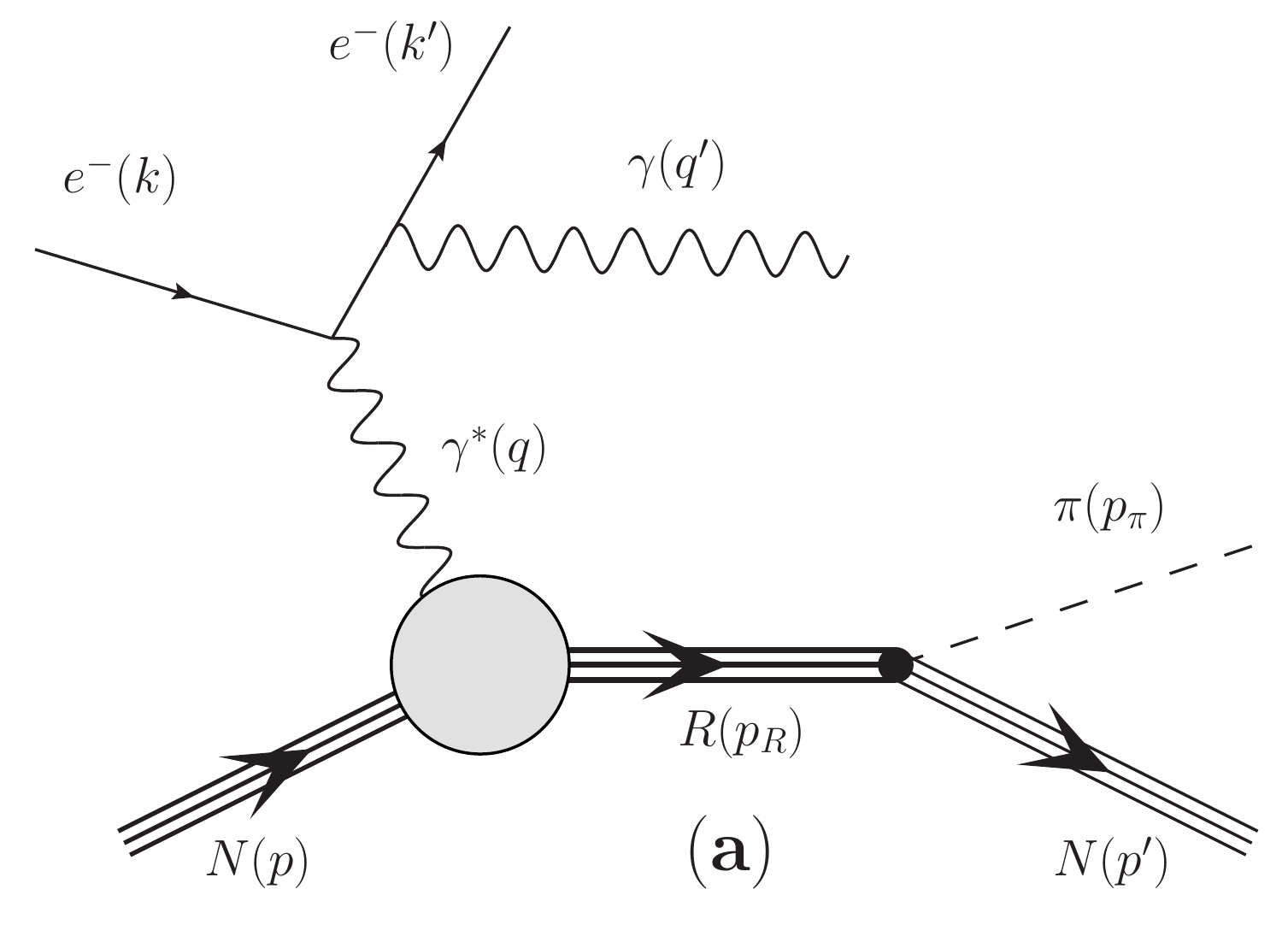}
    \includegraphics[width = 4cm]{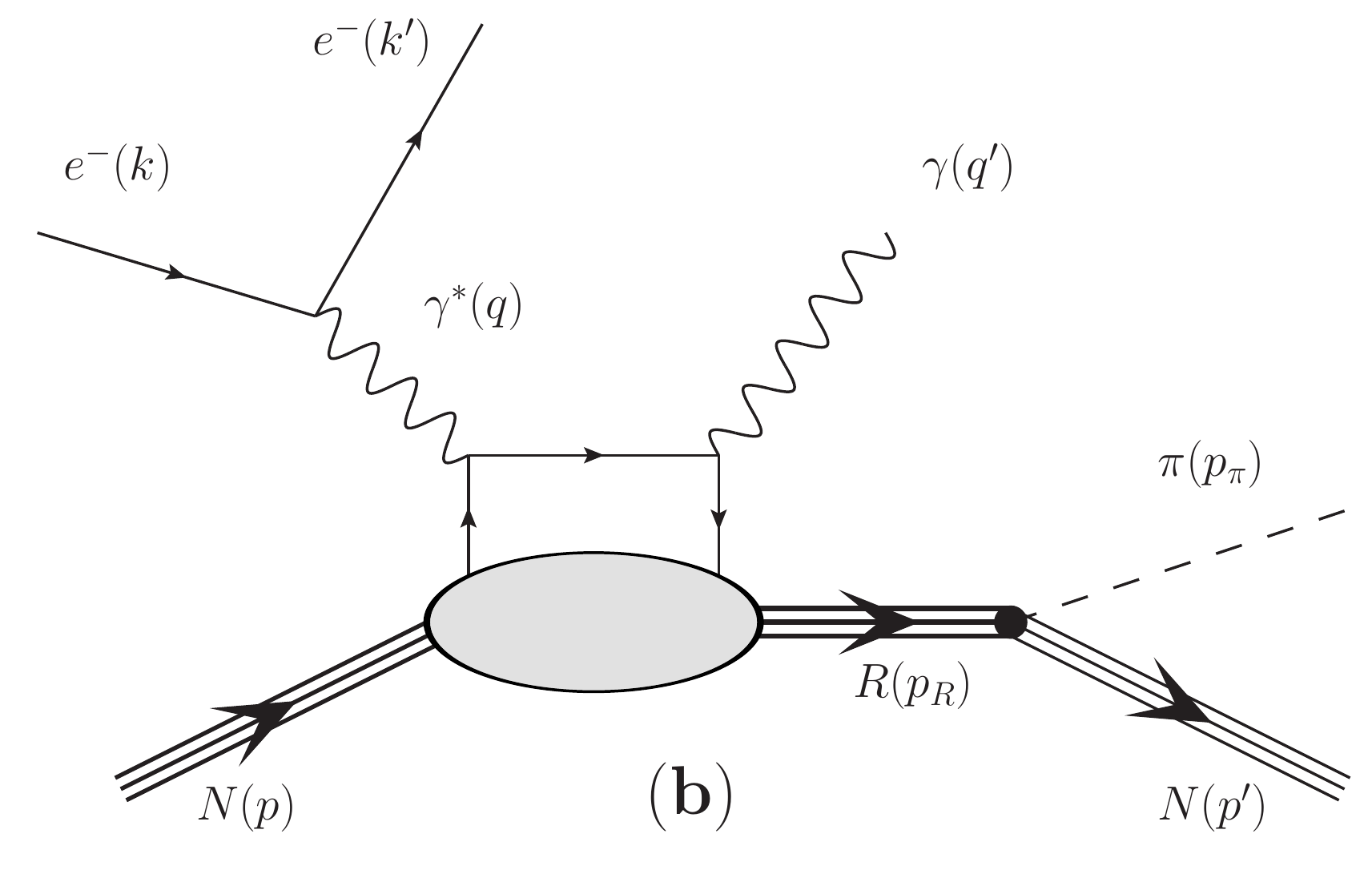}
    \caption{Amplitudes contributing to the $e^- N \to e^- \gamma R \to  e^- \gamma \pi N $ process. Diagram (a) shows the Bethe-Heitler process, Diagram (b) shows the deeply-virtual Compton scattering process. Diagrams in which the real and virtual photon lines are crossed are not shown explicitly.}
    \label{fig:bhdvcs}
\end{figure}

The amplitude for the process $e^- N \to e^- \gamma R$ gets contributions from two main mechanisms as shown in Fig.~\ref{fig:bhdvcs}: 
the Bethe-Heitler process (Fig.~\ref{fig:bhdvcs}a), in which the hard photon is emitted from an electron line, 
and the DVCS process (Fig.~\ref{fig:bhdvcs}b) in which the hard   
photon is emitted from the hadron side.  The Bethe-Heitler process can be calculated exactly given the input on the nucleon to nucleon resonance transition form factors. The DVCS process will be parameterized in the Bjorken limit 
at leading twist-2 level, in terms of transition GPDs. We subsequently discuss both contributions.  

\subsection{$e^- N \to e^- \gamma R$: Bethe-Heitler (BH) process}

The Bethe-Heitler (BH) amplitude for the $e^- N \to e^- \gamma R$ process can in general be expressed as:
\begin{eqnarray}
   && \mathcal{M}_\text{BH}(e^- N \to e^- \gamma R)  \nonumber \\
   &&= \frac{ie^3}{t} 
    \bar{u}\left( k', h' \right)\left[  \gamma^\mu \frac{\slashed{k}' + \slashed{q}' +m}{2 k' \cdot q'} \gamma_\nu 
     \right. \nonumber \\
     && \left. \hspace{2.25cm} + \, \gamma_\nu \frac{\slashed{k} - \slashed{q}' +m}{-2 k \cdot q'}\gamma^\mu\right] u (k, h) 
     \, \varepsilon^\ast_{\mu} (q', \lambda_\gamma)  \nonumber \\
    && \times \,  \langle R(p_R, s_R) | J_\text{em}^\nu (0)  | N(p, s_N) \rangle ,  
        \nonumber \\
    \label{eq:BH} 
\end{eqnarray}
where $u(k,h) (u(k',h') )$ denote the initial (final) electron spinors, with $h$ ($h'$) the corresponding electron helicities, and $m$ the electron mass. 
Furthermore, $\varepsilon_{\mu} (q', \lambda_\gamma) $ denotes the polarization vector of the outgoing photon with helicity $\lambda_\gamma$. 
The unit of electric charge is denoted by $e$ with $e^2 / (4 \pi) \approx 1/137$.  
The BH amplitude depends on the matrix element of the electromagnetic current operator $J_\text{em}^\nu (0)$ between a nucleon with helicity $s_N$  and a nucleon resonance with helicity $s_R$. We subsequently discuss the electromagnetic vertex for the $N \to R$ transition for the four lowest-lying nucleon resonances.

\subsubsection{Electromagnetic vertex  for $N \to \Delta(1232)$ transition}

We start by discussing the electromagnetic transition from the nucleon to its first excited state, the  $\Delta(1232)$ resonance, with 
$J^P = \frac{3}{2}^+$ and 
the Breit-Wigner mass $M_R$ = 1.232 GeV. For this transition, the matrix element appearing in Eq.~(\ref{eq:BH}) is given by:
\begin{eqnarray}
&&\langle R(p_R, s_R) | J_\text{em}^\nu (0)  | N(p, s_N) \rangle \nonumber \\
&&= C_{iso} \, \bar{R}_\beta \left(p_R, s_R \right) 
\Gamma_{\gamma N \Delta}^{\beta \nu}(p_R, p)  N\left(p, s_N \right),
\label{eq:NDELem1}
\end{eqnarray}
where it is standard convention in discussing the $N \to \Delta(1232)$ electromagnetic transition to explicitly separate the isospin factor $C_{iso}$. 
For the $p \to \Delta^+$ transition which we consider in this work, it takes on the value: $C_{iso} = \sqrt{2/3}$. 
Furthermore in Eq.~(\ref{eq:NDELem1}), $R_\beta(p_R, s_R)$ denotes the spin-3/2 Rarita-Schwinger spinor of the produced $\Delta(1232)$ resonance, and $N(p, s_N)$ denotes the initial nucleon Dirac spinor. 
For the vertex $\Gamma_{\gamma N \Delta}^{\beta \nu}$ appearing in Eq.~(\ref{eq:NDELem1}), we follow the parameterization 
given in~\cite{Pascalutsa:2006up}:
\begin{eqnarray}
\Gamma_{\gamma N \Delta}^{\beta \nu} (p_R, p)
&=& g_M(\Delta^2)  \Gamma_M^{\beta \nu}  
+ g_E(\Delta^2)  \Gamma_E^{\beta \nu}  
+  g_C(\Delta^2)  \Gamma_C^{\beta \nu}, \nonumber \\
\label{eq:NDELem2}
\end{eqnarray}
with tensors:
\begin{eqnarray}
\Gamma_M^{\beta \nu} &=& \frac{3}{2}\frac{(M_R +M_N)}{M_N \Delta_+^2} \, 
\left[  i \varepsilon^{\beta\nu\kappa\lambda}(p_R)_\kappa \Delta_\lambda \right], 
\nonumber \\
\Gamma_E^{\beta \nu} &=& \frac{3}{2}\frac{(M_R +M_N)}{M_N \Delta_+^2} \, 
\left[(\Delta \cdot p_R) \,  g^{\beta \nu} -  \Delta^\beta (p_R)^\nu \right] \,  \gamma_5, 
\nonumber \\
\Gamma_C^{\beta \nu} &=& \frac{3}{2}\frac{(M_R +M_N)}{M_N M_R \Delta_+^2} \, 
\left[- \gamma \cdot p_R \, \left( \Delta^\beta \Delta^\nu- \Delta^2 g^{\beta \nu} \right) \right. \nonumber \\
&&\left. \hspace{1.5cm} + \, \gamma^\beta \left( \Delta \cdot p_R \, \Delta^\nu- \Delta^2 (p_R)^\nu \right) \right] \, \gamma_5,
    \nonumber \\
\label{eq:NDELem3}
\end{eqnarray}
where $\Delta_\pm \equiv \sqrt{(M_R \pm M_N)^2 - \Delta^2}$, and where we adopt the convention $\varepsilon_{0123} = +1$.  
The vertices defined in Eq.~(\ref{eq:NDELem3}) satisfy the property:
\begin{eqnarray}
(p_R)_\beta \, \Gamma_{M, E, C}^{\beta \nu} = 0,
\end{eqnarray} 
which define them as ``consistent" couplings~\cite{Pascalutsa:2006up}, decoupling the unphysical spin-1/2 degrees of freedom present in the Rarita-Schwinger field. 
The corresponding form factors (FFs) $g_M$, $g_E$, and $g_C$ appearing in Eq.~(\ref{eq:NDELem2}) have spacelike virtuality. 
It is conventional to express these FFs 
at the resonance position, i.e. for $p_R^2 = M_R^2$, in terms of the magnetic dipole ($G_M^*$), electric quadrupole ($G_E^*$), and Coulomb quadrupole ($G_C^*$) transition FFs as:
\begin{align}
    g_M&=\frac{\Delta_+}{M_N+M_R}(G_M^*-G_E^*),\nonumber\\
    g_E&=-\frac{\Delta_+}{M_N+M_R}\frac{2}{\Delta_-^2}\{(M_\Delta^2-M_N^2+ \Delta^2)G_E^*- \Delta^2G_C^*\},\nonumber\\
    g_C&=\frac{\Delta_+}{M_N+M_R}\frac{1}{\Delta_-^2}\{4M_\Delta^2G_E^*-(M_R^2-M_N^2+ \Delta^2)G_C^*\},
    \label{eq:gMgEgC_related_to_Ash}
\end{align}
with the so-called Ash FFs parameterized~
\footnote{We are using the Ash parmeterization for the $N \to \Delta$ electromagnetic FFs in this work. Another often used parameterization was given in Ref.~\cite{Jones:1972ky}. Using the Jones-Scadron parameterization for the multipole FFs $G_{M,E,C}^*$ amounts to dropping the common factor $\Delta_+ / (M_N+M_R)$ in Eq.~(\ref{eq:gMgEgC_related_to_Ash}).}, for spacelike virtuality $\Delta^2 = t < 0$, through the MAID2007 analysis~\cite{Drechsel:2007if,Tiator:2011pw}:
\begin{eqnarray}
    G^{*}_{M}(t) &=& 3.00 \, (1- 0.01 \, t)e^{0.23 t}G_D(t),\nonumber\\
    G^{*}_{E}(t) &=& 0.064 \, (1+ 0.021 \, t)e^{0.16 t}G_D(t),\nonumber\\
    G^{*}_{C}(t) &=& 0.124 \,  \frac{1- 0.12 \, t}{1 - 4.9 \frac{t}{4M_N^2}} \frac{4M_R^2}{M_R^2-M_N^2} e^{0.23 t}G_D(t), \;
    \nonumber \\
    \label{eq:GMGEGC}
\end{eqnarray}
with $t$ in GeV$^2$ and the dipole FF $G_D(t) \equiv 1/(1 - t/0.71)^2$. 

\subsubsection{Electromagnetic vertex  for $N \to P_{11}(1440)$ transition}

We next discuss the electromagnetic transition from the nucleon to the $P_{11}(1440)$ resonance, with 
$J^P = \frac{1}{2}^+$ and 
the Breit-Wigner mass $M_R$ = 1.440 GeV. For this transition, the matrix element appearing in Eq.~(\ref{eq:BH}) is given by:
\begin{eqnarray}
&&\langle R(p_R, s_R) | J_\text{em}^\nu (0) | N(p, s_N) \rangle \nonumber \\
&&= \bar{R} \left(p_R, s_R \right) 
\Gamma_{\gamma N P_{11}}^{\nu}(p_R, p)  N\left(p, s_N \right), 
\label{eq:NP11em1}
\end{eqnarray}
where $R(p_R,s_R)$ denotes the Dirac spinor for the $P_{11}(1440)$ resonance.  
The vertex $\Gamma_{\gamma N P_{11}}^{\nu}$ is parameterized as~\cite{Tiator:2008kd}:
\begin{eqnarray}
\Gamma_{\gamma N P_{11}}^{\nu} (p_R, p)
&=& F_1^{NP_{11}} (\Delta^2) \, \left[ \gamma^\nu -  \frac{(\gamma \cdot \Delta) \Delta^\nu}{\Delta^2}  \right] \nonumber\\
&+& F_2^{NP_{11}} (\Delta^2) \, \frac{i \, \sigma^{\nu \kappa} \Delta_\kappa}{(M_R + M_N)}, 
\label{eq:NP11em}
\end{eqnarray}
with $F_1^{NP_{11}}$ and $F_2^{NP_{11}}$ the transition FFs, for which we use 
the empirical MAID2008 analysis for the proton and the MAID2007 analysis for the neutron, as detailed in Ref.~\cite{Tiator:2011pw}.

\subsubsection{Electromagnetic vertex  for $N \to D_{13}(1520)$ transition}

For the electromagnetic transition from the nucleon to the $D_{13}(1520)$ resonance, with 
$J^P = \frac{3}{2}^-$ and the Breit-Wigner mass $M_R$ = 1.520 GeV, the matrix element appearing in Eq.~(\ref{eq:BH}) is given by:
\begin{eqnarray}
&&\langle R(p_R, s_R) | J_\text{em}^\nu (0) | N(p, s_N) \rangle \nonumber \\
&&= \bar{R}_\beta \left(p_R, s_R \right) 
\Gamma_{\gamma N D_{13}}^{\beta \nu}(p_R, p)  N\left(p, s_N \right),
\label{eq:ND13em1}
\end{eqnarray}
where $R_\beta(p_R, s_R)$ denotes the spin-3/2 Rarita-Schwinger spinor of the produced $D_{13}(1520)$ resonance. 
The vertex $\Gamma_{\gamma N D_{13}}^{\beta \nu}$ is parametrized as:
\begin{eqnarray}
\Gamma_{\gamma N D_{13}}^{\beta \nu} (p_R, p) &=& 
F_1^{ND_{13}} (\Delta^2) \, \tilde \Gamma_{1}^{\beta \nu} 
+ F_2^{ND_{13}} (\Delta^2) \, \tilde \Gamma_{2}^{\beta \nu}  \nonumber \\
&+& F_3^{ND_{13}} (\Delta^2) \, \tilde \Gamma_{3}^{\beta \nu}, 
    \nonumber \\
\label{eq:ND13em2} 
\end{eqnarray}
with tensors:
\begin{eqnarray}
\tilde \Gamma_{1}^{\beta \nu}  &=&  \frac{M_R}{\Delta_+^2}  \, \bigg\{ 
\left[ \Delta^\beta \gamma^\nu - (\gamma \cdot \Delta) \, g^{\beta \nu} \right]  \nonumber\\
& - &  \left[ (\Delta \cdot p_R) \gamma^\nu - (\gamma \cdot \Delta) \, (p_R)^\nu  \right]  \frac{(p_R)^\beta}{p_R^2} \bigg\}, \nonumber \\
\tilde \Gamma_{2}^{\beta \nu} &=& \frac{1}{\Delta_+^2}  \left[ \Delta^\beta (p_R)^\nu - (\Delta \cdot p_R) \,  g^{\beta \nu} \right], \nonumber\\
\tilde \Gamma_{3}^{\beta \nu} &=& \frac{1}{\Delta_+^2}  \bigg\{ \left[\Delta^\beta \Delta^\nu- \Delta^2 \, g^{\beta \nu} \right] \nonumber \\
&-& \left[ (\Delta \cdot p_R) \Delta^\nu - \Delta^2 \, (p_R)^\nu \right] 
\frac{(p_R)^\beta}{p_R^2} \bigg\},    
    \nonumber \\
\label{eq:ND13em3}
\end{eqnarray}
where we have included terms proportional to $(p_R)_\beta$  to make the spin-3/2 couplings consistent, i.e. satisfying  
$(p_R)_\beta \, \tilde \Gamma_{1, 2, 3}^{\beta \nu} = 0$, analogously to the discussion for the $\gamma N \Delta$ couplings. 
For the transition FFs $F_{1,2,3}^{ND_{13}}$ appearing in Eq.~(\ref{eq:ND13em2}), we use the empirical MAID2008 analysis for the proton and the MAID2007 analysis for the neutron, as detailed in Ref.~\cite{Tiator:2011pw}.

\subsubsection{Electromagnetic vertex  for $N \to S_{11}(1535)$ transition}

For the electromagnetic transition from the nucleon to the $S_{11}(1535)$ resonance, with 
$J^P = \frac{1}{2}^-$ and 
the Breit-Wigner mass $M_R$ = 1.535 GeV, the matrix element appearing in Eq.~(\ref{eq:BH}) is given by:
\begin{eqnarray}
&&\langle R(p_R, s_R) | J_\text{em}^\nu (0) | N(p, s_N) \rangle \nonumber \\
&&= \bar{R} \left(p_R, s_R \right) 
\Gamma_{\gamma N S_{11}}^{\nu}(p_R, p)  N\left(p, s_N \right),
    \nonumber \\
\label{eq:NS11em1}
\end{eqnarray}
where $R(p_R,s_R)$ denotes the Dirac spinor for the $S_{11}(1535)$ resonance.  
The vertex $\Gamma_{\gamma N S_{11}}^{\nu}$ is parameterized as:
\begin{eqnarray}
\Gamma_{\gamma N S_{11}}^{\nu} (p_R, p)
&=& \bigg\{ F_1^{NS_{11}} (\Delta^2) \, \left[ \gamma^\nu -  \frac{(\gamma \cdot \Delta) \Delta^\nu}{\Delta^2}  \right] \nonumber\\
&+& \, F_2^{NS_{11}} (\Delta^2) \, \frac{i \, \sigma^{\nu \kappa} \Delta_\kappa}{(M_R + M_N)} \bigg\} \gamma_5, 
    \nonumber \\
\label{eq:NS11em}
\end{eqnarray}
with $F_1^{NS_{11}}$ and $F_2^{NS_{11}}$ the transition FFs, for which we use 
the empirical MAID2008 analysis for the proton and the MAID2007 analysis for the neutron, as detailed in Ref.~\cite{Tiator:2011pw}.

\subsection{$e^- N \to e^- \gamma R$: deeply-virtual Compton scattering (DVCS) process and generalized parton distributions (GPDs)}

\begin{figure}[h]
\centering
  \hspace{-0.9cm}\includegraphics[scale=0.55]{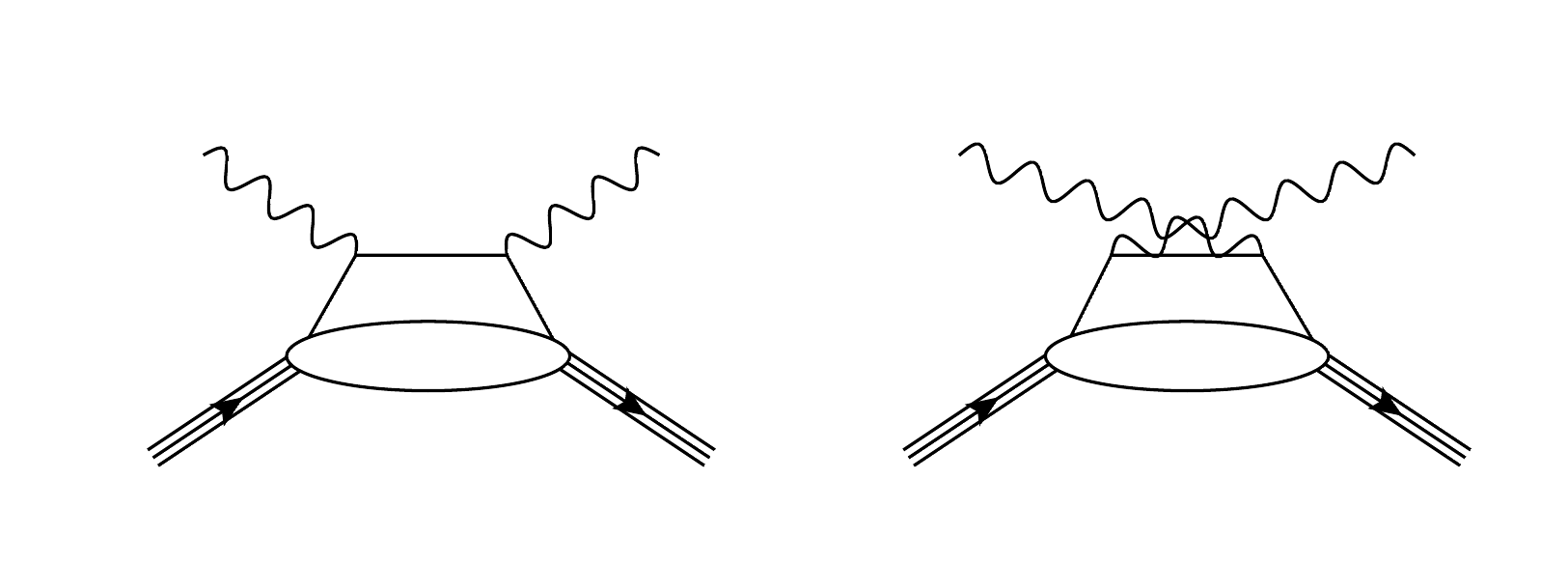}
  \caption{Handbag diagrams for the deeply virtual Compton amplitude. The 
  single (composite) lines represent quarks (baryons) respectively. The blobs represent the GPDs. 
  \label{dia:handbag}}
\end{figure}

The deeply-virtual Compton scattering (DVCS) amplitude for the $e^- N \to e^- \gamma R$ process can in general be expressed as:
\begin{eqnarray}
    \mathcal{M}_\text{DVCS}(e^- N \to e^- \gamma R) 
  &=& \frac{ie^3 q_l}{q^2} \, 
    \bar{u}\left( k', h' \right)  \gamma_\nu  u (k, h)  \, \nonumber \\
  &\times &   \varepsilon^\ast_{\mu} (q', \lambda_\gamma) \,   H^{\mu \nu}_{NR},
     \nonumber \\
   \label{eq:DVCS}
\end{eqnarray}
with lepton charge $q_l = -1$ for an electron beam, and 
where the virtual Compton tensor $H^{\mu \nu}_{NR}$ describing the $\gamma^\ast (q) + N (p_N) \to \gamma (q') + R (p_R)$ transition is defined as:
\begin{eqnarray}
H^{\mu \nu}_{NR} &=& -i \int d^4 y \, e^{-i q y} \nonumber \\
&\times& \langle R(p_R, s_R) | T\left[ J_\text{em}^\mu (0) J_\text{em}^\nu (y) \right] | N(p, s_N) \rangle. 
\nonumber \\
\label{eq:DVCStensor}
\end{eqnarray}
In this work, we will consider the Bjorken limit, $Q^2 \to \infty$ and $x_B$ constant. In this limit, the leading (twist-2) DVCS amplitude can be calculated from the handbag  diagrams shown in Fig. \ref{dia:handbag}. 
For the evaluation of these diagrams, one introduces the two light-like vectors $\tilde p^\mu$ and $n^\mu$ with $\tilde p \cdot n=1$, which are related to the four-momenta $\bar P^\mu \equiv (p + p_R)/2$ and $q^\mu$ as:
\begin{eqnarray}
\bar P^\mu &=& \tilde p^\mu + \frac{\bar P^2}{2} n^\mu, \label{eq:P}\\
q^\mu &=& (- 2 \xi^\prime) \, \tilde p^\mu + \frac{Q^2}{4 \xi^\prime} n^\mu \label{eq:q}, 
\end{eqnarray}
where $\bar P^2 = \frac{M_R^2 + M_N^2}{2} - \frac{t}{4}$, and 
\begin{eqnarray}
    2 \xi^\prime&=&\frac{q \cdot \bar P}{\bar P^2} \, \left[-1+\sqrt{1 + \frac{ Q^2 \bar P ^2}{(q \cdot \bar P)^2} } \right].
    \label{eq:xip}
\end{eqnarray}
Furthermore, the momentum transfer $\Delta$ between the initial nucleon and the nucleon resonance can be expressed as:
\begin{eqnarray}
\Delta^\mu &=& (- 2 \xi) \, \tilde p^\mu + \left[ \xi \bar P^2 + \frac{(M_R^2 - M_N^2)}{2} \right] n^\mu + \Delta^\mu_\perp, \quad
\nonumber \\
\label{eq:delta}
\end{eqnarray}
with 
\begin{eqnarray}
    2 \xi&=& 2 \xi^\prime \, \left( \frac{Q^2 - t -  2 \xi^\prime (M_R^2 - M_N^2)}{Q^2 +  (2 \xi^\prime)^2 {\bar P}^2} \right),
    \label{eq:xi}
\end{eqnarray}
and where $\Delta_\perp$ is the perpendicular component of the momentum transfer, i.e. $\tilde p \cdot \Delta_\perp = n \cdot \Delta_\perp = 0$.

In the Bjorken limit:
\begin{eqnarray}
\xi \approx \xi^\prime \to \frac{x_B / 2}{1 - x_B / 2}.
 \end{eqnarray}
In this limit, the leading (twist-2) DVCS tensor  $H^{\mu \nu}_{NR}$ of Eq.~(\ref{eq:DVCStensor}) 
for the $N \to R$ transition can be expressed as:
\begin{widetext}
\begin{eqnarray}
H^{\mu\nu}_{NR} &=& \frac{1}{2}(-g^{\mu\nu}_{\perp}) \int_{-1}^{1}dx
 \left[ \frac{1}{x - \xi + i \epsilon} + \frac{1}{x + \xi - i \epsilon}  \right] 
\int \frac{d \lambda}{2 \pi} e^{i \lambda x} \sum_{q}  e_q^2
\langle R(p_R, s_R) | \bar q \left(- \frac{\lambda n}{2} \right) \gamma \cdot n \, q \left(\frac{\lambda n}{2} \right)  | N(p, s_N) \rangle
\nonumber\\
&+&\frac{i}{2}(\varepsilon^{\mu\nu}_{\perp})\int_{-1}^{1}dx
\left[ \frac{1}{x - \xi + i \epsilon} - \frac{1}{x + \xi - i \epsilon}  \right]  
 \int \frac{d \lambda}{2 \pi} e^{i \lambda x} \sum_{q}  e_q^2 
\langle R(p_R, s_R) | \bar q \left(- \frac{\lambda n}{2} \right) \gamma \cdot n \gamma_5 \, q \left(\frac{\lambda n}{2} \right)  | N(p, s_N) \rangle, \nonumber \\
\label{eq:DVCStensor2}
\end{eqnarray}
\end{widetext}
where the sum goes over the quarks $q = u, d, s$ present in the $N$ and $R$ states. 
Furthermore, in Eq.~(\ref{eq:DVCStensor2}) we have defined:
\begin{align}
    -g^{\mu\nu}_{\perp}&=-g^{\mu\nu}+\tilde{p}^\mu n^\nu+\tilde{p}^\nu n^\mu,\nonumber\\
    \varepsilon^{\mu\nu}_\perp&=\varepsilon^{\mu\nu\alpha\beta}n_\alpha \tilde{p}_\beta,
\end{align}
where the lightlike four-vectors $\tilde p$ and $n$ are obtained from Eqs.~(\ref{eq:P},\ref{eq:q}).  

In the Bjorken limit, the DVCS tensor given by 
Eq.~(\ref{eq:DVCStensor2}) is expressed as a convolution integral over the average quark momentum fraction $x$ 
(defined as the average in the initial nucleon and the final nucleon resonance) of a perturbatively calculable hard kernel function representing the quark propagators between both photon vertices in Fig.~\ref{dia:handbag}, and a one-dimensional Fourier integral. The latter involves a quark bilinear operator along the light-cone direction $n^\mu$ sandwiched between initial nucleon state and final nucleon resonance state. 
This non-perturbative matrix element can in general be expressed in terms of generalized parton distributions (GPDs). In the following, we will parameterize this matrix element for the $N \to R$ transition for the four lowest-lying nucleon resonances. 

\subsubsection{GPDs for $N \to \Delta(1232)$ transition}

For the $N \to \Delta(1232)$ transition, the matrix element of the vector bilinear quark operator along the light-cone can be parameterized as: 
\begin{widetext}
\begin{eqnarray}
&&\int \frac{d \lambda}{2 \pi} e^{i \lambda x} \sum_{q}  e_q^2
\langle R(p_R, s_R) | \bar q \left(- \frac{\lambda n}{2} \right) \gamma \cdot n \, q \left(\frac{\lambda n}{2} \right)  | N(p, s_N) \rangle 
\nonumber  \\
&&= \left(\frac{C_{iso}}{6} \right) \bar{R}_\beta \left(p_R, s_R \right)  
 \bigg\{ h_M(x, \xi, \Delta^2) \, \Gamma_M^{\beta \nu}  
+ h_E(x, \xi, \Delta^2) \, \Gamma_E^{\beta \nu}  
+  \, h_C(x, \xi, \Delta^2) \, \Gamma_C^{\beta \nu} 
+  \, h_4(x, \xi, \Delta^2) \, \Gamma_4^{\beta \nu}   \bigg\} \, n_\nu \,  N\left(p, s_N \right), 
\nonumber \\
\label{eq:NDELgpdunpol}
\end{eqnarray}
\end{widetext}
with tensors $ \Gamma_M^{\beta \nu}$,  $ \Gamma_E^{\beta \nu}$, and  $ \Gamma_C^{\beta \nu}$ defined as in Eq.~(\ref{eq:NDELem3}), and 
where we defined the tensor $\Gamma_4^{\beta \nu}$ as:
\begin{eqnarray}
\Gamma_4^{\beta \nu} &=& \frac{1}{M_N M_R} \, 
\left[ \Delta^\beta - \frac{\Delta \cdot p_R}{p_R^2} \,  (p_R)^\beta \right] \, \Delta^\nu \, \gamma_5,
\label{eq:NDEL4}
\end{eqnarray}
which also satisfies the spin-3/2 consistency condition $(p_R)_\beta \Gamma_4^{\beta \nu} = 0$.
In Eq.~(\ref{eq:NDELgpdunpol}), we conventionally separated the isospin factor and the quadratic quark charge factor (1/6) for the vector $N \to \Delta(1232)$ transition in defining the unpolarized transition GPDs $h_M, h_E, h_C$ and $h_4$.   
Using this convention, the first moments of the unpolarized GPDs are then related to the transition FFs defined in 
Eqs.~(\ref{eq:NDELem1},\ref{eq:NDELem2}) as:
\begin{eqnarray} 
\int_{-1}^{+1} dx \, h_{M, E, C}(x, \xi, \Delta^2) &=& 2 \, g_{M, E, C} (\Delta^2), \label{eq:NDELgpdsr1}  
\end{eqnarray}
where the factor $2$ appears on the {\it rhs} of Eq.~(\ref{eq:NDELgpdsr1}) due to the definition of the isovector current operator 
$(\bar u \gamma^\nu u - \bar d \gamma^\nu d)/2$.  
Furthermore,  the first moment of the GPD $h_4$ vanishes: 
\begin{eqnarray} 
\int_{-1}^{+1} dx \, h_{4}(x, \xi, \Delta^2) &=& 0, 
\label{eq:NDELgpdsr2} 
\end{eqnarray}
as the corresponding term is absent in the matrix element of the local operator due to electromagnetic gauge invariance. 
As the first moment of the GPDs $h_E$ and $h_C$ are directly proportional to the electric quadrupole and Coulomb quadrupole FFs which at small momentum transfer are in the few percent range, see Eq.~(\ref{eq:GMGEGC}), we will in the following numerical analysis neglect the contribution of the GPDs $h_E$, $h_C$, as well as $h_4$ when evaluating the $N \to \Delta$ DVCS amplitude. 

For the $N \to \Delta(1232)$ transition, the matrix element of the axial-vector bilinear quark operator along the light-cone can be 
parameterized as: 
\begin{widetext}
\begin{eqnarray}
&&\int \frac{d \lambda}{2 \pi} e^{i \lambda x} \sum_{q}  e_q^2
\langle R(p_R, s_R) | \bar q \left(- \frac{\lambda n}{2} \right) \gamma \cdot n \, \gamma_5 \, q \left(\frac{\lambda n}{2} \right)  | N(p, s_N) \rangle 
\nonumber \\
&&= \frac{1}{6}  \bar{R}_\beta \left(p_R, s_R \right)  
 \bigg\{ C_1(x, \xi, \Delta^2) \, n^\beta 
+ C_2(x, \xi, \Delta^2) \, \left( \frac{\Delta \cdot n}{M_N^2} \right) \Delta^\beta \nonumber \\
&&\hspace{2.4cm} +  \, C_3(x, \xi, \Delta^2) \, \frac{1}{M_N} \left ( n^\beta \gamma \cdot \Delta - \Delta^\beta \gamma \cdot n \right)
+  \, C_4(x, \xi, \Delta^2) \, \frac{2}{M_N^2} \left( n^\beta \, \bar P \cdot \Delta - \Delta^\beta  \right)  \bigg\} \,   N\left(p, s_N \right). 
\nonumber \\
\label{eq:NDELgpdpol}
\end{eqnarray}
\end{widetext}
The first moments of the polarized GPDs $C_i$ (for $i = 1,..., 4$) are constrained through the FFs parameterizing the 
matrix element of the isovector axial current operator for the $N \to \Delta$ transition. 
The third component of the isovector axial current $ A_3^\nu$ is defined as:
\begin{eqnarray}
A_3^\nu(0) \equiv \frac{1}{2} \left[ \bar u(0) \gamma^\nu \gamma_5 u(0) - \bar d(0) \gamma^\nu \gamma_5 d(0) \right].
\end{eqnarray}
Following Adler~\cite{Adler:1968tw,Adler:1975tm},  
the matrix element of $A_3^\nu$ for the $N \to \Delta$ transition is then given by:
\begin{eqnarray}
&&\langle R(p_R, s_R) | A_3^\nu (0) | N(p, s_N) \rangle \nonumber \\
&&= \bar{R}_\beta \left(p_R, s_R \right)  
 \bigg\{ C_5^A(\Delta^2) \, g^{\beta \nu} 
+ \, C_6^A(\Delta^2) \, \frac{\Delta^\beta \Delta^\nu }{M_N^2}  \nonumber \\
&& + \, C_3^A(\Delta^2) \, \frac{1}{M_N} \left( g^{\beta \nu} \gamma \cdot \Delta - \Delta^\beta \gamma^\nu \right) \nonumber \\
&&+ \, C_4^A(\Delta^2) \, \frac{2}{M_N^2} \left( g^{\beta \nu} \, \bar P \cdot \Delta - \Delta^\beta \bar P^\nu \right)  \bigg\} \,   N\left(p, s_N \right). \quad
\nonumber \\
\label{eq:NDELax}
\end{eqnarray}
The relations between the first moments of the polarized GPDs defined in Eq.~(\ref{eq:NDELgpdpol}) and the axial transitions FFs $C^A_{5, 6, 3, 4}$ defined in Eq.~(\ref{eq:NDELax}) are then obtained as:
\begin{eqnarray} 
\int_{-1}^{+1} dx \, C_{1, 2, 3, 4}(x, \xi, \Delta^2) &=& 2 \, C^A_{5, 6, 3, 4} (\Delta^2).
\end{eqnarray}
For the determination of the dominant $N \to R$ axial FFs in this work, we will use PCAC relations, which relate:
\begin{eqnarray}
\partial_\nu \, A^\nu_3 = -f_\pi m_\pi^2 \, \Pi_3,
\label{eq:pcac}
\end{eqnarray}
with pion decay constant $f_\pi \simeq 92.4$~MeV, and where $\Pi_3$ stands for the $\pi^0$ field. 
Using the expression of Eq.~(\ref{eq:NDELax}), the PCAC relation of Eq.~(\ref{eq:pcac}) yields for the $p \to \Delta^+$ matrix element:
\begin{eqnarray}
&&i \,   \bar{R}_\beta \left(p_R, s_R \right)  
 \bigg\{ C_5^A(\Delta^2) 
+ C_6^A(\Delta^2)  \frac{ \Delta^2}{M_N^2} \bigg\} \Delta^{\beta}    N\left(p, s_N \right) \nonumber \\
&&= -f_\pi m_\pi^2 \langle \Delta^+(p_R, s_R) | \Pi_3 (0) | p(p, s_N) \rangle .
\label{eq:pcacNDEL}
\end{eqnarray}
The latter matrix element is obtained from the $\pi N \Delta$ coupling, which is given by the effective Lagrangian~\cite{Pascalutsa:2006up}:
\begin{eqnarray}
{\cal L}_{\pi N \Delta} = i \left( \frac{f_{\pi N \Delta}}{m_\pi M_R} \right) \left( \partial_\nu \bar R_\beta \right) \gamma^{\mu \nu \beta} T_i^\dagger N \left( \partial_\mu \Pi_i \right) + \mathrm{h.c.} , \nonumber \\
\label{eq:LpiNDel}
\end{eqnarray}
with $\pi N \Delta$ coupling constant $f_{\pi N \Delta} \simeq 2.08$ determined from the $\Delta \to \pi N$ decay width, 
where $T_i^\dagger$ (for $i = 1, 2, 3$) is the isospin $\frac{1}{2} \to \frac{3}{2}$ transition operator, and where we follow the 
notations of Ref.~\cite{Pascalutsa:2006up}: $\gamma^{\mu \nu \beta} \equiv  \frac{1}{2} \left\{\gamma^{\mu \nu}, \gamma^\beta \right\}$, with  
  $\gamma^{\mu \nu} \equiv  \frac{1}{2} \left[ \gamma^\mu, \gamma^\nu \right]$. 
  From the Lagrangian (\ref{eq:LpiNDel}) we can extract:
\begin{eqnarray}
&&  \langle \Delta^+(p_R, s_R) | \Pi_3 (0) | p(p, s_N) \rangle   \nonumber \\
&& =  \sqrt{\frac{2}{3}} \left( \frac{f_{\pi N \Delta}}{m_\pi} \right) \frac{-i }{-\Delta^2 + m_\pi^2}  \bar{R}_\beta \left(p_R, s_R \right)  \Delta^{\beta}    N\left(p, s_N \right).  \nonumber \\
\label{eq:LpiNDel2}
\end{eqnarray}
Combining Eqs.~(\ref{eq:pcacNDEL}) and (\ref{eq:LpiNDel2}) allows to extract at $\Delta^2 = 0$ a generalized Goldberger-Treiman relation as:
\begin{eqnarray}
C_5^A(0) = \sqrt{\frac{2}{3}} \left( \frac{f_{\pi N \Delta}}{m_\pi} \right) f_\pi.  
\label{eq:GTNDel}
\end{eqnarray} 
Using the phenomenological values on the {\it rhs} as input, we obtain $C_5^A(0) \simeq 1.16$. 
Furthermore, using the pion-pole dominance for the axial FF $C_6^A$ at small values of $\Delta^2$, we obtain from 
Eqs.~(\ref{eq:pcacNDEL}) and (\ref{eq:LpiNDel2}):
\begin{eqnarray}
C_6^A(\Delta^2) \approx\frac{M_N^2}{- \Delta^2 + m_\pi^2}  C_5^A(\Delta^2).  
\end{eqnarray} 
For the two remaining Adler form factors $C_{3}^{A} $ and $ C_{4}^{A}$,
a detailed comparison with experimental data for neutrino-induced
$\Delta ^{++}$ production led to the values \cite{Kitagaki:1990vs}~:
\begin{eqnarray}
C_{3}^{A}(0)\simeq 0.0, \quad 
C_{4}^{A}(0)\simeq -0.3 .
\end{eqnarray} 
Given the smallness of these values, we will neglect in the following
numerical analysis the contributions from the GPDs $ C_{3} $ and $ C_{4} $.

To provide estimates for the $ N \to \Delta  $
DVCS amplitudes, we need a model for the three dominant GPDs
which appear in Eqs.~(\ref{eq:NDELgpdunpol},\ref{eq:NDELgpdpol}) i.e. $ h_{M}$,
$C_{1}$ and $C_{2}$. Here we will be guided by the large-$ N_{c} $ relations discussed in~\cite{Frankfurt:1999xe,Goeke:2001tz}, which 
connect the $ N \to \Delta  $ GPDs $ h_{M}$, $ C_{1} $,
and $ C_{2} $ to the nucleon isovector GPDs $ E^{u}-E^{d} $,
$ \tilde{H}^{u}-\tilde{H}^{d}$, and $\tilde{E}^{u}-\tilde{E}^{d} $
respectively, as:
\begin{eqnarray}
h_{M}(x,\xi ,\Delta^2) & = & 
\sqrt{2} \left[ E^{u}(x,\xi ,\Delta^2)
-E^{d}(x,\xi ,\Delta^2)\right], \quad 
\label{eq:modelhM} 
\end{eqnarray}
\begin{eqnarray}
C_{1}(x,\xi ,\Delta^2) & = & 
\sqrt{3}\left[ \tilde{H}^{u}(x,\xi ,\Delta^2)
-\tilde{H}^{d}(x,\xi ,\Delta^2)\right], \quad
\label{eq:modelC1} \\
C_{2}(x,\xi ,\Delta^2) & = & 
\frac{\sqrt{3}}{4}\left[ \tilde{E}^{u}(x,\xi ,\Delta^2)
-\tilde{E}^{d}(x,\xi ,\Delta^2)\right]. \quad
\label{eq:modelC2} 
\end{eqnarray}
To provide an estimate of the accuracy of these relations, we can calculate the first moments of both sides of Eqs.~(\ref{eq:modelhM},\ref{eq:modelC1},\ref{eq:modelC2}), and compare their values at $\Delta^2 = 0$. 

For the unpolarized GPD $h_M$, the large-$N_c$ prediction for its normalization at $\Delta^2 = 0$ yields 
$G_M^\ast(0) = \kappa_V / \sqrt{2} \simeq 2.6$, with $\kappa_V = 3.70$ the  isovector nucleon anomalous magnetic moment. 
The large-$N_c$ prediction of  $G_M^\ast(0) $ is around 15\% smaller compared to its phenomenological value given in Eq.~(\ref{eq:GMGEGC}).  
For the purpose of providing realistic predictions, we therefore follow~\cite{Pascalutsa:2006up} and improve the large-$N_c$ relation by multiplying the {\it rhs} of Eq.~(\ref{eq:modelhM}) by $\sqrt{2} G_M^\ast(0) / \kappa_V$, and using the empirical value for $G_M^\ast(0)$. 

With the use of the model for nucleon GPD $E$ specified in Sec. 2.7.2 of Ref.~\cite{Pascalutsa:2006up},  the model of 
Eq.~(\ref{eq:modelhM}) for the GPD $h_M$ with the normalization specified above was found to provide a very good description of $G_M^\ast(\Delta^2)$ over a large range of $\Delta^2$. 
In the present paper we employ this model in our numerical applications.

For the GPD $ C_{1} $, the phenomenological
value of its first moment at $\Delta^2 = 0$ yields $ 2C_{5}(0)\simeq 2.32 $, using Eq.~(\ref{eq:GTNDel}),  
whereas the large-$ N_{c} $
prediction of Eq.~(\ref{eq:modelC1}) yields $2 C_5(0) =  \sqrt{3}g_{A}\simeq 2.21$, accurate within around 5\%. 
For the pion-pole contribution to $C_{6}$, we obtain
the same accuracy as for $C_{5}$.

\subsubsection{GPDs for $N \to P_{11}(1440)$ transition}

We next discuss the parameterization of the matrix elements entering the DVCS tensor of Eq.~(\ref{eq:DVCStensor2}) for the 
$N \to P_{11}(1440)$ transition. As the $P_{11}$ resonance has nucleon quantum numbers, the DVCS amplitude for the 
$N \to P_{11}$ transition has a similar form as for the nucleon case. 
Thus the matrix element of the vector bilinear quark operator along the light-cone can be parameterized as:
\begin{widetext}
\begin{eqnarray}
&&\int \frac{d \lambda}{2 \pi} e^{i \lambda x} \sum_{q}  e_q^2
\langle R(p_R, s_R) | \bar q \left(- \frac{\lambda n}{2} \right) \gamma \cdot n  \, q \left(\frac{\lambda n}{2} \right)  | N(p, s_N) \rangle 
\nonumber \\
&&=   H_1^{pP_{11}}(x, \xi, \Delta^2) \; \bar{R} \left(p_R, s_R \right) \left( n^\nu - \frac{n \cdot \Delta}{\Delta^2} \Delta^\nu \right) 
\gamma_\nu N\left(p, s_N \right) 
 +  H_2^{pP_{11}}(x, \xi, \Delta^2) \; \bar{R}  \left(p_R, s_R \right)  \frac{i \sigma_{\nu \kappa} n^\nu \Delta^\kappa}{M_R + M_N} N\left(p, s_N \right).
 \nonumber \\
 \label{eq:NP11gpdunpol} 
\end{eqnarray}
\end{widetext}
The GPDs which enter the $p \to P_{11}$ DVCS process are weighted by the quadratic quark charges as (for $i = 1,2$): 
\begin{eqnarray} 
H_i^{pP_{11}} = \frac{4}{9} H_i^{u, pP_{11}} + \frac{1}{9} H_i^{d, pP_{11}}  + \frac{1}{9} H_i^{s, pP_{11}},  
\label{eq:NP11gpdunpol2} 
\end{eqnarray}
in terms of $u$-, $d$-, and $s$-quark GPDs. 
In the following analysis, we will neglect the strange quark contributions which is expected to be small in the $p \to P_{11}$ DVCS process. 
As the $P_{11}$ resonance has isospin 1/2, the first moments of the unpolarized GPD quark flavor combination 
of Eq.~(\ref{eq:NP11gpdunpol2}) are related to the $N \to P_{11}$ electromagnetic transition FFs for proton ($N = p$) and neutron ($N = n$) target defined in Eq.~(\ref{eq:NP11em}) as:
\begin{eqnarray} 
\int_{-1}^{+1} dx \, H_{1, 2}^{pP_{11}}(x, \xi, \Delta^2) &=& F_{1, 2}^{pP_{11}} (\Delta^2) + \frac{2}{3} F_{1, 2}^{nP_{11}} (\Delta^2).  
\nonumber \\
\end{eqnarray}

Furthermore for the $N \to P_{11}$ transition, 
the matrix element of the axial-vector bilinear quark operator along the light-cone can be parameterized as:
\begin{widetext}
\begin{eqnarray}
&&\int \frac{d \lambda}{2 \pi} e^{i \lambda x} \sum_{q}  e_q^2
\langle R(p_R, s_R) | \bar q \left(- \frac{\lambda n}{2} \right) \gamma \cdot n \, \gamma_5 \, q \left(\frac{\lambda n}{2} \right)  | N(p, s_N) \rangle 
\nonumber \\
&&=   \tilde H_1^{pP_{11}}(x, \xi, \Delta^2) \; \bar{R} \left(p_R, s_R \right) \gamma \cdot n \gamma_5  N\left(p, s_N \right) 
 +  \tilde H_2^{pP_{11}}(x, \xi, \Delta^2) \; \bar{R}  \left(p_R, s_R \right) \frac{\Delta \cdot n}{M_R + M_N} \gamma_5  N\left(p, s_N \right). 
 \nonumber \\
 \label{eq:NP11gpdpol} 
\end{eqnarray}
\end{widetext}

The first moments of the polarized GPDs $\tilde H_{1,2}^{pP_{11}}$ are constrained through the FFs parameterizing the 
matrix elements of the axial-vector current operator for the $N \to P_{11}(1440)$ transition. 
The matrix element of the third component of the isovector axial current $A_3^\nu$ is given by:
\begin{eqnarray}
&& \langle R(p_R, s_R) | A_3^\nu (0) | N(p, s_N) \rangle 
\nonumber \\
&&= \bar{R} \left(p_R, s_R \right)  
 \bigg\{ G_A^{NP_{11}}(\Delta^2) \, \gamma^\nu \gamma_5 \nonumber \\
 &&\hspace{1.25cm}+ \,  
 G_P^{NP_{11}}(\Delta^2) \, \frac{\Delta^\nu \gamma_5}{(M_R + M_N)}  \bigg\} \,  \frac{\tau_3}{2} N\left(p, s_N \right), \quad \quad
 \nonumber \\
 \label{eq:NP11isovax}
\end{eqnarray}
with $\tau_3$ the third isospin Pauli matrix, 
and where $G_A^{NP_{11}}$ and $G_P^{NP_{11}}$ are the corresponding isovector axial FFs.  
We will also need the matrix element of the isoscalar axial current $A_0^\nu$ to fully characterize the $N \to P_{11}$ axial transition. With the definition of the isoscalar axial current $A_0^\nu$:
\begin{eqnarray}
A_0^\nu(0) \equiv \frac{1}{2} \left[ \bar u(0) \gamma^\nu \gamma_5 u(0) + \bar d(0) \gamma^\nu \gamma_5 d(0) \right],
\label{eq:A0def}
\end{eqnarray}
we can parameterize the $N \to P_{11}$ isoscalar axial transition as:
\begin{eqnarray}
&&\langle R(p_R, s_R) | A_0^\nu (0) | N(p, s_N) \rangle 
 \nonumber \\
&&= \bar{R} \left(p_R, s_R \right)  
 \bigg\{ G_{A, 0}^{NP_{11}}(\Delta^2) \, \gamma^\nu \gamma_5 \nonumber \\
 &&\hspace{1.25cm}+ \,  
 G_{P, 0}^{NP_{11}}(\Delta^2) \, \frac{\Delta^\nu \gamma_5}{(M_R + M_N)}  \bigg\} \,  \frac{1}{2} N\left(p, s_N \right), \quad \quad  
 \nonumber \\
 \label{eq:NP11isosax}
\end{eqnarray}
with $G_{A, 0}^{NP_{11}}$ and $G_{P, 0}^{NP_{11}}$ the corresponding isoscalar axial FFs.

The relations between the first moments of the polarized GPDs for a given quark flavor entering the definition in Eq.~(\ref{eq:NP11gpdpol}) and the axial transitions FFs defined in Eqs.~(\ref{eq:NP11isovax}) and (\ref{eq:NP11isosax}) are then obtained as:
\begin{eqnarray} 
\int_{-1}^{+1} dx  \tilde H_{1}^{u, pP_{11}}(x, \xi, \Delta^2) &=& \frac{1}{2} \left( G_{A}^{NP_{11}} + G_{A, 0}^{NP_{11}} \right) (\Delta^2),
\nonumber \\
\int_{-1}^{+1} dx  \tilde H_{1}^{d, pP_{11}}(x, \xi, \Delta^2) &= &\frac{1}{2} \left( - G_{A}^{NP_{11}} + G_{A, 0}^{NP_{11}} \right) (\Delta^2),
\nonumber \\
\label{eq:NP11gpdpolsr}
\end{eqnarray}
and analogous relations for $\tilde H_2^{u, pP_{11}}$ and $\tilde H_2^{d, pP_{11}}$ in terms of $G_{P}^{NP_{11}}$ and $G_{P, 0}^{NP_{11}}$.

For the determination of the $N \to P_{11}(1440)$ isovector axial FFs, we will again use the PCAC relation of Eq.~(\ref{eq:pcac}), which yields:
\begin{eqnarray}
&&i \, (M_R + M_N)  \bar{R} \left(p_R, s_R \right)  
 \bigg\{ G_A^{NP_{11}}(\Delta^2)  \nonumber \\
 &&\hspace{1.5cm}+  \, 
 G_P^{NP_{11}}(\Delta^2) \, \frac{\Delta^2}{(M_R + M_N)^2}  \bigg\}  \gamma_5 \,  \frac{\tau_3}{2} N\left(p, s_N \right)  \nonumber \\
&&= - f_\pi m_\pi^2 \langle P_{11}(p_R, s_R) | \Pi_3 (0) | N(p, s_N) \rangle .
 \label{eq:pcacNP11}
\end{eqnarray}
The latter matrix element is obtained from the $\pi N P_{11}$ coupling, which is given by the effective Lagrangian:
\begin{eqnarray}
{\cal L}_{\pi N P_{11}} =  \left( \frac{f_{\pi N P_{11}}}{m_\pi} \right) \bar R  \gamma^{\mu} \gamma_5 \tau_i N \left( \partial_\mu \Pi_i \right) + \mathrm{h.c.} , 
\label{eq:LpiNP11}
\end{eqnarray}
with $\pi N P_{11}$ coupling constant $f_{\pi N P_{11}}$.  
  From the Lagrangian of Eq.~(\ref{eq:LpiNP11}) we can then extract:
\begin{eqnarray}
&&  \langle P_{11}(p_R, s_R) | \Pi_3 (0) | N(p, s_N) \rangle   \nonumber \\
&& =  (M_R + M_N) \left( \frac{f_{\pi N P_{11}}}{m_\pi} \right) \frac{-i}{-\Delta^2 + m_\pi^2}  \nonumber \\
&&\times \bar{R}\left(p_R, s_R \right)  \gamma_5 \tau_3 
  N\left(p, s_N \right).  
  \nonumber \\
  \label{eq:LpiNP112}
\end{eqnarray}
Combining Eqs.~(\ref{eq:pcacNP11}) and (\ref{eq:LpiNP112}) allows to extract at $\Delta^2 = 0$ a generalized Goldberger-Treiman relation 
for the $N \to P_{11}$ isovector axial FF as:
\begin{eqnarray}
G_A^{NP_{11}}(0) = \left( \frac{f_{\pi N P_{11}}}{m_\pi} \right) 2 f_\pi.  
\end{eqnarray} 
Furthermore, using the pion-pole dominance for the axial FF $G_P^{NP_{11}}$ at small values of $\Delta^2$, 
Eq.~(\ref{eq:pcacNP11}) yields:
\begin{eqnarray}
G_P^{NP_{11}}(\Delta^2) \approx\frac{(M_R + M_N)^2}{- \Delta^2 + m_\pi^2}  G_A^{NP_{11}}(\Delta^2).  
\end{eqnarray} 
As the isoscalar $N \to P_{11}$ axial FFs $G_{A, 0}^{NP_{11}}$ and $G_{P, 0}^{NP_{11}}$ are not known, we 
will use for $G_{A, 0}^{NP_{11}}$ the same quark model relation as for the nucleon isoscalar axial FF:   
\begin{eqnarray}
G_{A, 0}^{NP_{11}}(\Delta^2) \approx \frac{3}{5} G_{A}^{NP_{11}}(\Delta^2), 
\end{eqnarray} 
and parameterize $G_A^{NP_{11}}$ by a dipole form as in the nucleon case:
\begin{eqnarray}
 G_A^{NP_{11}}(\Delta^2) = 1/(1 - \Delta^2 / M_A^2)^2,
 \end{eqnarray} 
with dipole mass $M_A \simeq 1.0$~GeV.  
Furthermore, we set the isoscalar FF 
\begin{eqnarray}
G_{P, 0}^{NP_{11}} \approx 0, 
\end{eqnarray} 
in line with pion-pole dominance.
  It may be useful to check such estimates in future by calculating these FFs within dynamical 
quark / di-quark approaches, which have proven to yield a good understanding of the data for the vector FFs $F_{1, 2}^{NP_{11}}$ for the $N \to P_{11}$ transition~\cite{Burkert:2017djo}. 

In order to provide estimates for the $p \to P_{11}$ DVCS amplitude, 
we will need a parameterization of the $p \to P_{11}$ GPDs. 
For this purpose, we will use the simplest factorized parameterization, consisting of a product of the transition FF, and a $\xi$-independent 
valence quark parameterization, as:
\begin{eqnarray}
H_{1, 2}^{pP_{11}}(x, \xi, \Delta^2) &=& \left[ F_{1, 2}^{pP_{11}} (\Delta^2) + \frac{2}{3} F_{1, 2}^{nP_{11}} (\Delta^2) \right] \nonumber \\
&\times& N  x^r (1 - x)^s, \quad \quad (x > 0)  
\nonumber \\ 
\tilde H_{1}^{pP_{11}}(x, \xi, \Delta^2) &=&  \frac{1}{3} G_{A}^{NP_{11}} (\Delta^2) N  x^r (1 - x)^s, \quad     (x > 0) \nonumber \\
\tilde H_{2}^{pP_{11}}(x, \xi, \Delta^2) &=&  \frac{1}{6} G_{P}^{NP_{11}} (\Delta^2) N  x^r (1 - x)^s, \quad  (x > 0) \nonumber \\
\label{eq:NP11gpdpolmod} 
\end{eqnarray}
with normalization $N$ chosen as:
\begin{eqnarray}
N^{-1} = \int_0^1d x \, x^r (1 - x)^s, 
\end{eqnarray}
ensuring that the first moments of the corresponding GPDs are correctly normalized to the corresponding transition FFs, given our 
assumptions for the isoscalar axial transition FFs. 
The parameters $r$ and $s$ in the valence type parameterization can be adjusted separately in the transition GPDs to forthcoming data. 
To provide a guidance for these experiments, we will use in the following numerical analyses $r = -0.5$, and $s = 3$, which are typical values for a nucleon valence quark distribution, resulting in the constraint $N = 1.094$. 

\subsubsection{GPDs for $N \to D_{13}(1520)$ transition}

For the $N \to D_{13}$ transition, the matrix element of the vector bilinear quark operator along the light-cone can be parameterized 
through four tensor structures as:
\begin{widetext}
\begin{eqnarray}
&&\int \frac{d \lambda}{2 \pi} e^{i \lambda x} \sum_{q}  e_q^2
\langle R(p_R, s_R) | \bar q \left(- \frac{\lambda n}{2} \right) \gamma \cdot n \, q \left(\frac{\lambda n}{2} \right)  | N(p, s_N) \rangle 
=  \bar{R}_\beta \left(p_R, s_R \right)  
 \bigg\{ H_1^{pD_{13}}(x, \xi, \Delta^2) \, \tilde \Gamma_1^{\beta \nu}  \nonumber \\
&&\hspace{4.75cm}+ H_2^{pD_{13}}(x, \xi, \Delta^2)  \tilde \Gamma_2^{\beta \nu}  
+  H_3^{pD_{13}}(x, \xi, \Delta^2)  \tilde \Gamma_3^{\beta \nu} 
+  H_4^{pD_{13}}(x, \xi, \Delta^2)  \tilde \Gamma_4^{\beta \nu}   \bigg\}  n_\nu   N\left(p, s_N \right), \quad \quad
\nonumber \\
\label{eq:ND13gpdunpol1}
\end{eqnarray}
\end{widetext}
with tensors $ \tilde \Gamma_{1,2,3}^{\beta \nu}$ as in Eq.~(\ref{eq:ND13em3}), 
and tensor $ \tilde \Gamma_4^{\beta \nu}$ defined as:
\begin{eqnarray}
\tilde \Gamma_{4}^{\beta \nu} &=& g^{\beta \nu} -  
\frac{ (p_R)^\beta (p_R)^\nu}{p_R^2}  .    
\end{eqnarray}
The GPDs which enter the $p \to D_{13}$ DVCS process are weighted by the quadratic quark charges analogous 
as in~(\ref{eq:NP11gpdunpol2}). 
We will again neglect the strange quark contributions which is expected to be small in the $p \to D_{13}$ DVCS process. 
The first moments of the unpolarized GPD quark flavor combination entering the $p \to D_{13}$ transition are then related to the 
$N \to D_{13}$ electromagnetic transition FFs for proton ($N = p$) and neutron ($N = n$) target defined in Eq.~(\ref{eq:ND13em2}) as:
\begin{eqnarray} 
\int_{-1}^{+1} dx \, H_{1, 2, 3}^{pD_{13}}(x, \xi, \Delta^2) &=& F_{1, 2, 3}^{pD_{13}} (\Delta^2) + \frac{2}{3} F_{1, 2, 3}^{nD_{13}} (\Delta^2),  
\nonumber \\
\end{eqnarray}
The first moment of the GPD $ H_{4}^{pD_{13}}$ vanishes, analogous as for the $N \to \Delta$ transition:
\begin{eqnarray} 
\int_{-1}^{+1} dx \, H_{4}^{pD_{13}}(x, \xi, \Delta^2) &=& 0,
\end{eqnarray}
as the corresponding tensor structure is absent in the electromagnetic (local) current operator, as a consequence of electromagnetic gauge invariance.  

For the $N \to D_{13}$ transition, 
the matrix element of the axial-vector bilinear quark operator along the light-cone can be parameterized as:
\begin{widetext}
\begin{eqnarray}
&&\int \frac{d \lambda}{2 \pi} e^{i \lambda x} \sum_{q}  e_q^2
\langle R(p_R, s_R) | \bar q \left(- \frac{\lambda n}{2} \right) \gamma \cdot n \, \gamma_5 \, q \left(\frac{\lambda n}{2} \right)  | N(p, s_N) \rangle 
\nonumber \\
&&=  \bar{R}_\beta \left(p_R, s_R \right)  
 \bigg\{ \tilde H_1^{pD_{13}}(x, \xi, \Delta^2) \, n^\beta 
+ \tilde H_2^{pD_{13}}(x, \xi, \Delta^2) \, \left( \frac{\Delta \cdot n}{M_N^2} \right) \Delta^\beta \nonumber \\
&&\hspace{2.1cm} +  \, \tilde H_3^{pD_{13}}(x, \xi, \Delta^2) \, \frac{1}{M_N} \left ( n^\beta \gamma \cdot \Delta - \Delta^\beta \gamma \cdot n \right)
+  \, \tilde H_4^{pD_{13}}(x, \xi, \Delta^2) \, \frac{2}{M_N^2} \left( n^\beta \, \bar P \cdot \Delta - \Delta^\beta  \right)  \bigg\} \, 
\gamma_5 \,  N\left(p, s_N \right). 
\nonumber \\
\label{eq:ND13gpdpol}
\end{eqnarray}
\end{widetext}
The first moments of the polarized GPDs $\tilde H_i^{pD_{13}}$ (for $i = 1,...,4$) are constrained through the FFs parameterizing the 
axial $N \to D_{13}$ transition. As the $D_{13}$ resonance has isospin $1/2$, both isoscalar and isovector transitions are possible. 
For the isovector transition, the matrix element of the third component of the isovector axial current is parameterized as:
\begin{eqnarray}
&&\langle R(p_R, s_R) | A_3^\nu (0) | N(p, s_N) \rangle \nonumber \\
&&= \bar{R}_\beta \left(p_R, s_R \right)  
 \bigg\{ C_{1,V}^{ND_{13}}(\Delta^2) \, g^{\beta \nu} 
+ C_{2,V}^{ND_{13}}(\Delta^2) \frac{\Delta^\beta \Delta^\nu }{M_N^2} \nonumber \\
&&\hspace{1.75cm}+  \, C_{3,V}^{ND_{13}}(\Delta^2) \frac{1}{M_N} \left( g^{\beta \nu} \gamma \cdot \Delta - \Delta^\beta \gamma^\nu \right)  \nonumber \\
&&\hspace{1.75cm}+  \, C_{4,V}^{ND_{13}}(\Delta^2) \, \frac{2}{M_N^2} \left( g^{\beta \nu} \, \bar P \cdot \Delta - \Delta^\beta \bar P^\nu \right)  \bigg\} 
 \nonumber \\
&&\hspace{1.75cm}\times \gamma_5  \frac{\tau_3}{2} N\left(p, s_N \right), 
\nonumber \\
\label{eq:ND13axisov}
\end{eqnarray}
with isovector axial FFs $C_{i,V}^{ND_{13}}$ for $i = 1,...,4$. 
Analogous relations hold for the matrix element of the isoscalar axial current, with the replacement $\tau_3 \to 1$ in Eq.~(\ref{eq:ND13axisov}), 
and defining isoscalar FFs $C_{i,S}^{ND_{13}}$. 
The relations between the first moments of the polarized GPDs for a given quark flavor entering the definition in Eq.~(\ref{eq:ND13gpdpol}) and the axial transitions FFs are then obtained as (for $i = 1,...,4$):
\begin{eqnarray} 
\int_{-1}^{+1} dx  \tilde H_{i}^{u, pD_{13}}(x, \xi, \Delta^2) &=& \frac{1}{2} \left( C_{i,V}^{ND_{13}} + C_{i,S}^{ND_{13}} \right) (\Delta^2),
\nonumber \\
\int_{-1}^{+1} dx  \tilde H_{i}^{d, pP_{11}}(x, \xi, \Delta^2) &= &\frac{1}{2} \left( - C_{i,V}^{NP_{11}} + C_{i,S}^{ND_{13}} \right) (\Delta^2).
\nonumber \\
\label{eq:ND13gpdpolsr}
\end{eqnarray}

For the determination of the dominant $N \to D_{13}(1520)$ isovector axial FFs, we will use PCAC relation of Eq.~(\ref{eq:pcac}), which yields:
\begin{eqnarray}
&&i \,   \bar{R}_\beta \left(p_R, s_R \right)  
 \bigg\{ C_{1,V}^{ND_{13}}(\Delta^2) 
+ C_{2,V}^{ND_{13}}(\Delta^2)  \frac{ \Delta^2}{M_N^2} \bigg\} \Delta^{\beta}  \nonumber \\
&&\hspace{2cm} \times \gamma_5 \frac{\tau_3}{2}  N\left(p, s_N \right) \nonumber \\
&&= - f_\pi m_\pi^2 \langle D_{13}(p_R, s_R) | \Pi_3 (0) | N(p, s_N) \rangle .
\nonumber \\
\label{eq:pcacND13}
\end{eqnarray}
The latter matrix element is obtained from the $\pi N D_{13}$ coupling, which is given by the effective Lagrangian:
\begin{eqnarray}
{\cal L}_{\pi N D_{13}} &=& i \left( \frac{f_{\pi N D_{13}}}{m_\pi M_R} \right) \left( \partial_\nu \bar R_\beta \right) \gamma^{\mu \nu \beta} \gamma_5 \tau_i N \left( \partial_\mu \Pi_i \right) \nonumber \\
&+& \mathrm{h.c.} , 
\label{eq:LpiND13}
\end{eqnarray}
with $\pi N D_{13}$ coupling constant $f_{\pi N D_{13}}$.  
  From the Lagrangian of Eq.~(\ref{eq:LpiND13}) we can then extract:
\begin{eqnarray}
&&  \langle D_{13}(p_R, s_R) | \Pi_3 (0) | N(p, s_N) \rangle   
\nonumber \\
&& =   \left( \frac{f_{\pi N D_{13}}}{m_\pi} \right) \frac{-i }{-\Delta^2 + m_\pi^2}  \bar{R}_\beta \left(p_R, s_R \right) \gamma_5  \Delta^{\beta} \tau_3   N\left(p, s_N \right).  \nonumber \\
\label{eq:LpiND132}
\end{eqnarray}
Combining Eqs.~(\ref{eq:pcacND13}) and (\ref{eq:LpiND132}) allows to extract at $\Delta^2 = 0$ a generalized Goldberger-Treiman relation as:
\begin{eqnarray}
C_{1,V}^{N D_{13}}(0) =  \left( \frac{f_{\pi N D_{13}}}{m_\pi} \right) 2 f_\pi.  
\label{eq:GTND13}
\end{eqnarray} 
Furthermore, using the pion-pole dominance for the axial FF $C_{2,V}^{N D_{13}}$ at small values of $\Delta^2$, we obtain from 
Eqs.~(\ref{eq:pcacND13}, \ref{eq:LpiND132}):
\begin{eqnarray}
C_{2,V}^{N D_{13}}(\Delta^2) \approx\frac{M_N^2}{- \Delta^2 + m_\pi^2}  C_{1,V}^{N D_{13}}(\Delta^2).  
\end{eqnarray} 
In this work, we will parameterize $C_{1,V}^{N D_{13}}$ by a dipole form as in the nucleon case:
 \begin{eqnarray}
 C_{1,V}^{N D_{13}}(\Delta^2) = 1/(1 - \Delta^2 / M_A^2)^2, 
 \end{eqnarray} 
 with dipole mass $M_A \simeq 1.0$~GeV.  
 
For the corresponding unknown isoscalar axial FFs 
$C_{1,S}^{N D_{13}} $ and $C_{2,S}^{N D_{13}}$, we will make the same type of approximations as 
discussed for the $N \to P_{11}$ isoscalar axial FFs:  \begin{eqnarray}
C_{1,S}^{N D_{13}}(\Delta^2) &\approx& \frac{3}{5} C_{1,V}^{N D_{13}}(\Delta^2), \nonumber \\   
C_{2,S}^{N D_{13}}(\Delta^2) &\approx& 0.
\end{eqnarray}
The remaining FFs, which are not constrained by PCAC or the pion-pole dominance, will be neglected in the following numerical analysis, i.e. 
  \begin{eqnarray}
C_{3,V}^{N D_{13}} = C_{4,V}^{N D_{13}} &\approx& 0, 
\nonumber \\  
C_{3,S}^{N D_{13}} = C_{4,S}^{N D_{13}} &\approx& 0.
\end{eqnarray} 

To provide estimates for the $p \to D_{13}$ DVCS amplitude in the following, 
we will also make the simplest factorized parameterization for the GPDs as product of the transition FF, and a $\xi$-independent 
valence quark parameterization, as:
\begin{eqnarray}
H_{1, 2, 3}^{pD_{13}}(x, \xi, \Delta^2) &=& \left[ F_{1, 2, 3}^{pD_{13}} (\Delta^2) + \frac{2}{3} F_{1, 2, 3}^{nD_{13}} (\Delta^2) \right] \nonumber \\
&\times& N  x^r (1 - x)^s, \quad \quad (x > 0)  
\nonumber \\ 
\tilde H_{1}^{pD_{13}}(x, \xi, \Delta^2) &=&  \frac{1}{3} C_{1,V}^{ND_{13}} (\Delta^2) N  x^r (1 - x)^s, \quad     (x > 0) \nonumber \\ 
\tilde H_{2}^{pD_{13}}(x, \xi, \Delta^2) &=&  \frac{1}{6} C_{2,V}^{ND_{13}} (\Delta^2) N  x^r (1 - x)^s, \quad  (x > 0) \nonumber \\
\label{eq:ND13gpdpolmod} 
\end{eqnarray}
and use for the parameters $r$ and $s$ in the valence type quark parameterizations the values  $r = -0.5$, and $s = 3$, with constraint $N = 1.094$. 
The remaining GPDs are neglected, i.e. 
\begin{eqnarray}
H_{4}^{pD_{13}} = \tilde H_{3}^{pD_{13}} = \tilde H_{4}^{pD_{13}} \approx 0, 
\end{eqnarray}
in the following numerical analysis.

\subsubsection{GPDs for $N \to S_{11}(1535)$ transition}

For the $N \to S_{11}$ transition, the matrix element of the vector bilinear quark operator along the light-cone can be parameterized as:
\begin{widetext}
\begin{eqnarray}
&&\int \frac{d \lambda}{2 \pi} e^{i \lambda x} \sum_{q}  e_q^2
\langle R(p_R, s_R) | \bar q \left(- \frac{\lambda n}{2} \right) \gamma \cdot n  \, q \left(\frac{\lambda n}{2} \right)  | N(p, s_N) \rangle 
\nonumber \\
&&=   H_1^{pS_{11}}(x, \xi, \Delta^2) \, \bar{R} \left(p_R, s_R \right) \left(  n^\nu - \frac{n \cdot \Delta}{\Delta^2} \Delta^\nu  \right) 
\gamma_\nu \gamma_5  N\left(p, s_N \right) 
 +  H_2^{pS_{11}}(x, \xi, \Delta^2) \, \bar{R}  \left(p_R, s_R \right) \frac{i \sigma_{\nu \kappa}  n^\nu \Delta^\kappa}{M_R + M_N} \, 
 \gamma_5  N\left(p, s_N \right).
 \nonumber \\
\label{eq:NS11gpdunpol} 
\end{eqnarray}
\end{widetext}
Neglecting again the strange quark contribution in the $p \to S_{11}$ DVCS process,  
the first moments of the unpolarized GPD quark flavor combination entering Eq.~(\ref{eq:NS11gpdunpol}) are related to the $N \to S_{11}$  
electromagnetic FFs for proton ($N = p$) and neutron ($N = n$) target defined in Eq.~(\ref{eq:NS11em}) as:
\begin{eqnarray} 
\int_{-1}^{+1} dx \, H_{1, 2}^{pS_{11}}(x, \xi, \Delta^2) &=& F_{1, 2}^{pS_{11}} (\Delta^2) + \frac{2}{3} F_{1, 2}^{nS_{11}} (\Delta^2).  
\nonumber \\
\end{eqnarray}

Furthermore for the $N \to S_{11}$ transition, 
the matrix element of the axial-vector bilinear quark operator along the light-cone can be parameterized as:
\begin{widetext}
\begin{eqnarray}
&&\int \frac{d \lambda}{2 \pi} e^{i \lambda x} \sum_{q}  e_q^2
\langle R(p_R, s_R) | \bar q \left(- \frac{\lambda n}{2} \right) \gamma \cdot n \, \gamma_5 \, q \left(\frac{\lambda n}{2} \right)  | N(p, s_N) \rangle 
\nonumber \\
&&=   \tilde H_1^{pS_{11}}(x, \xi, \Delta^2) \; \bar{R} \left(p_R, s_R \right) \gamma \cdot n   N\left(p, s_N \right) 
 +  \tilde H_2^{pS_{11}}(x, \xi, \Delta^2) \; \bar{R}  \left(p_R, s_R \right) \frac{\Delta \cdot n}{M_R + M_N}   N\left(p, s_N \right). 
 \nonumber \\
\label{eq:NS11gpdpol} 
\end{eqnarray}
\end{widetext}
The first moments of the polarized GPDs $\tilde H_{1,2}^{pS_{11}}$ are constrained through the FFs parameterizing the 
matrix elements of the axial current operator for the $N \to S_{11}(1535)$ transition. 
The matrix element of the third component of the isovector axial current $A_3^\nu$ is given by:
\begin{eqnarray}
&& \langle R(p_R, s_R) | A_3^\nu (0) | N(p, s_N) \rangle 
\nonumber \\
&&= \bar{R} \left(p_R, s_R \right)  
 \bigg\{ G_A^{NS_{11}}(\Delta^2) \, \gamma^\nu \nonumber \\
 &&\hspace{1.25cm}+ \,  
 G_P^{NS_{11}}(\Delta^2) \, \frac{\Delta^\nu}{(M_R + M_N)}  \bigg\} \,  \frac{\tau_3}{2} N\left(p, s_N \right), \quad \quad
\nonumber \\
\label{eq:NS11isovax}
\end{eqnarray}
with $G_A^{NS_{11}}$ and $G_P^{NS_{11}}$ the corresponding isovector axial FFs.  
Analogous relations hold for the matrix element of the isoscalar axial current, with the replacement $\tau_3 \to 1$ in Eq.~(\ref{eq:NS11isovax}), 
and defining isoscalar FFs $G_{A, 0}^{NS_{11}}$ and $G_{P, 0}^{NS_{11}}$. 

The relations between the first moments of the polarized GPDs for a given quark flavor entering the definition in Eq.~(\ref{eq:NS11gpdpol}) and the axial transitions FFs are then obtained as:
\begin{eqnarray} 
\int_{-1}^{+1} dx  \tilde H_{1}^{u, pS_{11}}(x, \xi, \Delta^2) &=& \frac{1}{2} \left( G_{A}^{NS_{11}} + G_{A, 0}^{NS_{11}} \right) (\Delta^2),
\nonumber \\
\int_{-1}^{+1} dx  \tilde H_{1}^{d, pS_{11}}(x, \xi, \Delta^2) &= &\frac{1}{2} \left( - G_{A}^{NS_{11}} + G_{A, 0}^{NS_{11}} \right) (\Delta^2),
\nonumber \\
\label{eq:NS11gpdpolsr}
\end{eqnarray}
and analogous relations for $\tilde H_2^{u, pS_{11}}$ and $\tilde H_2^{d, pS_{11}}$ in terms of $G_{P}^{NS_{11}}$ and $G_{P, 0}^{NS_{11}}$.

For the determination of the $N \to S_{11}(1535)$ isovector axial FFs, we will again use the PCAC relation of Eq.~(\ref{eq:pcac}), which yields:
\begin{eqnarray}
&&i \, (M_R - M_N)  \bar{R} \left(p_R, s_R \right)  
 \bigg\{ G_A^{NS_{11}}(\Delta^2)  \nonumber \\
 &&\hspace{1.5cm}+  \, 
 G_P^{NS_{11}}(\Delta^2) \, \frac{\Delta^2}{(M_R^2 - M_N^2)}  \bigg\}  \,  \frac{\tau_3}{2} N\left(p, s_N \right)  \nonumber \\
&&= - f_\pi m_\pi^2 \langle S_{11}(p_R, s_R) | \Pi_3 (0) | N(p, s_N) \rangle .
\nonumber \\
\label{eq:pcacNS11}
\end{eqnarray}
The latter matrix element is obtained from the $\pi N S_{11}$ coupling, which is given by the effective Lagrangian:
\begin{eqnarray}
{\cal L}_{\pi N S_{11}} =  \left( \frac{f_{\pi N S_{11}}}{m_\pi} \right) \bar R  \gamma^{\mu} \tau_i N \left( \partial_\mu \Pi_i \right) + \mathrm{h.c.} , 
    \nonumber \\
\label{eq:LpiNS11}
\end{eqnarray}
with $\pi N S_{11}$ coupling constant $f_{\pi N S_{11}}$.  
  From the Lagrangian of Eq.~(\ref{eq:LpiNS11}) we can then extract:
\begin{eqnarray}
&&  \langle S_{11}(p_R, s_R) | \Pi_3 (0) | N(p, s_N) \rangle   \nonumber \\
&& =  (M_R - M_N) \left( \frac{f_{\pi N S_{11}}}{m_\pi} \right) \frac{-i}{-\Delta^2 + m_\pi^2}  \nonumber \\
&&\times \, \bar{R}\left(p_R, s_R \right)  \tau_3 
  N\left(p, s_N \right).  
\nonumber \\
  \label{eq:LpiNS112}
\end{eqnarray}
Combining Eqs.~(\ref{eq:pcacNS11}) and (\ref{eq:LpiNS112}) allows to extract at $\Delta^2 = 0$ a generalized Goldberger-Treiman relation 
for the $N \to S_{11}$ isovector axial FF as:
\begin{eqnarray}
G_A^{NS_{11}}(0) = \left( \frac{f_{\pi N S_{11}}}{m_\pi} \right) 2 f_\pi.  
\end{eqnarray} 
Furthermore, using the pion-pole dominance for the axial FF $G_P^{NS_{11}}$ at small values of $\Delta^2$, we obtain:
\begin{eqnarray}
G_P^{NS_{11}}(\Delta^2) \approx\frac{(M_R^2 - M_N^2)}{- \Delta^2 + m_\pi^2}  G_A^{NS_{11}}(\Delta^2).  
\end{eqnarray} 
For the unknown isoscalar axial FFs $G_{A, 0}^{NS_{11}}$ and $G_{P, 0}^{NS_{11}}$, we will make the same type of approximations as 
discussed for the $N \to P_{11}$ isoscalar axial FFs:  \begin{eqnarray}
G_{A, 0}^{NS_{11}}(\Delta^2) &\approx& \frac{3}{5} G_{A}^{NS_{11}}(\Delta^2), \nonumber \\ 
G_{P, 0}^{NS_{11}}(\Delta^2) &\approx& 0, 
\end{eqnarray} 
and 
parameterize $G_A^{NS_{11}}$ by a dipole form as in the nucleon case:
 \begin{eqnarray}
 G_A^{NS_{11}}(\Delta^2) = 1/(1 - \Delta^2 / M_A^2)^2, 
 \end{eqnarray} 
 with dipole mass $M_A \simeq 1.0$~GeV.  

To provide estimates for the $p \to S_{11}$ DVCS amplitude in the following, 
we will again make the simplest factorized parameterization for the GPDs, consisting of a product of the transition FF, and a $\xi$-independent 
valence quark parameterization, as:
\begin{eqnarray}
H_{1, 2}^{pS_{11}}(x, \xi, \Delta^2) &=& \left[ F_{1, 2}^{pS_{11}} (\Delta^2) + \frac{2}{3} F_{1, 2}^{nS_{11}} (\Delta^2) \right] \nonumber \\
&\times& N  x^r (1 - x)^s, \quad \quad (x > 0)  
\nonumber \\ 
\tilde H_{1}^{pS_{11}}(x, \xi, \Delta^2) &=&  \frac{1}{3} G_{A}^{NS_{11}} (\Delta^2) N  x^r (1 - x)^s, \quad     (x > 0) \nonumber \\
\tilde H_{2}^{pS_{11}}(x, \xi, \Delta^2) &=&  \frac{1}{6} G_{P}^{NS_{11}} (\Delta^2) N  x^r (1 - x)^s, \quad  (x > 0) \nonumber \\
\label{eq:NS11gpdpolmod} 
\end{eqnarray}
and use for the parameters $r$ and $s$ in the valence type quark parameterizations the values  $r = -0.5$, and $s = 3$, with constraint $N = 1.094$.

\section{$e^- N \to e^- \gamma R \to e^- \gamma \pi N$ amplitude in the nucleon resonance region}
\label{sec4}

Having discussed the $e^- N \to e^- \gamma R$ amplitude in Section~\ref{sec3}, we will next discuss the $R \to \pi N$ decay to calculate the full amplitude for the $e^- N \to e^- \gamma \pi N$ process. As the pion decay angular distribution depends on the quantum numbers of the resonance considered (or more generally on the partial wave of the $\pi N$ system), we will subsequently discuss the four prominent resonances in the first and second nucleon resonance regions. 
In this work, we will follow an isobar approach and adopt a Breit-Wigner resonance parameterization~\cite{Workman:2022ynf}. This will serve as a first step towards a full partial-wave analysis of the $\pi N$ final state in a future work.

\subsection{$R = \Delta(1232)$}

We start by writing the $e^- N \to e^- \gamma \Delta$ amplitude of Eqs.~(\ref{eq:BH}, \ref{eq:NDELem1}) for the BH process and of 
Eqs.~(\ref{eq:DVCS}, \ref{eq:DVCStensor2}) for the DVCS process as:
\begin{eqnarray}
\mathcal{M}(e^- N \to e^- \gamma \Delta)  
\equiv \bar{R}_\beta \left(p_R, s_R \right) 
 \mathcal{M}_R^\beta(e^- N \to e^- \gamma \Delta),  
 \nonumber \\
\label{eq:NDelamplbeta}
\end{eqnarray}
defining $\mathcal{M}_R^\beta(e^- N \to e^- \gamma \Delta)$ for both BH and DVCS processes.

For the $e^- N \to e^- \gamma \Delta \to e^- \gamma \pi N$ process, the invariant amplitude entering 
the cross section of Eq.~(\ref{eq:cross1}) is then given by:
\begin{eqnarray}
&&\mathcal{M}(e^- N \to e^- \gamma \Delta \to e^- \gamma \pi N) = \bar C_{iso} \frac{f_{\pi N\Delta}}{m_\pi} 
\left(- p_\pi\right)^\alpha 
\nonumber \\
&&\times \, \bar{N} \left(p^\prime, s^\prime_N \right) 
\frac{i P^{(3/2)}_{\alpha \beta}(p_R)}{M_{\pi N}^2 - M_R^2 + i M_R \Gamma_\Delta(M_{\pi N})}
\nonumber \\
&& \times \, \mathcal{M}_R^\beta(e^- N \to e^- \gamma \Delta),  
\nonumber \\
\label{eq:NDelamplfull}
\end{eqnarray}
where $f_{\pi N\Delta}$ is the $\pi N \Delta$ coupling constant introduced in Eq.~(\ref{eq:LpiNDel}), and $\bar C_{iso}$ is the isospin factor for the $\Delta \to \pi N$ decay, which takes on the values for the $\Delta^+$ state:
\begin{eqnarray}
\bar C_{iso} &=& \sqrt{\frac{2}{3}}, \quad \mathrm{for}~\Delta^+ \to \pi^0 p, \nonumber \\
\bar C_{iso} &=& - \sqrt{\frac{1}{3}}, \quad 
\mathrm{for}~\Delta^+ \to \pi^+ n.
\nonumber \\
\label{eq:isoDELpiN}
\end{eqnarray}
Furthermore in Eq.~(\ref{eq:NDelamplfull}),  $P^{(3/2)}_{\alpha \beta}(p_R)$ denotes the spin-3/2 projector, defined as:
\begin{eqnarray}
P^{(3/2)}_{\alpha \beta}(p_R) 
&=& \sum_{s_R} R_\alpha(p_R, s_R) 
\bar R_\beta(p_R, s_R)
\nonumber \\
&=& (\gamma \cdot p_R + M_{\pi N}) 
\bigg\{- g_{\alpha \beta} + \frac{1}{3} 
\gamma_\alpha \gamma_\beta \nonumber \\
&+&  \frac{1}{3 p_R^2} 
\bigg( \gamma \cdot p_R \,  \gamma_\alpha (p_R)_\beta + (p_R)_\alpha \gamma_\beta \, \gamma \cdot p_R \bigg) 
\bigg\}, 
\nonumber \\
\label{eq:spin32proj}
\end{eqnarray}
$M_R$ is the resonance mass, and $\Gamma_\Delta(M_{\pi N})$ denotes the energy-dependent total $\Delta$-width, which to very good approximation is given by the $\Delta \to \pi N$ partial width: 
\begin{eqnarray}
\Gamma_\Delta(M_{\pi N}) 
&=& \Gamma_{\Delta \to \pi N}(M_{\pi N}).  
\end{eqnarray}
For a resonance $R$ which decays into the $\pi N$ state in an $l$-wave, the partial width is given by~\cite{Drechsel:2007if}:
\begin{eqnarray}
\Gamma_R(M_{\pi N}) &=& \beta_{\pi N} \Gamma_R 
\left( \frac{| \vec p^{\,*}_\pi (M_{\pi N}) |}{| \vec p^{\,*}_\pi (M_R) |} \right)^{2 l +1} \frac{M_R}{M_{\pi N}} 
\nonumber \\
&\times& \left( \frac{X_R^2 + | \vec p^{\,*}_\pi (M_R) |^2}{X_R^2 + | \vec p^{\,*}_\pi (M_{\pi N}) |^2} \right)^l,
\nonumber \\
\label{eq:widthpiN}
\end{eqnarray}
with functional form of the pion momentum in the resonance rest frame $| \vec p^{\,*}_\pi(M_{\pi N})|$
given in Eq.~(\ref{eq:pimomstar}), 
where $\beta_{\pi N}$ is the $\pi N$ branching fraction, $\Gamma_R$ the resonance width, and $X_R$ the cut-off parameter in the Blatt-Weisskopf form factor for the Breit-Wigner resonance~\cite{Workman:2022ynf}.
The $\Delta(1232)$ resonance production corresponds with the partial wave $l = 1$, and its parameters in the present analysis are given in Table~\ref{tab1}.

As we will show in the following predictions for the pion decay angular distribution in the resonance rest frame, it is convenient to work out the spinor structure in Eq.~(\ref{eq:NDelamplfull}) as:  
\begin{widetext}
\begin{eqnarray}
\mathcal{M}(e^- N \to e^- \gamma \Delta \to e^- \gamma \pi N) 
&=& \bar C_{iso} 
\frac{f_{\pi N\Delta}}{m_\pi} 
 \sqrt{\frac{4 \pi}{3}} 
 \,    
\frac{i \, | \vec p^{\,*}_\pi(M_{\pi N})| \, \left[(M_{\pi N} + M_N)^2 - m_\pi^2 \right]^{1/2}}{M_{\pi N}^2 - M_R^2 + i M_R \Gamma_\Delta(M_{\pi N})}
\nonumber \\
&\times& \sum_{s_R} \mathcal{M}(e^- N \to e^- \gamma \, \Delta (M_{\pi N}, s_R)) 
\,  \sum_{\lambda^\prime} 
\, \langle \, 1 \lambda^\prime, \frac{1}{2} s_N^\prime  \, | \, \frac{3}{2} s_R \, \rangle \, Y_{1 \lambda^\prime}(\Omega^\ast_\pi),  
\nonumber \\
\label{eq:NDelamplfull2}
\end{eqnarray}
\end{widetext}
where $Y_{l m_l}(\Omega^\ast_\pi)$ denotes the spherical harmonic function and $\langle \, s s_z, l m_l | J M \rangle$
the Clebsch-Gordan coefficient for the $\Delta \to \pi N$ decay. 

We can show the usefulness of Eq.~(\ref{eq:NDelamplfull2}) for calculating the resonance decay angular distributions in a few special cases. Firstly, when summing over the final nucleon helicities and integrating over the full pion solid angle, the squared amplitude entering the cross section of Eq.~(\ref{eq:cross1}) becomes:
\begin{eqnarray}
&&\sum_{s_N^\prime} \int  d \Omega^\ast_\pi | \mathcal{M}(e^- N \to e^- \gamma \Delta \to e^- \gamma \pi N) |^2 
\nonumber \\
&&= \left( \bar C_{iso} 
 \frac{f_{\pi N\Delta}}{m_\pi} \right)^2
\frac{| \vec p^{\,*}_\pi(M_{\pi N})|^2 \left[(M_{\pi N} + M_N)^2 - m_\pi^2 \right]}{(M_{\pi N}^2 - M_R^2)^2 +  \left( M_R \Gamma_\Delta(M_{\pi N}) \right)^2}
\nonumber \\
&&\times  \frac{4 \pi}{3} \sum_{s_R} 
| \mathcal{M}(e^- N \to e^- \gamma \, \Delta (M_{\pi N}, s_R)) |^2.  
\nonumber \\
\label{eq:NDelangdistr1}
\end{eqnarray}
We can check that in the limit of a narrow resonance \mbox{$\Gamma_R \ll M_R$}, Eq.~(\ref{eq:NDelangdistr1}) is equivalent to:
\begin{eqnarray}
&&\sum_{s_N^\prime} \int  d \Omega^\ast_\pi | \mathcal{M}(e^- N \to e^- \gamma \Delta \to e^- \gamma \pi N) |^2 
\nonumber \\
&&\approx (2 \pi)^3 \frac{4 M_{\pi N}}{| \vec p^{\,*}_\pi|} \, \sum_{s_R} 
| \mathcal{M}(e^- N \to e^- \gamma \, \Delta (M_{\pi N}, s_R)) |^2 \nonumber \\
&&\times \frac{1}{\pi} \frac{M_{\pi N} \Gamma_{\Delta \to \pi N}}{(M_{\pi N}^2 - M_R^2)^2 +  \left( M_R \Gamma_\Delta(M_{\pi N}) \right)^2},  
\nonumber \\
\label{eq:NDelangdistr2}
\end{eqnarray}
where for the case of the $\Delta^+$ resonance the sum over both isospin channels $\pi^+ n$ and $\pi^0 p$ is included in the $\Delta \to \pi N$ width $\Gamma_{\Delta \to \pi N}$, using Eq.~(\ref{eq:isoDELpiN}). 
For the case of a narrow resonance, we can furthermore perform the integration over the invariant mass $M_{\pi N}^2$ of the Breit-Wigner spectral function in Eq.~(\ref{eq:NDelangdistr2})  analytically to obtain as unpolarized cross section formula from Eq.~(\ref{eq:cross1}):
\begin{eqnarray}
&&
\int d M^2_{\pi N} \int  d \Omega^\ast_\pi 
\frac{d \sigma}{d Q^2 d x_B d t d \Phi d M^2_{\pi N} d \Omega^*_\pi} \nonumber \\
&&\approx \frac{1}{(2 \pi)^4} \frac{x_B \, y^2}{32 \, Q^4 \sqrt{ 1 + \frac{4 M_N^2 x_B^2}{Q^2} }}\nonumber \\
&&\times 
 \overline{\sum_i}\sum_f\left|\mathcal{M}(e^- N \to e^- \gamma \Delta(M_R, s_R)) \right|^2, 
\nonumber \\
\label{eq:crossnarrow1}
\end{eqnarray}
where the sum over the final helicities is for the $e^-$, $\gamma$, and $\Delta$ states. 
Eq.~(\ref{eq:crossnarrow1}) is found to be fully consistent with the result for a stable particle~\cite{Goeke:2001tz}. 

It is also instructive to work out the pion polar angular distribution  using Eq.~(\ref{eq:NDelamplfull2}) for the case of a narrow resonance, by integrating over the azimuthal angle $\phi^\ast_\pi$:
\begin{eqnarray}
&&
\int d M^2_{\pi N}  
\frac{d \sigma}{d Q^2 d x_B d t d \Phi d M^2_{\pi N} d \cos \theta^*_\pi} \nonumber \\
&&\approx \frac{1}{(2 \pi)^4} \frac{x_B \, y^2}{32 \, Q^4 \sqrt{ 1 + \frac{4 M_N^2 x_B^2}{Q^2} }} \, \,   
 \overline{\sum_{e, N}}\sum_{e, \gamma} (\bar C_{iso})^2 \nonumber \\
&&\times \bigg\{ \frac{1}{4} \left(1 + 3  \cos^2 \theta^\ast_\pi \right) \sum_{s_R = \pm 1/2} 
 \left|\mathcal{M}(e^- N \to e^- \gamma \Delta \right|^2 
 \nonumber \\
&& \quad + \frac{3}{4} \sin^2 \theta^\ast_\pi  \sum_{s_R = \pm 3/2} 
 \left|\mathcal{M}(e^- N \to e^- \gamma \Delta \right|^2  \bigg\},
\nonumber \\
\label{eq:crossnarrow}
\end{eqnarray}
highlighting the different polar angular distributions for the $\Delta$ helicity states 
$s_R = \pm 1/2$ versus $s_R = \pm 3/2$. 

\begin{table}[h]
\begin{tabular}{c|cccccc}
\hline\hline
R & $M_R$ & $\Gamma_R$ & $\beta_{\pi N}$ & $\beta_{\eta N}$ &  $f_{\pi N R}$ & $X_R$
\\
& [GeV] & [GeV] & & & & [GeV] \\
\hline
\quad $\Delta(1232)$ \quad & \quad $1.232$ \quad  & \quad $0.117$ \quad & \quad $1.0$ \quad  & 
\quad $0$ \quad  & 
\quad 2.08 \quad  & \quad $0.5$  \quad \\
\quad $P_{11}(1440)$ \quad & \quad $1.440$ \quad & \quad  $0.350$ \quad & \quad $0.7$ \quad & 
\quad $0$ \quad  & 
\quad 0.385 \quad  & \quad $0.5$ \quad \\
\quad $D_{13}(1520)$ \quad & \quad $1.520$ \quad & \quad  $0.130$ \quad & \quad $0.6$ \quad & 
\quad $0$ \quad  & 
\quad 1.59 \quad  & \quad $0.5$  \quad \\
\quad $S_{11}(1535)$ \quad & \quad  $1.535$ \quad & \quad  $0.100$ \quad & \quad  $0.4$ \quad & 
\quad $0.5$ \quad  & 
\quad 0.119 \quad  & \quad  $0.5$ \quad  \\
\hline \hline
\end{tabular}
\caption{Resonance parameters used in this work: mass $M_R$, width $\Gamma_R$, 
$\pi N$ branching fraction $\beta_{\pi N}$, 
$\eta N$ branching fraction $\beta_{\eta N}$, 
$\pi N R$ coupling constant $f_{\pi N R}$, 
and cut-off parameter $X_R$ in the Blatt-Weisskopf form factor.}
\label{tab1}
\end{table}

\subsection{$R = P_{11}(1440)$}

For the $P_{11}(1440)$ production process, we proceed in analogous way as for the $\Delta$-resonance by writing the $e^- N \to e^- P_{11}$ amplitude of Eqs.~(\ref{eq:BH}, \ref{eq:NP11em1}) for the BH process and of 
Eqs.~(\ref{eq:DVCS}, \ref{eq:DVCStensor2}) for the DVCS process as:
\begin{eqnarray}
&&\mathcal{M}(e^- N \to e^- \gamma P_{11})  
\nonumber \\
&&\equiv \bar{R} \left(p_R, s_R \right) 
 \mathcal{M}_R(e^- N \to e^- \gamma P_{11}),      
\label{eq:NP11ampltensor}
\end{eqnarray}
defining $\mathcal{M}_R(e^- N \to e^- \gamma P_{11})$ for both BH and DVCS processes. 

For the $e^- N \to e^- \gamma P_{11} \to e^- \gamma \pi N$ process, the invariant amplitude 
entering the cross section of Eq.~(\ref{eq:cross1}) is then given by:
\begin{eqnarray}
&&\mathcal{M}(e^- N \to e^- \gamma P_{11} \to e^- \gamma \pi N) = \bar C_{iso} \frac{f_{\pi N P_{11}}}{m_\pi} 
\nonumber \\
&&\times \, \bar{N} \left(p^\prime, s^\prime_N \right) 
\gamma_5 
\frac{i \left(M_{\pi N} + M_N\right) \, (\gamma \cdot p_R + M_{\pi N})}{M_{\pi N}^2 - M_R^2 + i M_R \Gamma_{P_{11}}(M_{\pi N})}
\nonumber \\
&& \times \, \mathcal{M}_R(e^- N \to e^- \gamma P_{11}), 
\nonumber \\
\label{eq:NP11amplfull}
\end{eqnarray}
where $f_{\pi N P_{11}}$ is the $\pi N P_{11}$ coupling constant introduced in Eq.~(\ref{eq:LpiNP11}). 
For an isospin-1/2 resonance as $P_{11}$, 
the isospin factor $\bar C_{iso}$ for the $R^+ \to \pi N$ decay takes on the values:
\begin{eqnarray}
\bar C_{iso} &=& 1, \quad \mathrm{for}~R^+ \to \pi^0 p, \nonumber \\
\bar C_{iso} &=& \sqrt{2}, \quad 
\mathrm{for}~R^+ \to \pi^+ n.
\nonumber \\
\label{eq:iso12piN}
\end{eqnarray}
Furthermore in Eq.~(\ref{eq:NP11amplfull}), 
$\Gamma_{P_{11}}(M_{\pi N})$ denotes the energy-dependent total $P_{11}$-width, which is given by~\cite{Drechsel:1998hk}:
\begin{eqnarray}
\Gamma_{P_{11}}(M_{\pi N}) 
= \Gamma_{P_{11} \to \pi N}(M_{\pi N})  + \Gamma_{P_{11} \to \pi \pi N}(M_{\pi N}),  
\nonumber \\
\end{eqnarray}
with the $P_{11} \to \pi N$ partial width given by Eq.~(\ref{eq:widthpiN}) for $l = 1$, and where  
$\Gamma_{P_{11} \to \pi \pi N}(M_{\pi N})$ denotes the partial width for the 
$P_{11} \to \pi \pi N$ decay.  
For a resonance $R$ which decays into the $\pi N$ state in an $l$-wave, we approximate 
the $R \to \pi \pi N$ partial width as in the MAID analysis~\cite{Drechsel:1998hk}:
\begin{eqnarray}
\Gamma_{R \to \pi \pi N}(W_R) 
&=& (1 - \beta_{\pi N} - \beta_{\eta N}) \Gamma_R 
\left( \frac{| \vec p^{\,*}_{2 \pi} (W_R) |}{| \vec p^{\,*}_{2 \pi} (M_R) |} \right)^{2 l + 4} 
\nonumber \\
&\times&
 \left( \frac{X_R^2 + | \vec p^{\,*}_{2 \pi} (M_R) |^2}{X_R^2 + | \vec p^{\,*}_{2 \pi} (W_R) |^2} \right)^{l + 2},  
\nonumber \\
\label{eq:width2piN}
\end{eqnarray}
with $| \vec p^{\,*}_{2 \pi}(W_R) |$ defined as~:
\begin{eqnarray}
| \vec p^{\,*}_{2 \pi}(W_R) | = \frac{1}{2 W_R} \lambda^{1/2}(W^2_R, M_N^2, (2 m_\pi)^2).
\nonumber \\
\label{eq:2pimomstar}
\end{eqnarray}
The $P_{11}(1440)$ resonance production corresponds with the case $l = 1$, and its parameters in the present analysis are given in Table~\ref{tab1}.

\subsection{$R = D_{13}(1520)$}

For the $D_{13}(1520)$ production process, we analogously write the $e^- N \to e^- D_{13}$ amplitude of Eqs.~(\ref{eq:BH}, \ref{eq:ND13em1}) for the BH process and of 
Eqs.~(\ref{eq:DVCS}, \ref{eq:DVCStensor2}) for the DVCS process as:
\begin{eqnarray}
&&\mathcal{M}(e^- N \to e^- \gamma D_{13})  
\nonumber \\
&&\equiv \bar{R}_\beta \left(p_R, s_R \right) 
 \mathcal{M}^\beta_R(e^- N \to e^- \gamma D_{13}),  
\nonumber \\
\label{eq:ND13amplbeta}
\end{eqnarray}
defining $\mathcal{M}^\beta_R(e^- N \to e^- \gamma D_{13})$ for both BH and DVCS processes.

For the $e^- N \to e^- \gamma D_{13} \to e^- \gamma \pi N$ process, the invariant amplitude 
entering the cross section of Eq.~(\ref{eq:cross1}) is then given by:
\begin{eqnarray}
&&\mathcal{M}(e^- N \to e^- \gamma D_{13} \to e^- \gamma \pi N) = \bar C_{iso} \frac{f_{\pi N D_{13}}}{m_\pi} 
\left(p_\pi\right)^\alpha 
\nonumber \\
&&\times \, \bar{N} \left(p^\prime, s^\prime_N \right) 
\gamma_5 
\frac{i P^{(3/2)}_{\alpha \beta}(p_R)}{M_{\pi N}^2 - M_R^2 + i M_R \Gamma_{D_{13}}(M_{\pi N})}
\nonumber \\
&& \times \, \mathcal{M}^\beta_R(e^- N \to e^- \gamma D_{13}),  
\nonumber \\
\label{eq:ND13amplfull}
\end{eqnarray}
where $f_{\pi N D_{13}}$ is the $\pi N D_{13}$ coupling constant introduced in Eq.~(\ref{eq:LpiND13}), and the isospin factor $\bar C_{iso}$ is defined as in Eq.~(\ref{eq:iso12piN}). 
Furthermore in Eq.~(\ref{eq:ND13amplfull}),  $P^{(3/2)}_{\alpha \beta}(p_R)$ denotes the spin-3/2 projector defined in Eq.~(\ref{eq:spin32proj}),  
$M_R$ is the resonance mass, and $\Gamma_{D_{13}}(M_{\pi N})$ denotes the energy-dependent total $D_{13}$-width, which is given by~\cite{Drechsel:1998hk}:
\begin{eqnarray}
\Gamma_{D_{13}}(M_{\pi N}) 
= \Gamma_{D_{13} \to \pi N}(M_{\pi N})  + \Gamma_{D_{13} \to \pi \pi N}(M_{\pi N}),  
\nonumber \\
\end{eqnarray}
with the $\Gamma_{D_{13} \to \pi N}$ and 
$\Gamma_{D_{13} \to \pi \pi N}$ 
partial widths given by Eqs.~(\ref{eq:widthpiN}) 
and~(\ref{eq:width2piN}) respectively for $l = 2$. 
The $D_{13}(1520)$ resonance parameters used in the present analysis are given in Table~\ref{tab1}.

For the pion decay angular distribution for the case of $l = 2$ in the resonance rest frame, it is again convenient to work out the spinor structure in Eq.~(\ref{eq:ND13amplfull}) as:  
\begin{widetext}
\begin{eqnarray}
\mathcal{M}(e^- N \to e^- \gamma D_{13} \to e^- \gamma \pi N) 
&=& \bar C_{iso} 
\frac{f_{\pi N D_{13}}}{m_\pi} 
\sqrt{\frac{4 \pi}{3}} \left( \frac{4 M^2_{\pi N}}{(M_{\pi N} + M_N)^2 - m_\pi^2 }\right)^{1/2}
 \,    
\frac{i \, | \vec p^{\,*}_\pi(M_{\pi N})|^2 }{M_{\pi N}^2 - M_R^2 + i M_R \Gamma_{D_{13}}(M_{\pi N})}
\nonumber \\
&\times& \sum_{s_R} \mathcal{M}(e^- N \to e^- \gamma \, D_{13} (M_{\pi N}, s_R)) 
\,  \sum_{m} 
\, \langle \, 2 m, \frac{1}{2} s_N^\prime \, | \, \frac{3}{2} s_R \, \rangle \, Y_{2 m}(\Omega^\ast_\pi).  
\nonumber \\
\label{eq:ND13amplfull2}
\end{eqnarray}
\end{widetext}

\subsection{$R = S_{11}(1535)$}

Finally, for the $S_{11}(1535)$ production process, we write the $e^- N \to e^- S_{11}$ amplitude as:
\begin{eqnarray}
&&\mathcal{M}(e^- N \to e^- \gamma S_{11})  
\nonumber \\
&&\equiv \bar{R} \left(p_R, s_R \right) 
 \mathcal{M}_R(e^- N \to e^- \gamma S_{11}), 
\nonumber \\
\label{eq:NS11ampltensor}
\end{eqnarray}
defining $\mathcal{M}_R(e^- N \to e^- \gamma S_{11})$ for both BH and DVCS processes. 

For the $e^- N \to e^- \gamma S_{11} \to e^- \gamma \pi N$ process, the invariant amplitude 
entering the cross section of Eq.~(\ref{eq:cross1}) is then given by:
\begin{eqnarray}
&&\mathcal{M}(e^- N \to e^- \gamma S_{11} \to e^- \gamma \pi N) = - \bar C_{iso} \frac{f_{\pi N S_{11}}}{m_\pi} 
\nonumber \\
&&\times \, \bar{N} \left(p^\prime, s^\prime_N \right) 
\frac{i \left(M_{\pi N} - M_N\right) \, (\gamma \cdot p_R + M_{\pi N})}{M_{\pi N}^2 - M_R^2 + i M_R \Gamma_{S_{11}}(M_{\pi N})}
\nonumber \\
&& \times \, \mathcal{M}_R(e^- N \to e^- \gamma S_{11}), 
\nonumber \\
\label{eq:NS11amplfull}
\end{eqnarray}
where $f_{\pi N S_{11}}$ is the $\pi N S_{11}$ coupling constant introduced in Eq.~(\ref{eq:LpiNS11}), and the isospin factor $\bar C_{iso}$ is defined as in Eq.~(\ref{eq:iso12piN}). 
Furthermore in Eq.~(\ref{eq:NS11amplfull}), $M_R$ is the resonance mass, and $\Gamma_{S_{11}}(M_{\pi N})$ denotes the energy-dependent total $S_{11}$-width, which we take from the $\eta$-MAID analysis~\cite{Chiang:2001as}:
\begin{eqnarray}
\Gamma_{S_{11}}(M_{\pi N}) 
&=& \Gamma_{S_{11} \to \pi N}(M_{\pi N})  +  
\Gamma_{S_{11} \to \eta N}(M_{\pi N})
\nonumber \\
&+& \Gamma_{S_{11} \to \pi \pi N}(M_{\pi N}),
\nonumber \\
\end{eqnarray}
with the $\Gamma_{S_{11} \to \pi N}$ and 
$\Gamma_{S_{11} \to \pi \pi N}$ 
partial widths given by Eqs.~(\ref{eq:widthpiN}) 
and~(\ref{eq:width2piN}) respectively for $l = 0$,  
and with the $S_{11} \to \eta N$ partial width $\Gamma_{S_{11} \to \eta N}$ given by:
\begin{eqnarray}
\Gamma_{S_{11} \to \eta N}(W_R) &=& \beta_{\eta N} \Gamma_R 
\left( \frac{| \vec p^{\,*}_\eta (W_R) |}{| \vec p^{\,*}_\eta (M_R) |} \right) \frac{M_R}{W_R},
\nonumber \\
\label{eq:widthetaN}
\end{eqnarray}
with $| \vec p^{\,*}_{\eta}(W_R) |$ defined as:
\begin{eqnarray}
| \vec p^{\,*}_{\eta}(W_R) | = \frac{1}{2 W_R} \lambda^{1/2}(W^2_R, M_N^2, m_\eta^2).
\nonumber \\
\label{eq:etamomstar}
\end{eqnarray}
The $S_{11}(1535)$ resonance parameters used in the present analysis are given in Table~\ref{tab1}.

\section{Results and discussion}
\label{sec5}

\subsection{Results in the $\Delta(1232)$ resonance region}

\begin{figure*}[!]
 \centering
\includegraphics[width=0.31\textwidth]{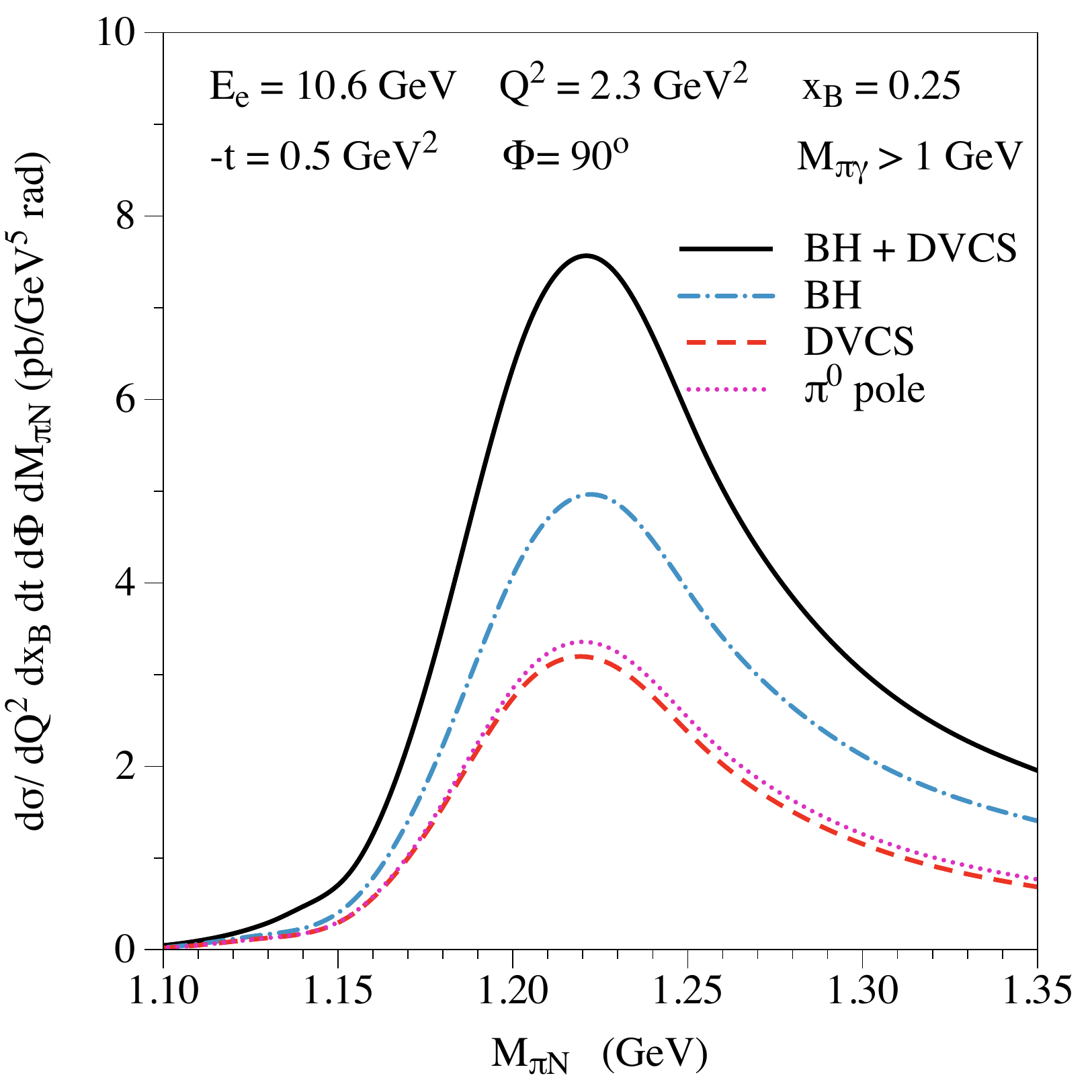}
\includegraphics[width=0.31\textwidth]{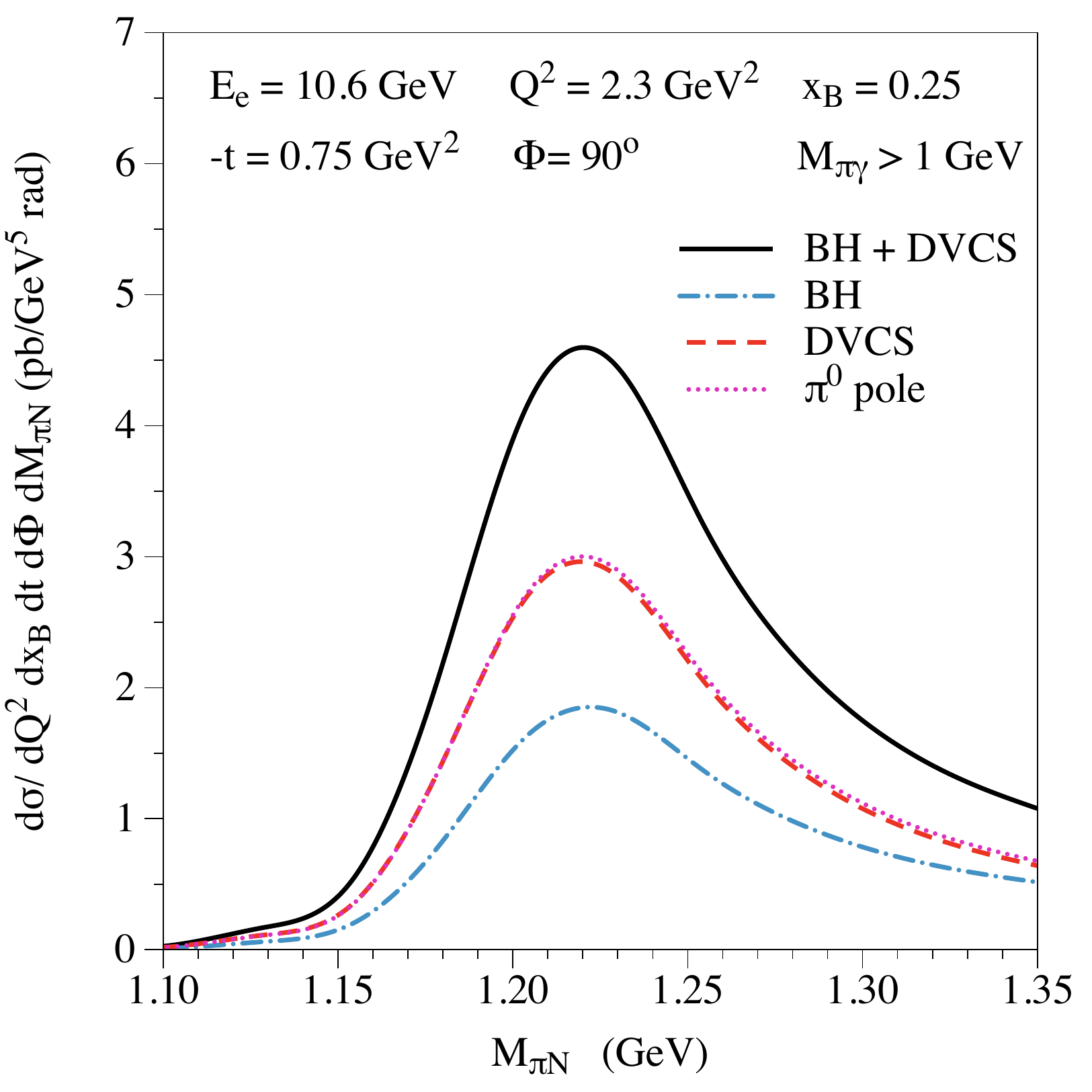}
\includegraphics[width=0.31\textwidth]{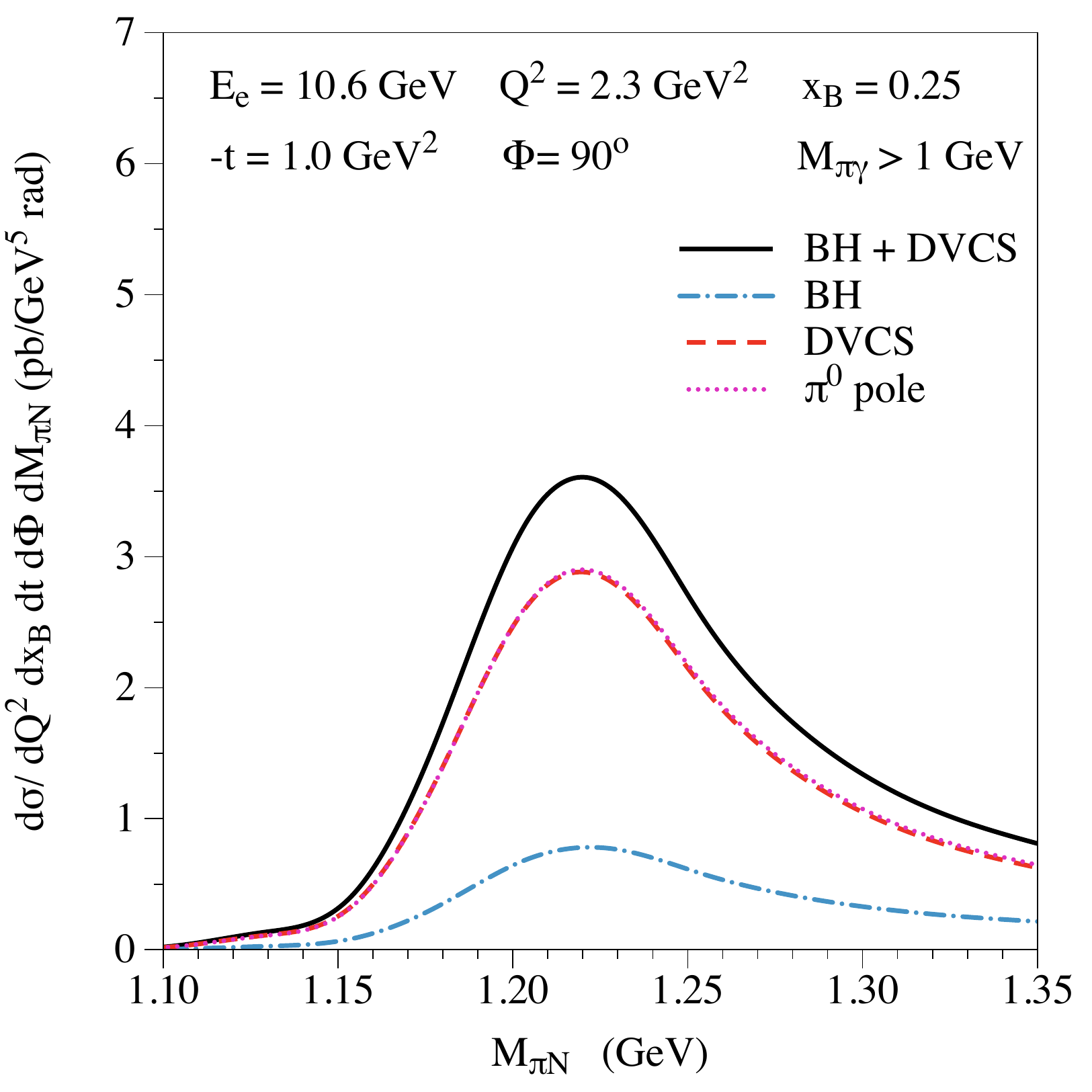}
\includegraphics[width=0.31\textwidth]{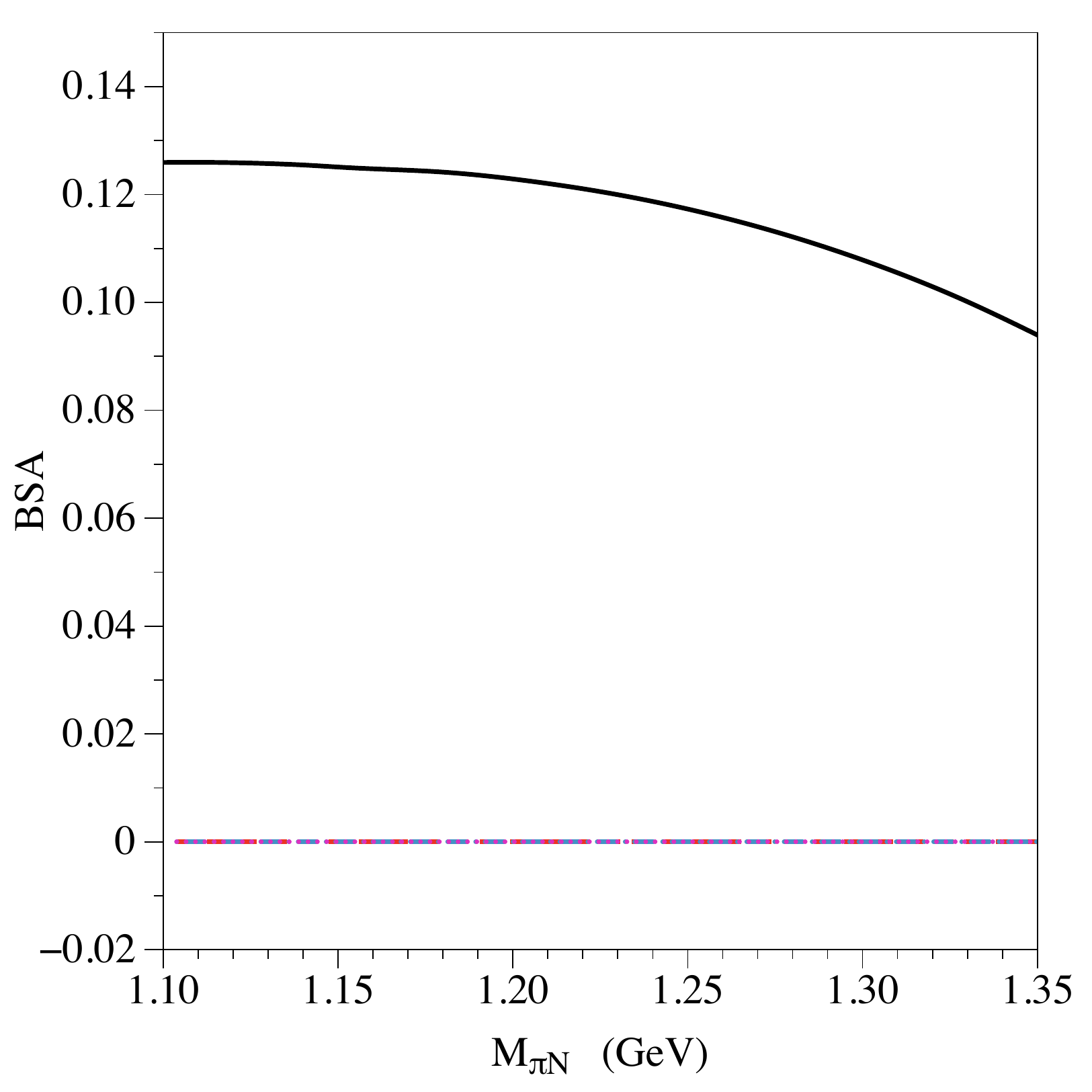}
\includegraphics[width=0.31\textwidth]{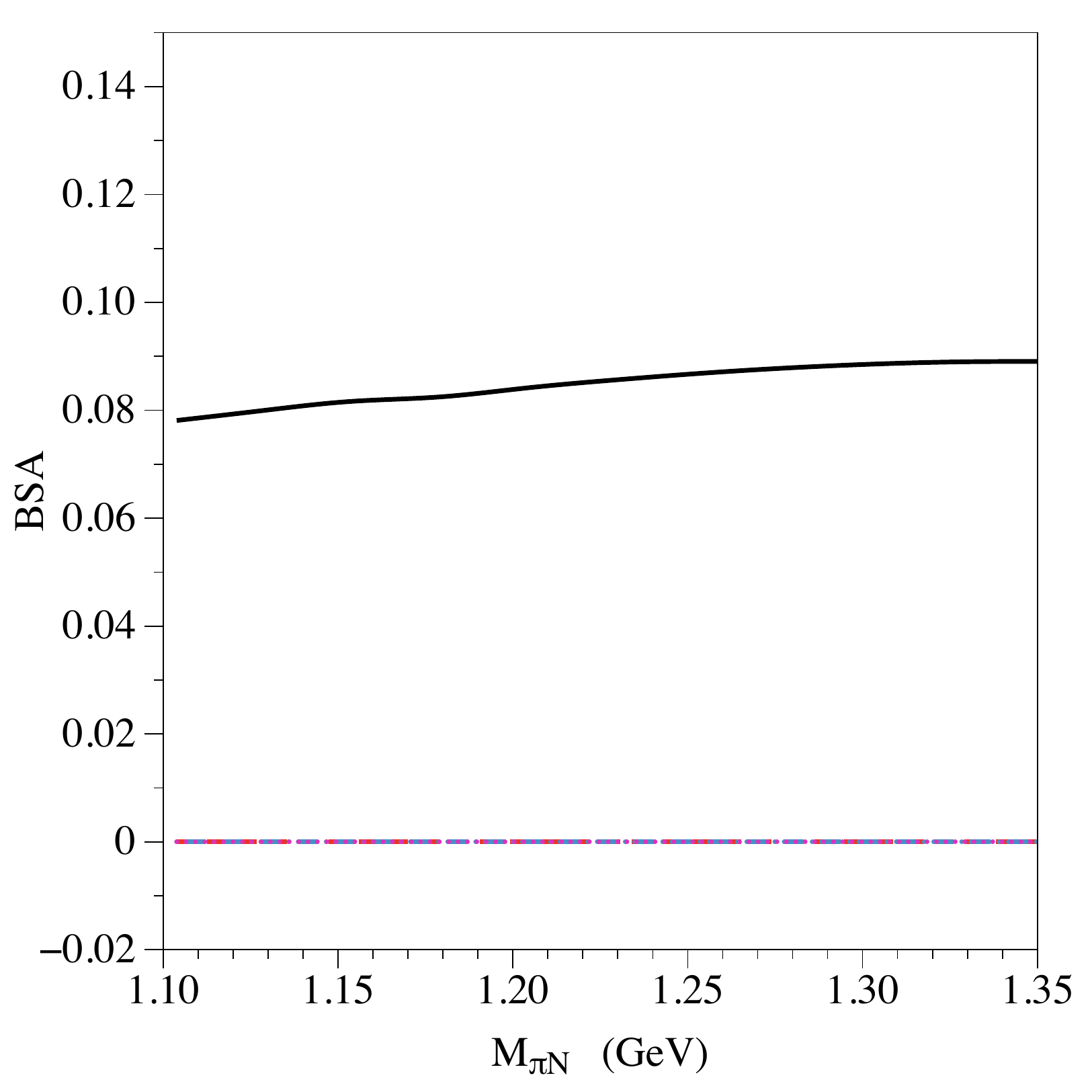}
\includegraphics[width=0.31\textwidth]{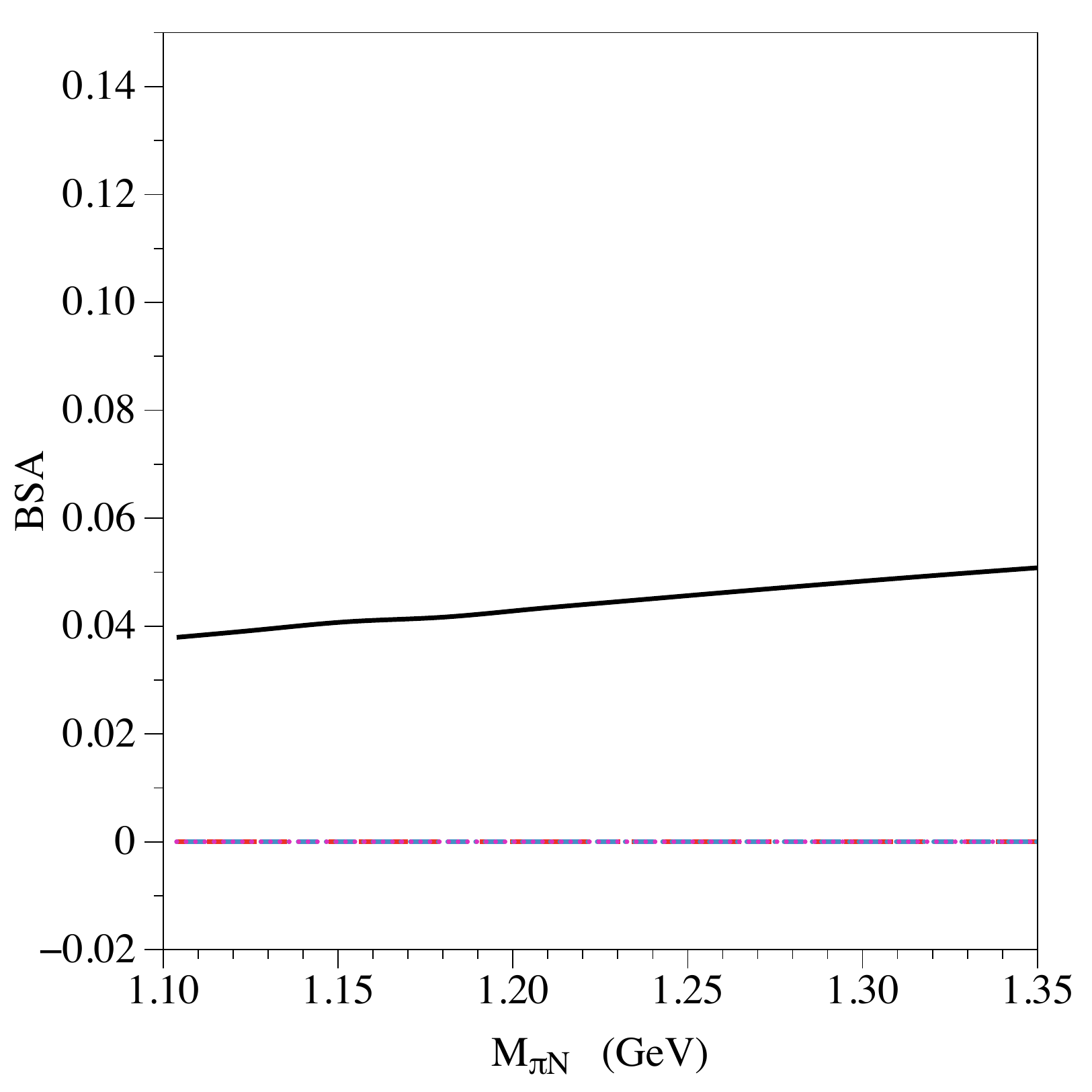}
\caption[]{Dependence on the invariant mass of the $\pi^+ n$ system ($M_{\pi N}$) of the $e^- p \to e^- \gamma \Delta(1232) \to e^- \gamma \pi^+ n$ 
cross section (upper panels) and 
corresponding beam-spin asymmetry (lower panels), integrated over the pion solid angle, 
with the cut $M_{\pi \gamma} > 1$~GeV,  for three values of $-t$.   
Blue dashed-dotted curves: $p \to \Delta(1232)$ Bethe-Heitler (BH) process; 
red dashed curves: $p \to \Delta(1232)$ DVCS process; 
black solid curves: BH + DVCS processes.  
The magenta dotted curves show the $\pi^0$-pole contribution to the $p \to \Delta(1232)$ DVCS process separately. 
}
\label{fig:5fcut}
\end{figure*}

We start our investigations by the results for the $N \to \Delta(1232)$ DVCS process. As we are interested in providing a theoretical guidance for ongoing analyses of CLAS12 data from JLab, we are choosing kinematical settings close to those of forthcoming data for the 
$e^- p \to e^- \gamma R \to e^- \gamma \pi^+ n$ reaction. 

For a realistic comparison with experiment, it is important to also determine the main physics  backgrounds contributing to the 
$e^- p \to e^- \gamma \pi^+ n$ reaction.
For invariant masses $M_{\pi N}$ around $\Delta(1232)$ energies, apart from the $e^- p \to e^- \gamma \Delta^+(1232) \to e^- \gamma \pi^+ n$ signal process, we expect the main physics background process to this reaction to originate from the $e^- p \to e^- \rho^+ n \to e^- \gamma \pi^+ n$ channel. In the latter process, a $\rho^+(770)$ is produced on the proton, which subsequently decays into $\pi^+ \gamma$. Due to the relatively large width of the $\rho$-meson, a full description of the $e^- p \to e^- \gamma \pi^+ n$ reaction in this region will require the coherent sum of both the $\gamma \Delta^+$ and $\rho^+ n$ channels, as both decay into the same final state $\gamma \pi^+ n$.

Since the main interest in this reaction is to extract information on the $N \to \Delta$ GPDs, we pursue in this work the first step towards a theoretical interpretation of forthcoming $e^- p \to e^- \gamma \pi^+ n$ data, by calculating the $e^- p \to e^- \gamma \Delta^+(1232) \to e^- \gamma \pi^+ n$ contribution. We aim to minimize the contribution arising from the $e^- p \to e^- \rho^+ n \to e^- \gamma \pi^+ n$ background process, which is expected to yield a peaked structure around $M_{\pi \gamma} \simeq 770$~MeV with a width around $150$~MeV. 
Therefore, we show in Fig.~\ref{fig:5fcut} the results for the $M_{\pi N}$ invariant mass dependence in the $\Delta(1232)$ region of the $e^- p \to e^- \gamma \pi^+ n$ cross section and 
corresponding beam-spin asymmetry (BSA) in CLAS12 kinematics, with the additional cut $M_{\pi \gamma} > 1$~GeV. The latter is chosen to ensure that one is above the $\rho^+$ production region. 
Furthermore, we choose the angle between the lepton plane and the $\gamma^\ast \gamma$ production plane in Fig.~\ref{fig:kin_plane} to be $\Phi = 90^\circ$, where the BSA becomes maximal. By comparing the $t$-dependence between $-t = 0.5$~GeV$^2$ and $-t = 1.0$~GeV$^2$, we notice that in the lower $t$-range, the BH process dominates the cross section. In the BH amplitude, the virtual photon propagator has a $1/t$ behavior, which leads at fixed value of $Q^2$ and $x_B$ to a much faster decrease in the cross section, with increasing values of $-t$, as compared to the DVCS process. We also note from Fig.~\ref{fig:5fcut} that the DVCS process in the $\Delta$-region is dominated by the $\pi^0$-pole contribution to the $N \to \Delta$ GPD $C_2$. For the corresponding BSA, which is obtained by flipping the helicity of the electron beam, we notice that in the lower $-t$ range, the interference of the imaginary part of the DVCS amplitude with the BH process leads to a BSA in the range of 10 \%. With increasing  values of $-t$, due to the decrease of the BH process relative to the DVCS process, we notice that the BSA also gradually decreases.

\begin{figure}[!]
\includegraphics[width=0.49\columnwidth]{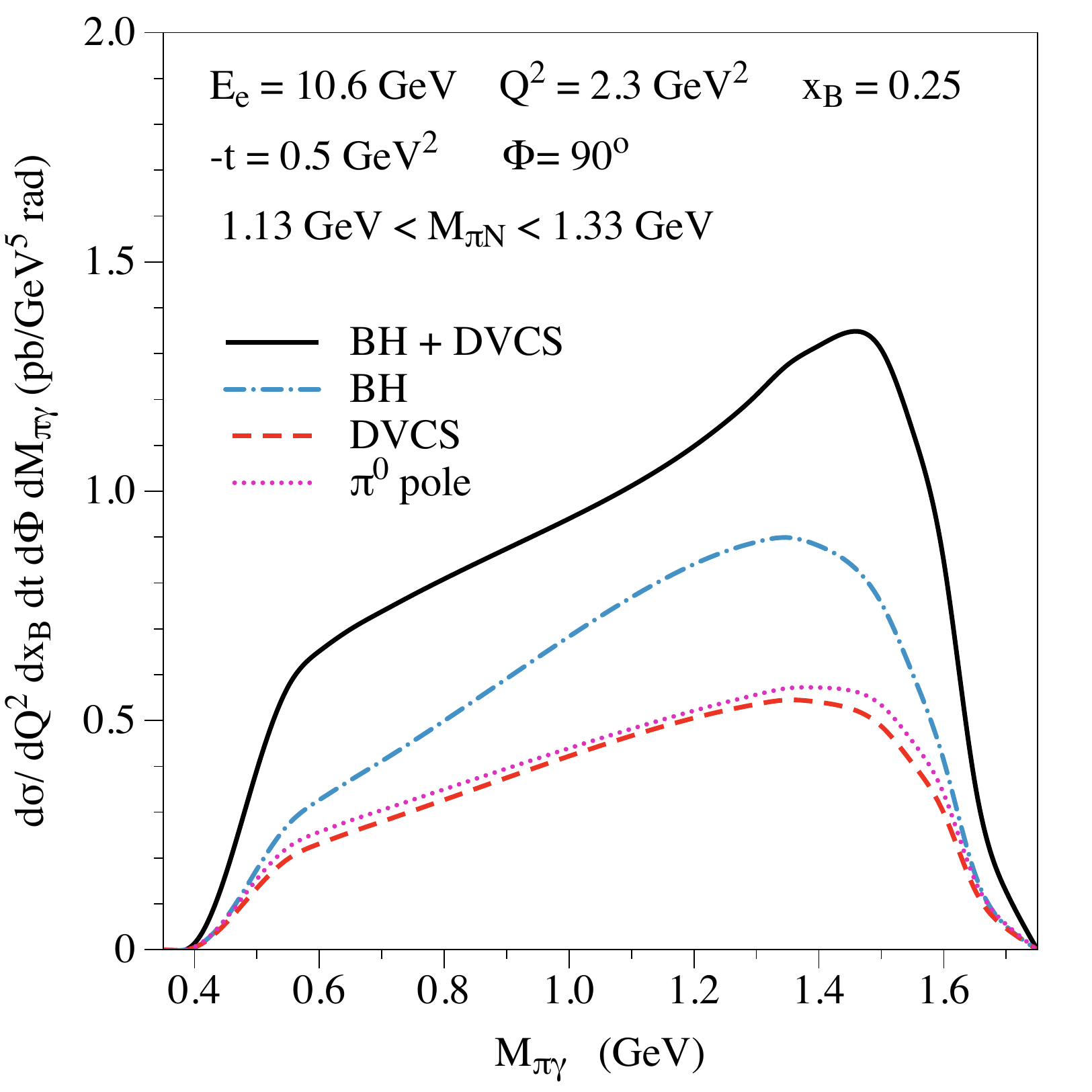}
\includegraphics[width=0.49\columnwidth]{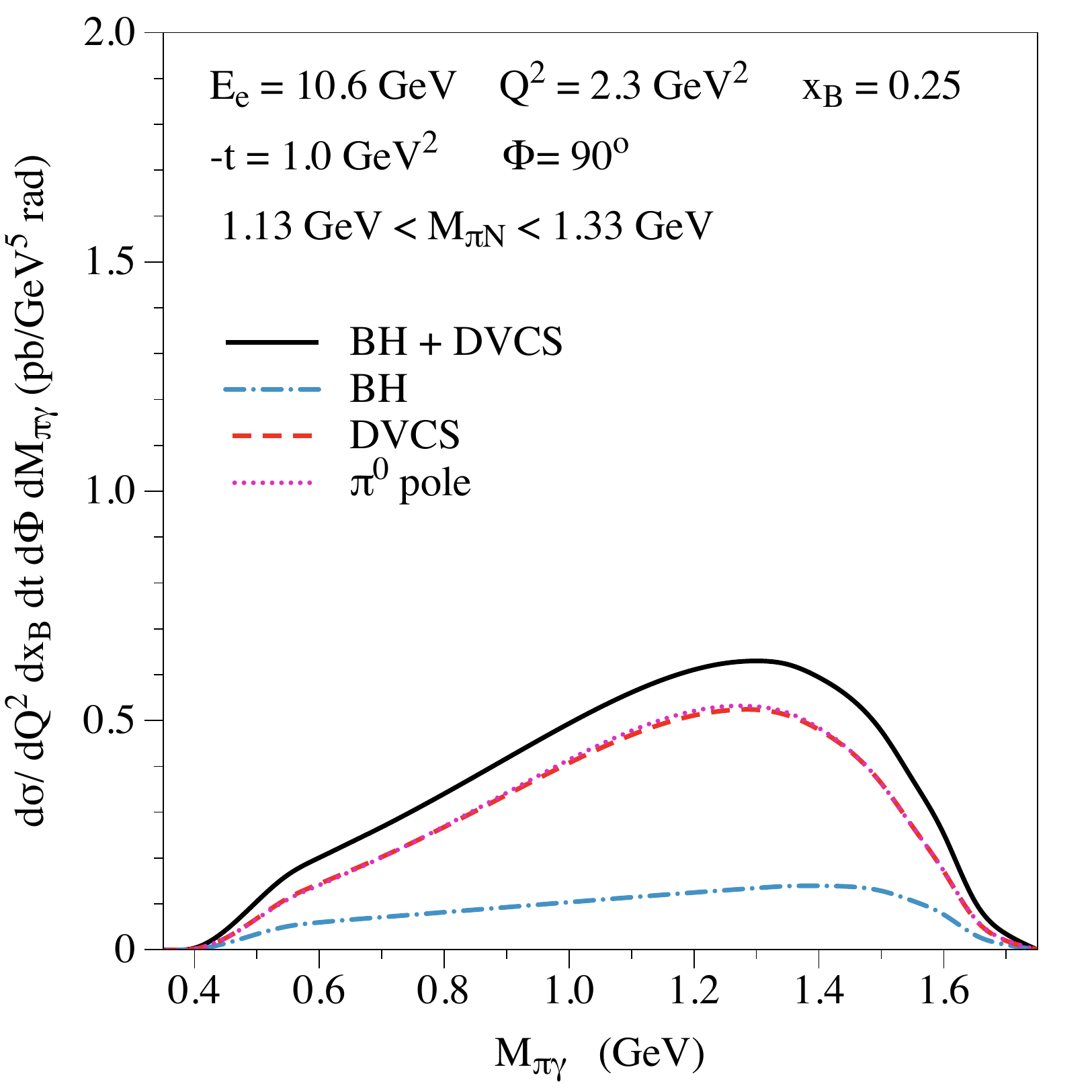}
\includegraphics[width=0.49\columnwidth]{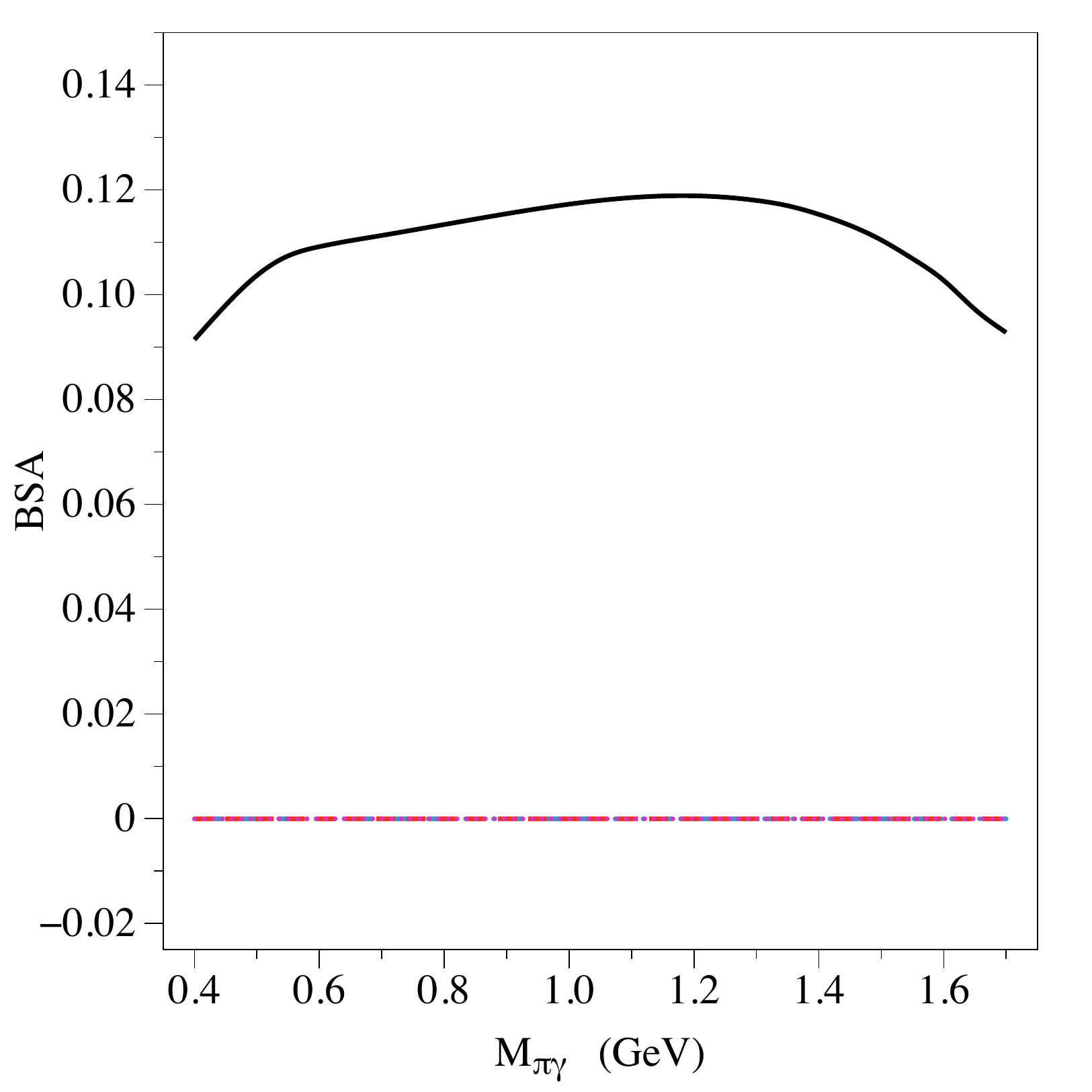}
\includegraphics[width=0.49\columnwidth]{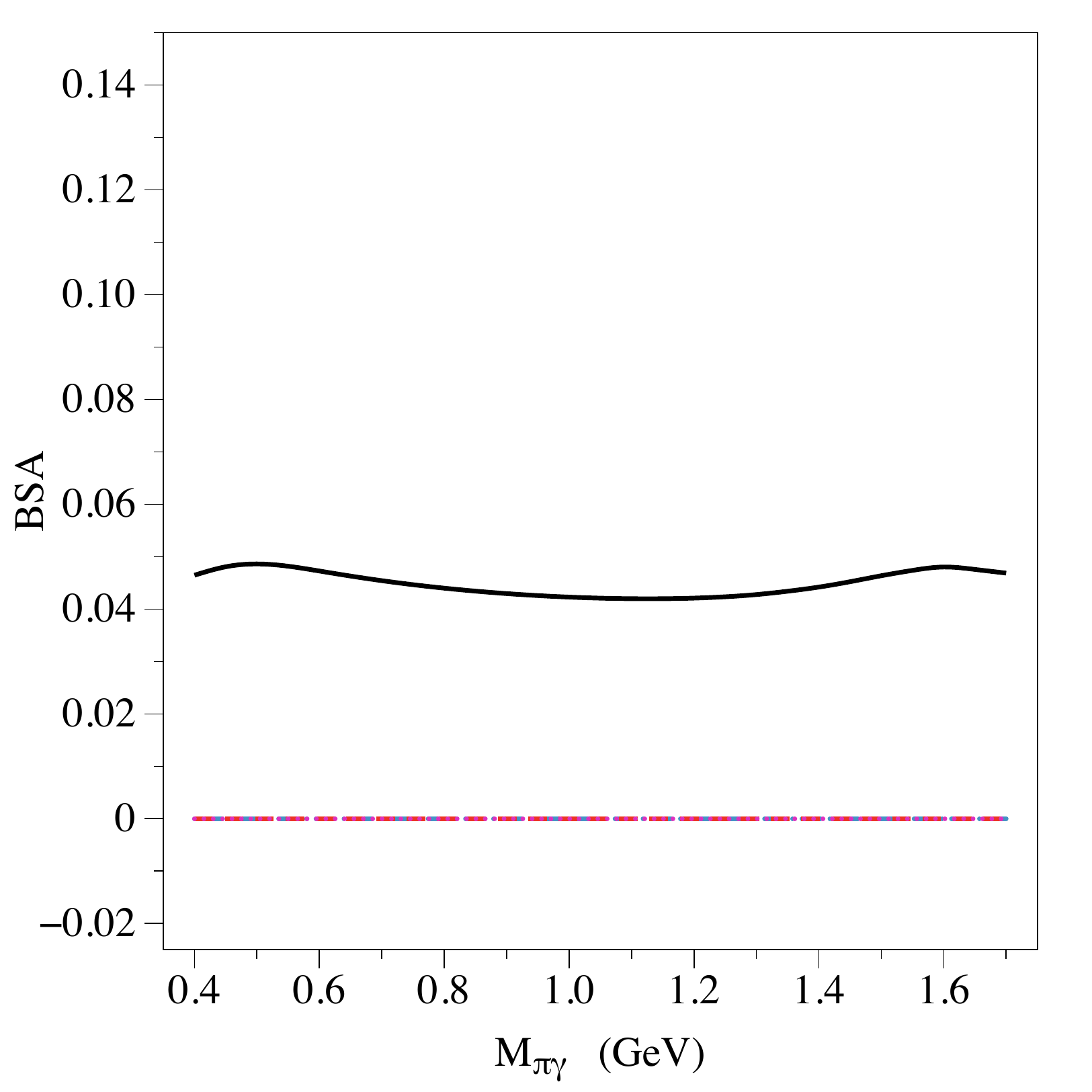}
\caption[]{Dependence on the invariant mass of the $\gamma \pi^+$ system ($M_{\pi \gamma}$) 
of the $e^- p \to e^- \gamma \Delta(1232) \to e^- \gamma \pi^+ n$ 
cross section (upper panels) and 
corresponding beam-spin asymmetry (lower panels), integrated over the bin: 1.13~GeV $\leq M_{\pi N} \leq$ 1.33~GeV, 
for two values of $-t$. 
Curve conventions as in Fig.~\ref{fig:5fcut}. 
}
\label{fig:5fmpiga}
\end{figure}

\begin{figure}[!]
    \centering
\includegraphics[width=0.49\columnwidth]{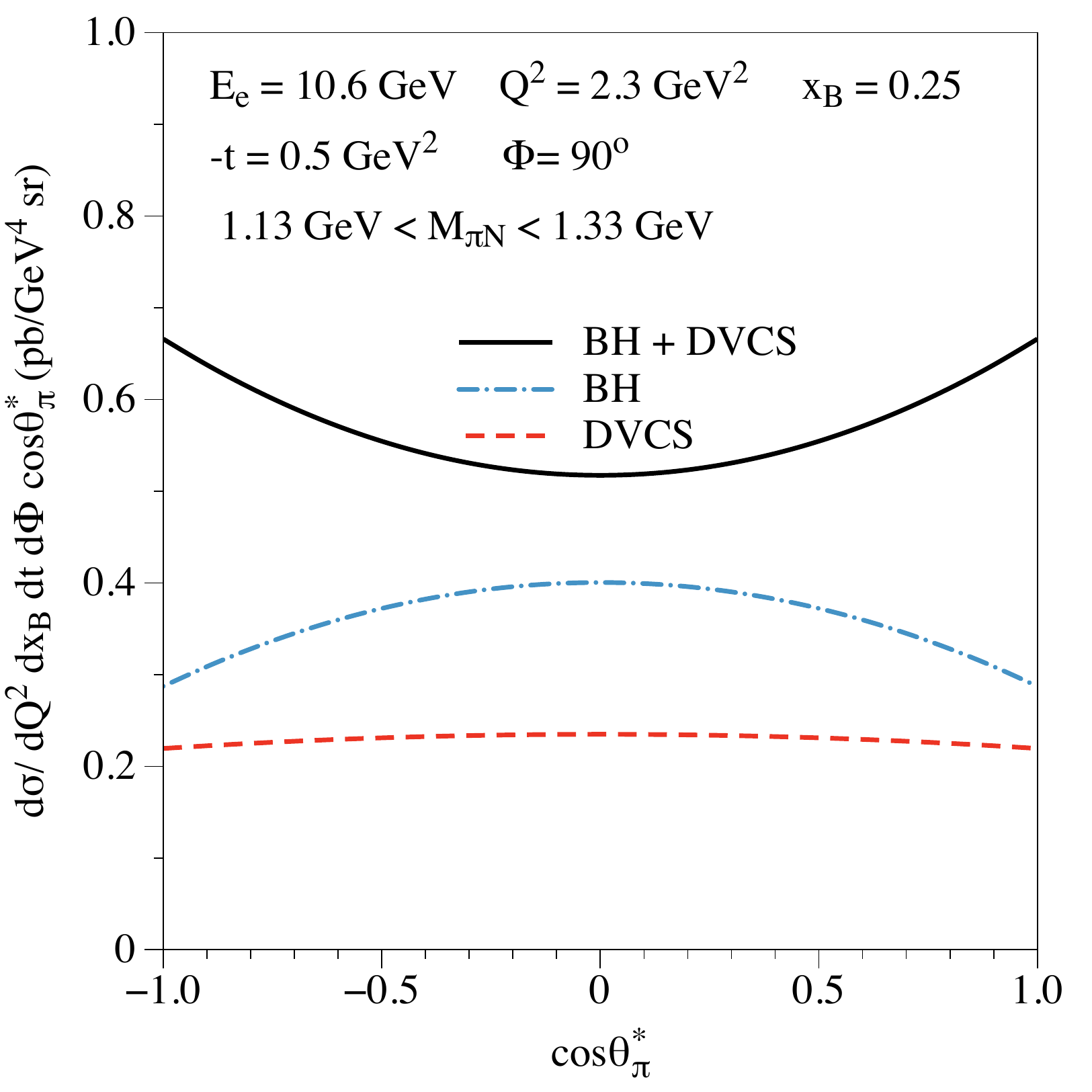}
\includegraphics[width=0.49\columnwidth]{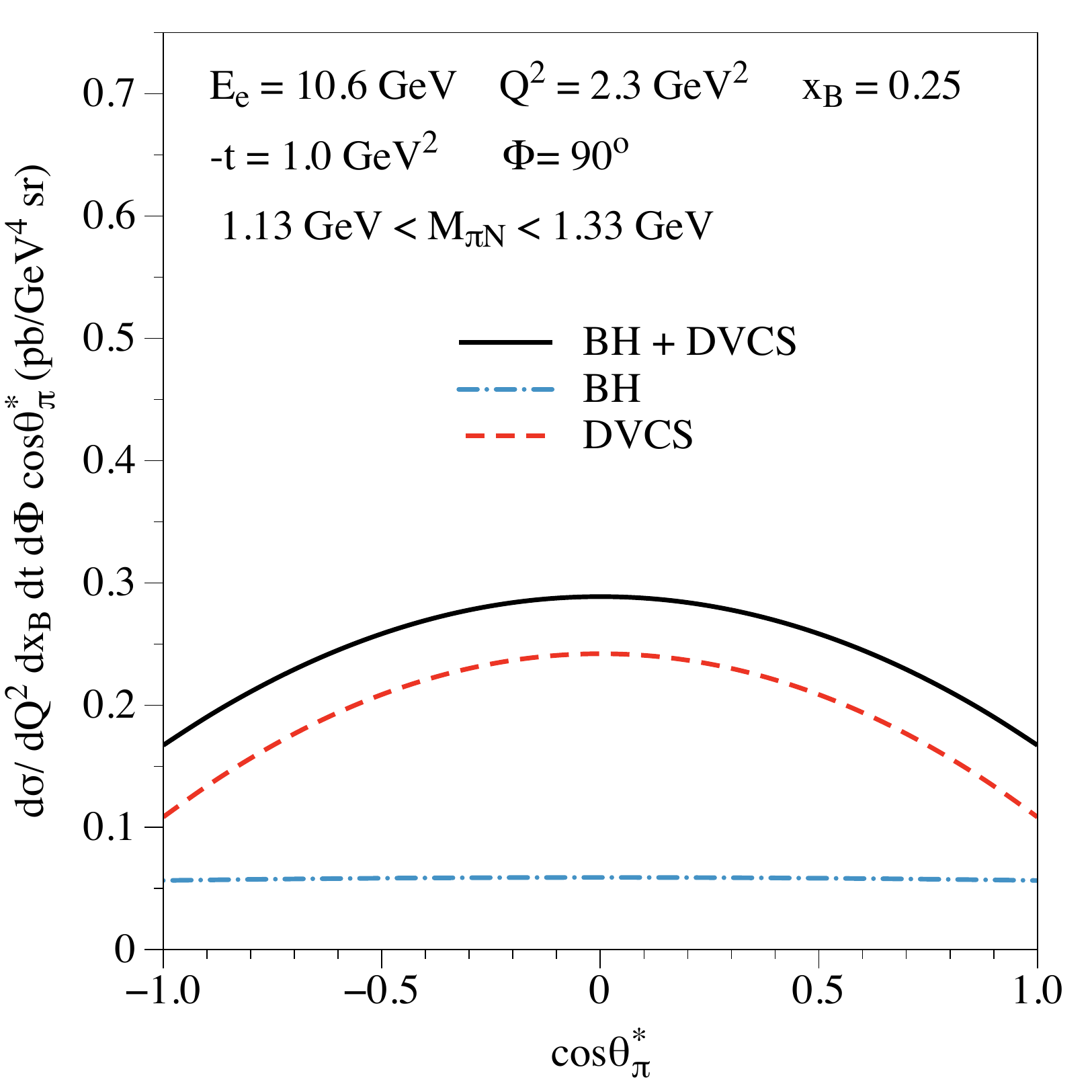}
\includegraphics[width=0.49\columnwidth]{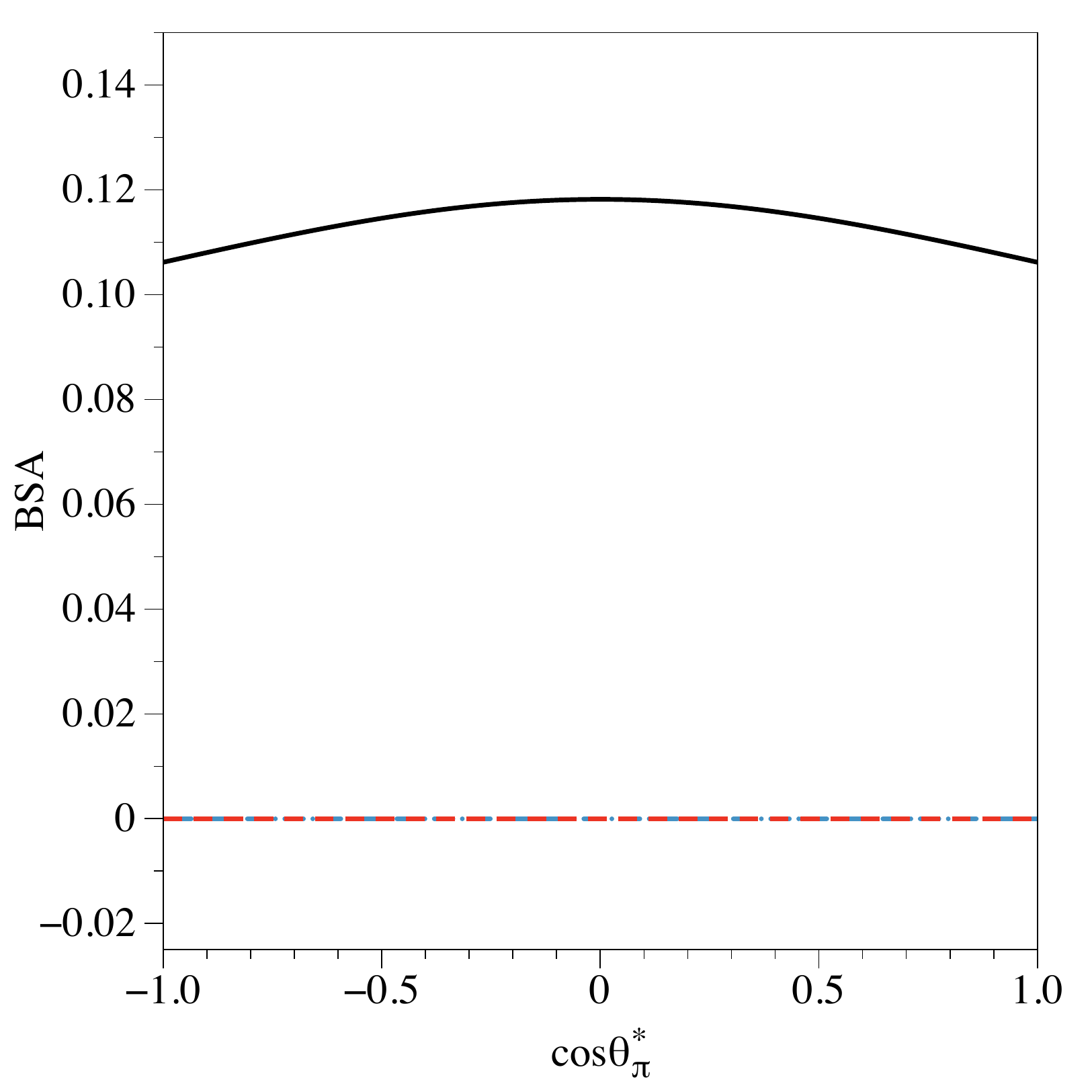}
\includegraphics[width=0.49\columnwidth]{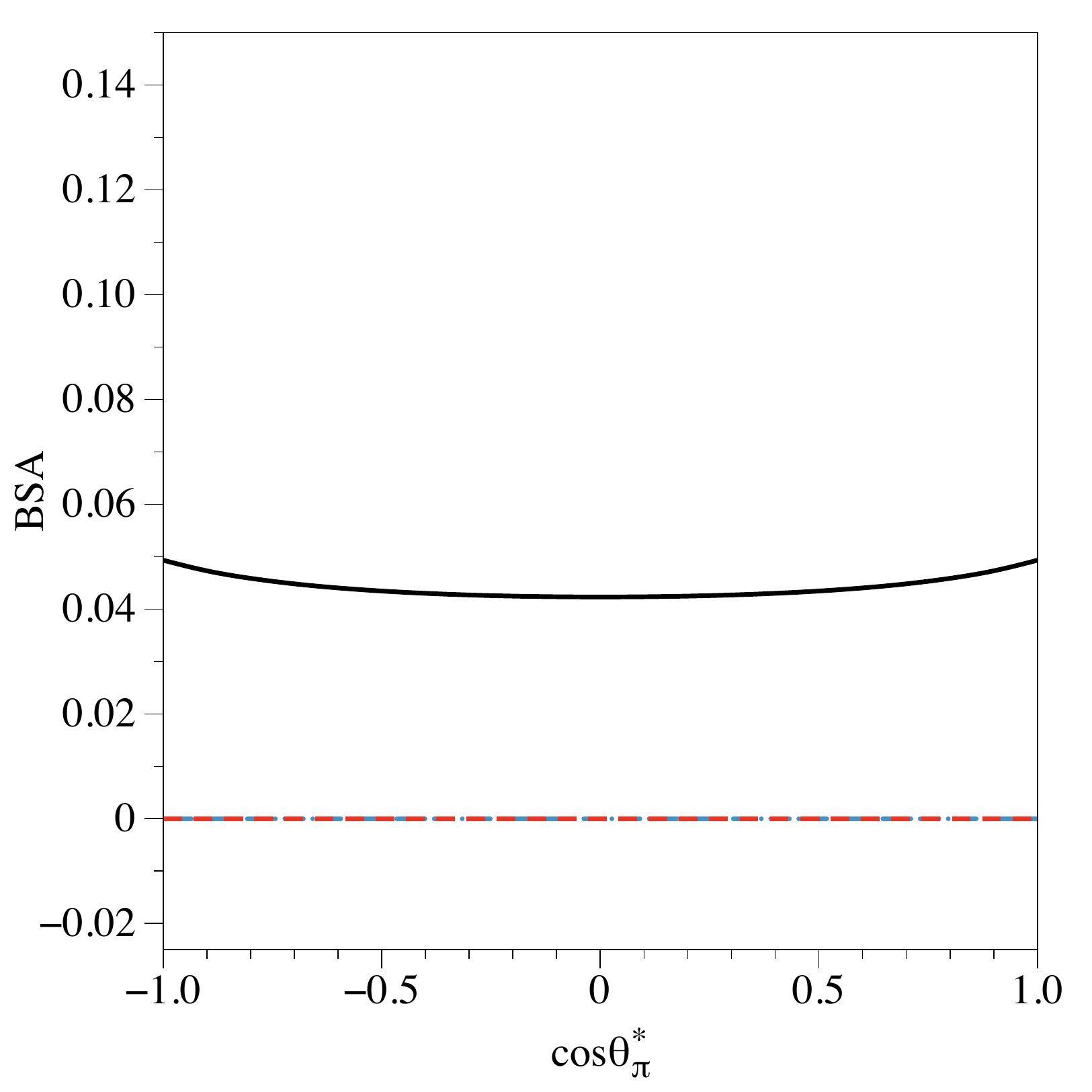}
\caption[]{Decay pion angular distribution (defined in the $\pi N$ rest frame) 
 of the $e^- p \to e^- \gamma \Delta(1232) \to e^- \gamma \pi^+ n$ 
cross section (upper panels) and 
corresponding beam-spin asymmetry (lower panels), integrated over the bin: 1.13~GeV $\leq M_{\pi N} \leq$ 1.33~GeV, corresponding with the $\Delta(1232)$ resonance region, 
for two values of $-t$. 
Curve conventions as in Fig.~\ref{fig:5fcut}. 
}
\label{fig:piang}
\end{figure}

In Fig.~\ref{fig:5fmpiga}, we show the 
$M_{\pi \gamma}$ invariant mass dependence of the 
$e^- p \to e^- \gamma \Delta^+(1232) \to e^- \gamma \pi^+ n$ cross section contribution when integrating over the $\Delta^+(1232)$ peak, i.e. for 1.13~GeV $\leq M_{\pi N} \leq$ 1.33~GeV. We note that the $\Delta^+(1232)$ production process yields a dependence which is rising with increasing value of $M_{\pi \gamma}$, with the dominant strength located in the region $M_{\pi \gamma} > 1$~GeV. It thus displays a distinctive difference from an expected $e^- p \to e^- \rho^+(770) n \to e^- \gamma \pi^+ n$ contribution, which is peaked around $M_{\pi \gamma} \simeq 770$~MeV, and has a strength  largely located in the region $M_{\pi \gamma} < 1$~GeV.

In Fig.~\ref{fig:piang}, we show the 
decay pion angular distribution of the 
$e^- p \to e^- \gamma \Delta^+(1232) \to e^- \gamma \pi^+ n$ process integrating over the $\Delta^+(1232)$ peak, i.e. for 1.13~GeV $\leq M_{\pi N} \leq$ 1.33~GeV. 
We note from Eq.~(\ref{eq:crossnarrow}) that a flat dependence in $\cos \theta_\pi^\ast$ results from a $\Delta^+$ produced with same probability in helicity $s_\Delta = \pm 3/2$ and helicity $s_\Delta = \pm 1/2$ states. Furthermore, Eq.~(\ref{eq:crossnarrow}) shows that the production of $s_\Delta = \pm 1/2$ states results in a distribution proportional to $\sim (1 + 3~\cos^2 \theta_\pi^\ast)/4$, whereas the production of $s_\Delta = \pm 3/2$ states results in a distribution proportional to 
$\sim 3 \sin^2 \theta_\pi^\ast /4$. One sees from  
Fig.~\ref{fig:piang} that the cross section for the BH + DVCS process at $-t = 0.5$~GeV$^2$corresponds with a $\Delta^+$ produced dominantly with $s_\Delta = \pm 1/2$, resulting in a $\cos \theta_\pi^\ast$ distribution with positive curvature. On the other hand, at $-t = 1.0$~GeV$^2$ 
the $\Delta^+$ is dominantly produced with $s_\Delta = \pm 3/2$, resulting in a $\cos \theta_\pi^\ast$ cross section distribution with negative curvature.

\begin{figure}[!]
    \centering
\includegraphics[width=0.49\columnwidth]{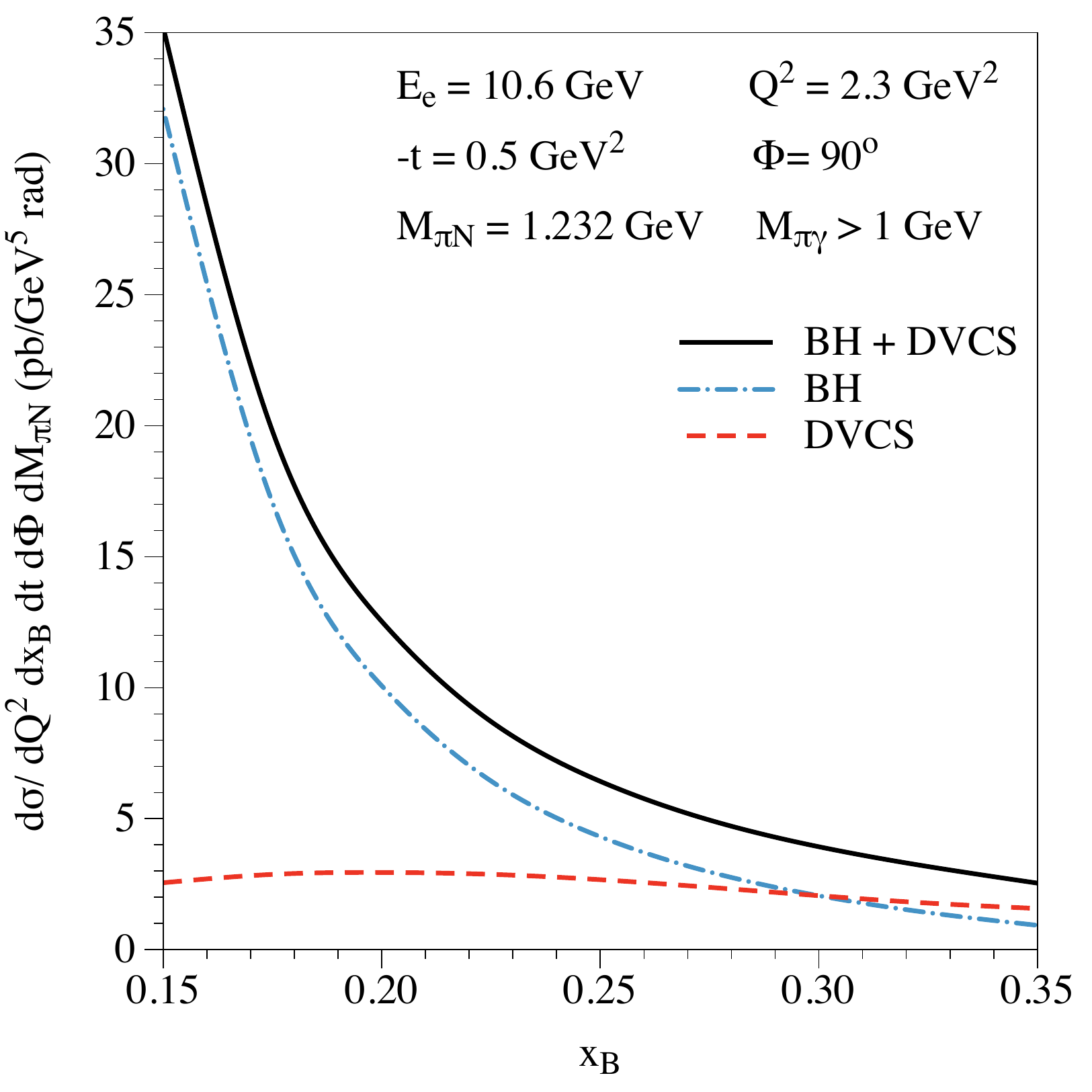}
\includegraphics[width=0.49\columnwidth]{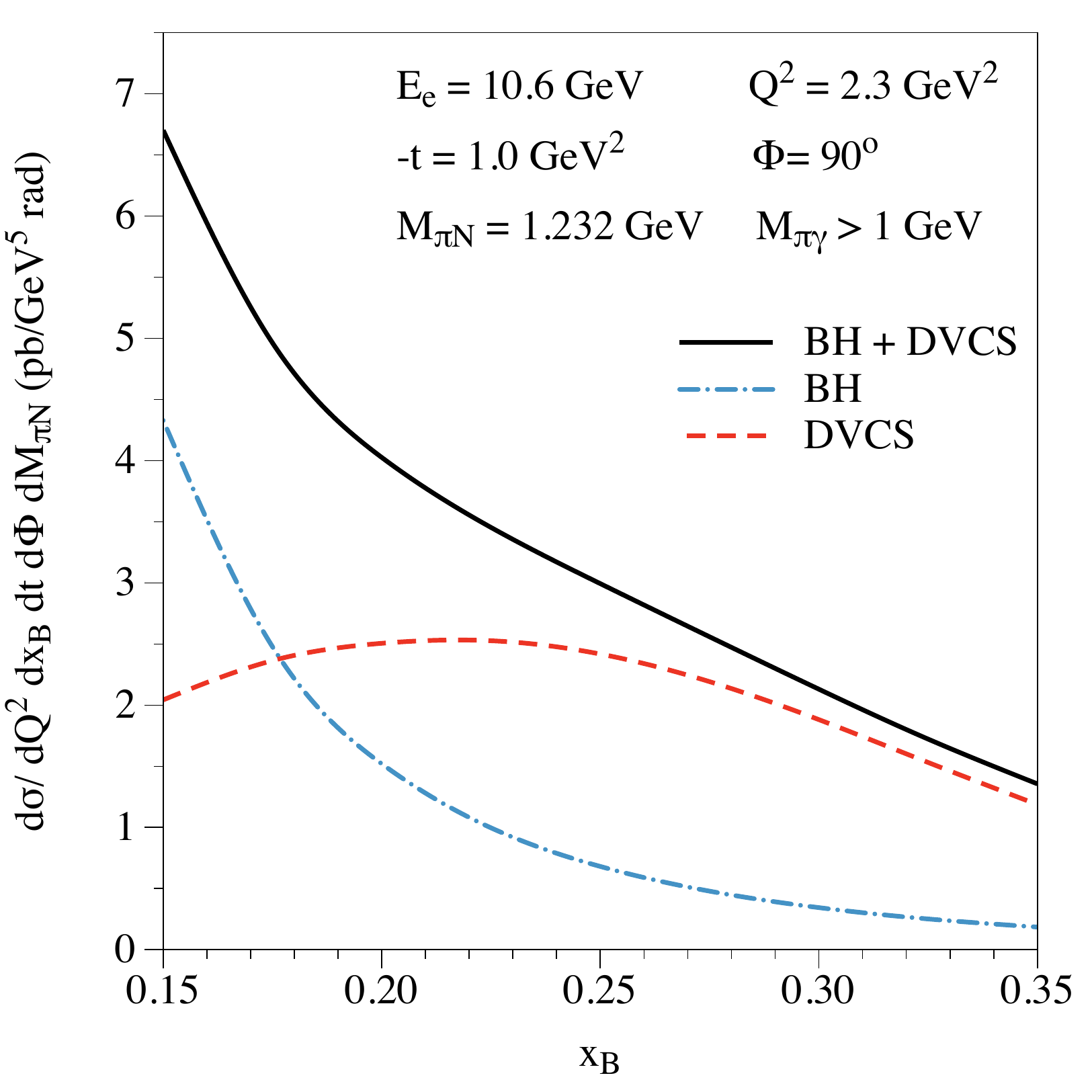}
\includegraphics[width=0.49\columnwidth]{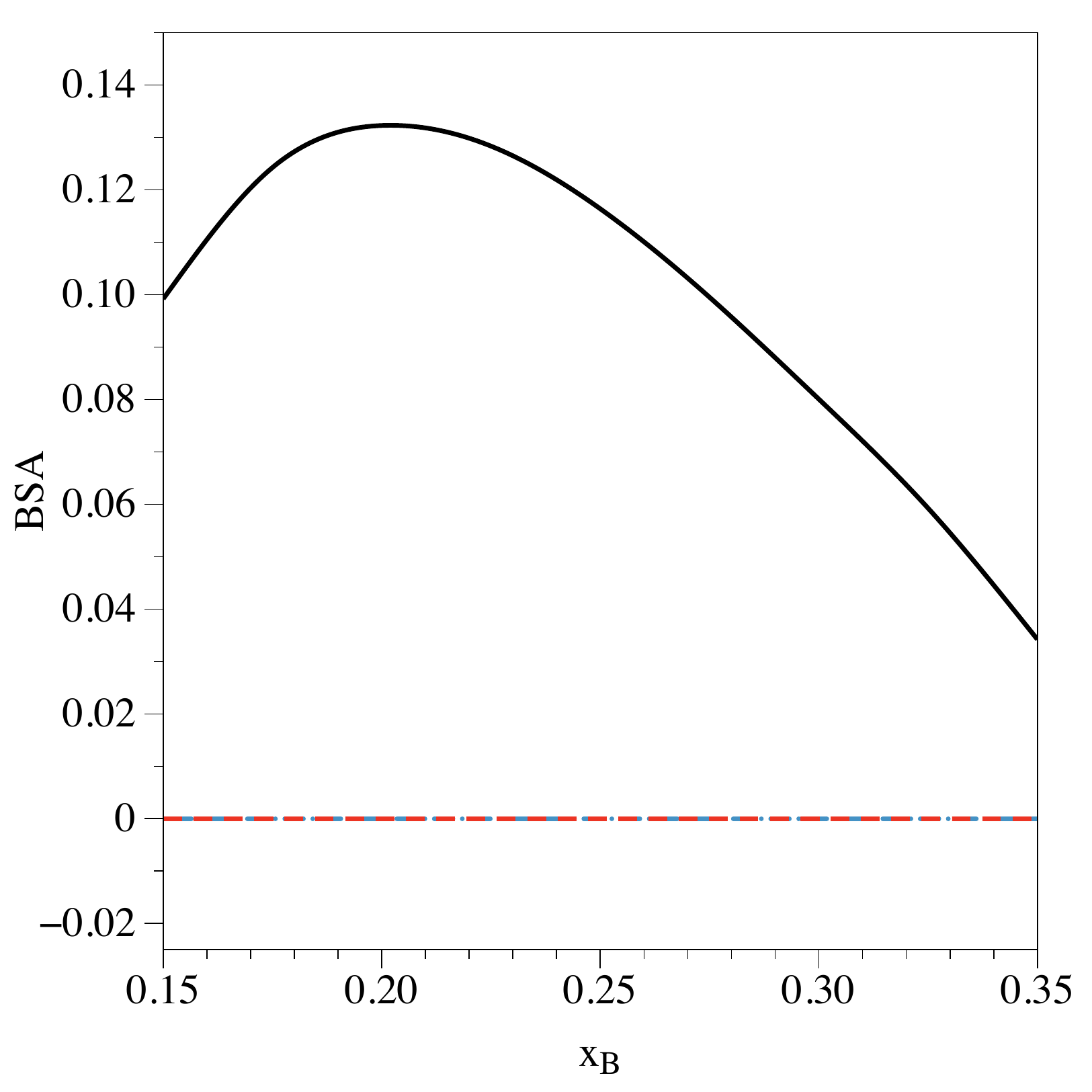}
\includegraphics[width=0.49\columnwidth]{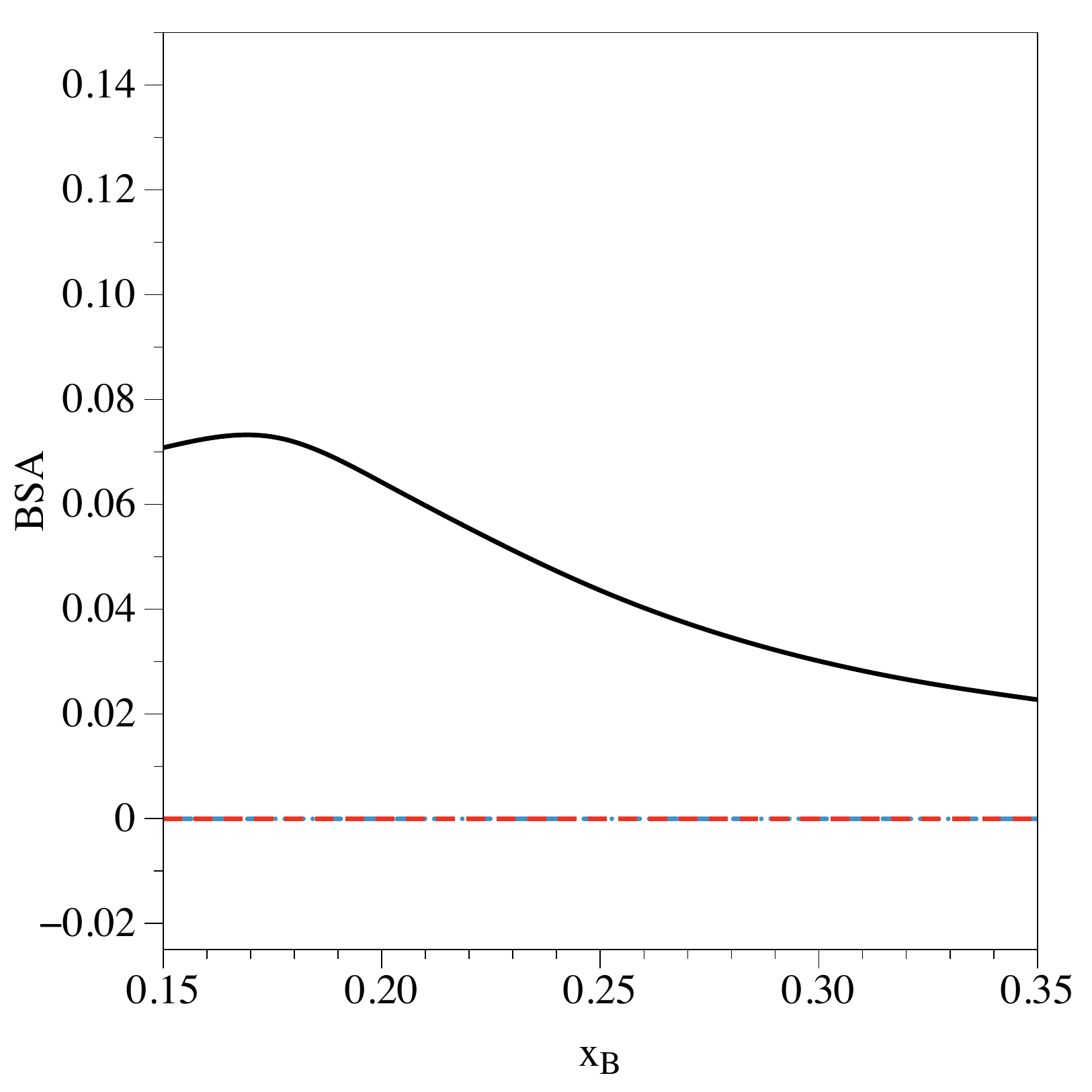}
\caption[]{$x_B$ dependence of the $e^- p \to e^- \gamma \Delta(1232) \to e^- \gamma \pi^+ n$ 
cross section (upper panels) and 
corresponding beam-spin asymmetry (lower panels) for $M_{\pi N} = 1.232$~GeV, 
integrated over the decay pion solid angle, with the cut $M_{\pi \gamma} > 1$~GeV, for two values of $-t$.  
Blue dashed-dotted curves: $p \to \Delta(1232)$ Bethe-Heitler (BH) process; 
red dashed curves: $p \to \Delta^+$ DVCS process; 
black solid curves: BH + DVCS processes. The $p \to \Delta^+$ DVCS process is calculated using the dominant GPDs for the vector and axial-vector transitions (GPDs $H_M$ and $C_1, C_2$ respectively). 
}
\label{fig:5fxb}
\end{figure}

In Fig.~\ref{fig:5fxb}, we show the $x_B$ dependence of the cross section and BSA for the $e^- p \to e^- \gamma \Delta(1232) \to e^- \gamma \pi^+ n$ process at the $\Delta$-resonance peak, with the cut $M_{\pi \gamma} > 1$~GeV. We notice that the 
BH process shows a strong rise for decreasing $x_B$ values. On the other hand the DVCS process, which is  estimated using a valence type parameterization for the $N \to \Delta$ GPDs, shows a much flatter $x_B$ dependence. This valence type behavior is also reflected in the BSA, which is due to the interference between the BH and DVCS amplitudes. 

\begin{figure*}[!]
    \centering
\includegraphics[width=0.45\textwidth]{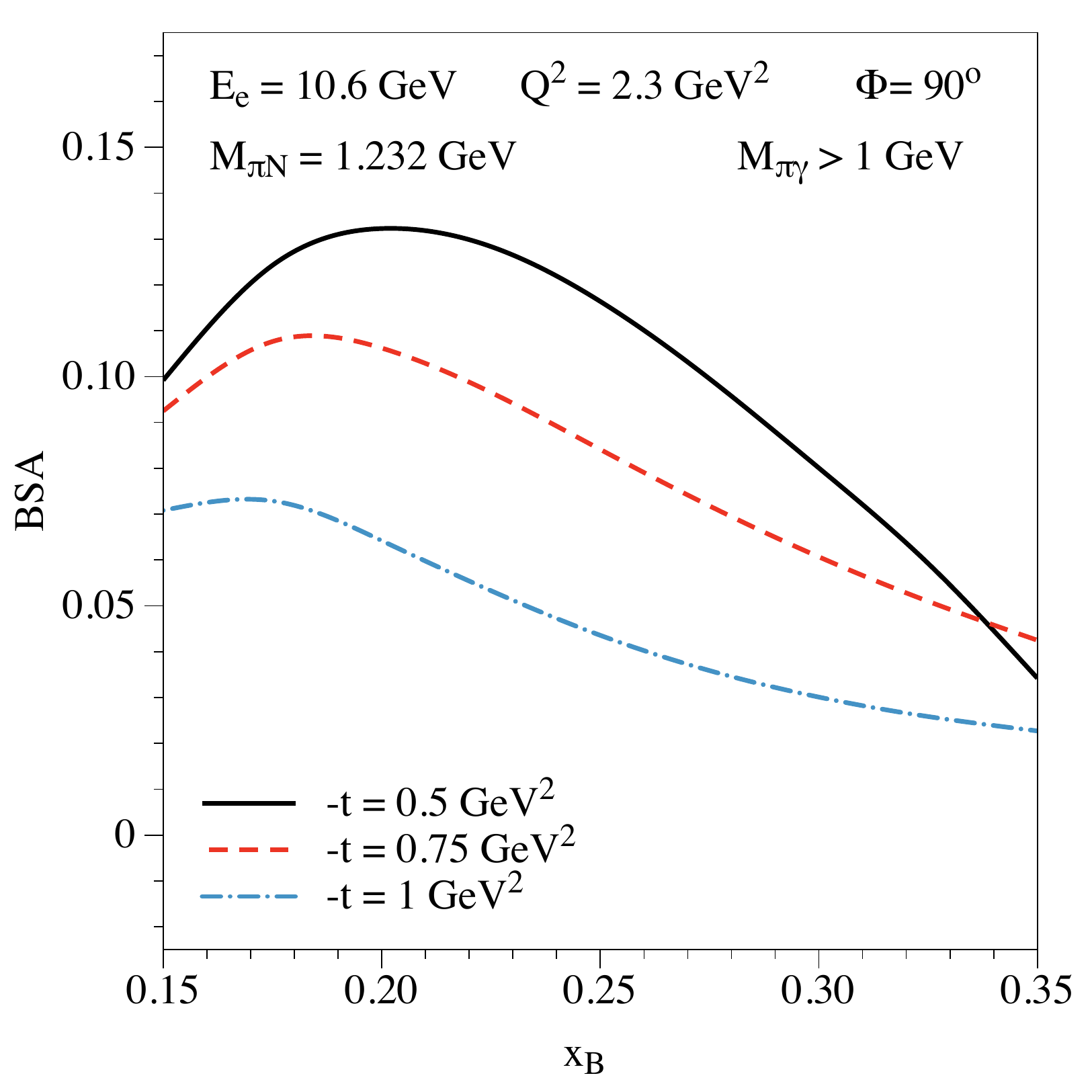}
\includegraphics[width=0.45\textwidth]{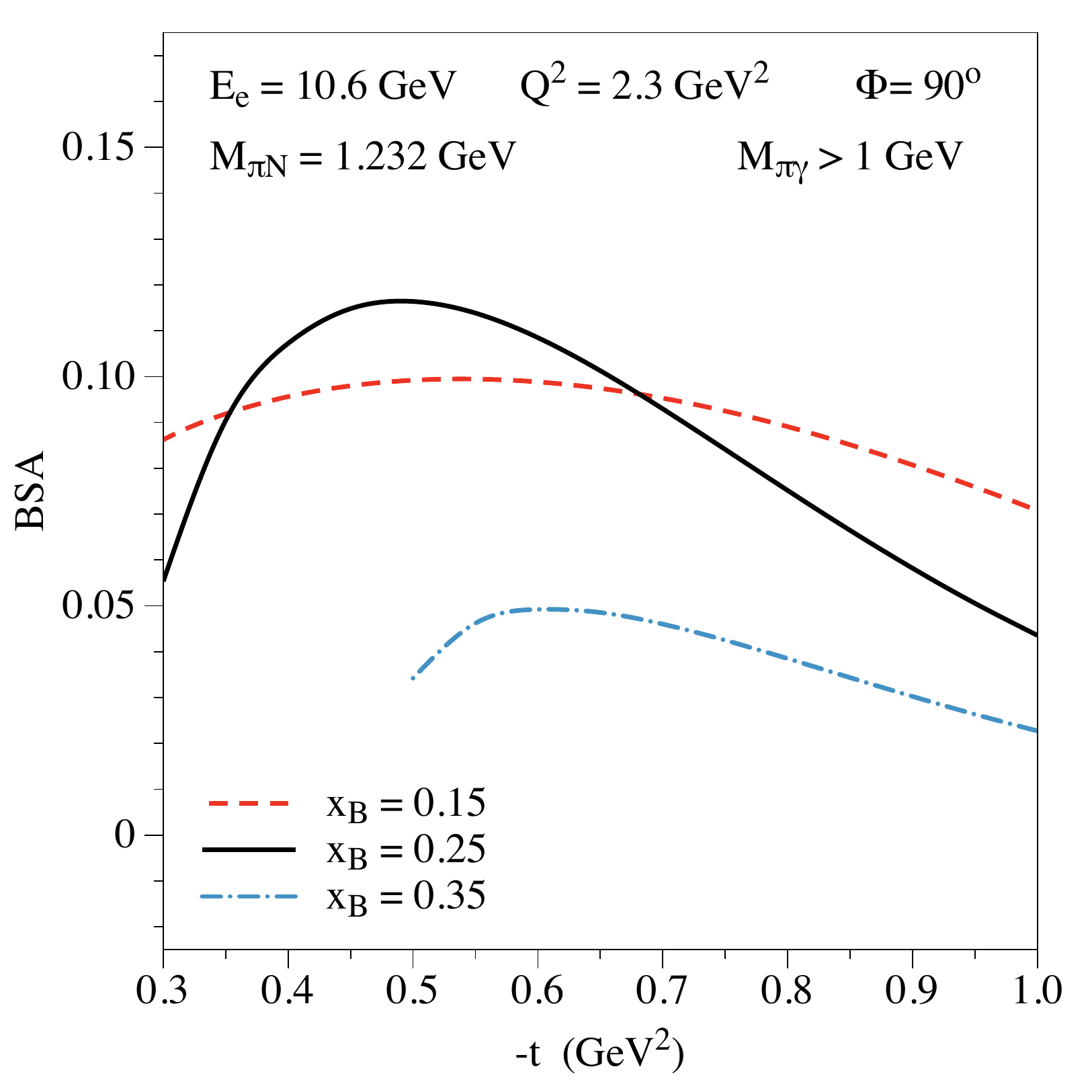}
\caption[]{Dependence on $x_B$ (left panel) and on $-t$ (right panel) of the $e^- p \to e^- \gamma \Delta(1232) \to e^- \gamma \pi^+ n$ 
beam-spin asymmetry for $M_{\pi N} = 1.232$~GeV, integrated over the decay pion solid angle, with the cut $M_{\pi \gamma} > 1$~GeV.  
}
\label{fig:5fbsa}
\end{figure*}

In Fig.~\ref{fig:5fbsa} (left panel), we compare the $x_B$ dependence of the BSA for three values of $-t$. The right panel of Fig.~\ref{fig:5fbsa} shows the $t$-dependence for three values of $x_B$ in the valence quark region. Forthcoming CLAS12 data on this observable have the strong potential to test the large-$N_c$ relations for the $N \to \Delta$ GPDs which underlie our predictions. 

\subsection{Results in the second nucleon resonance region}

\begin{figure*}[!]
\centering
\includegraphics[width=0.45\textwidth]{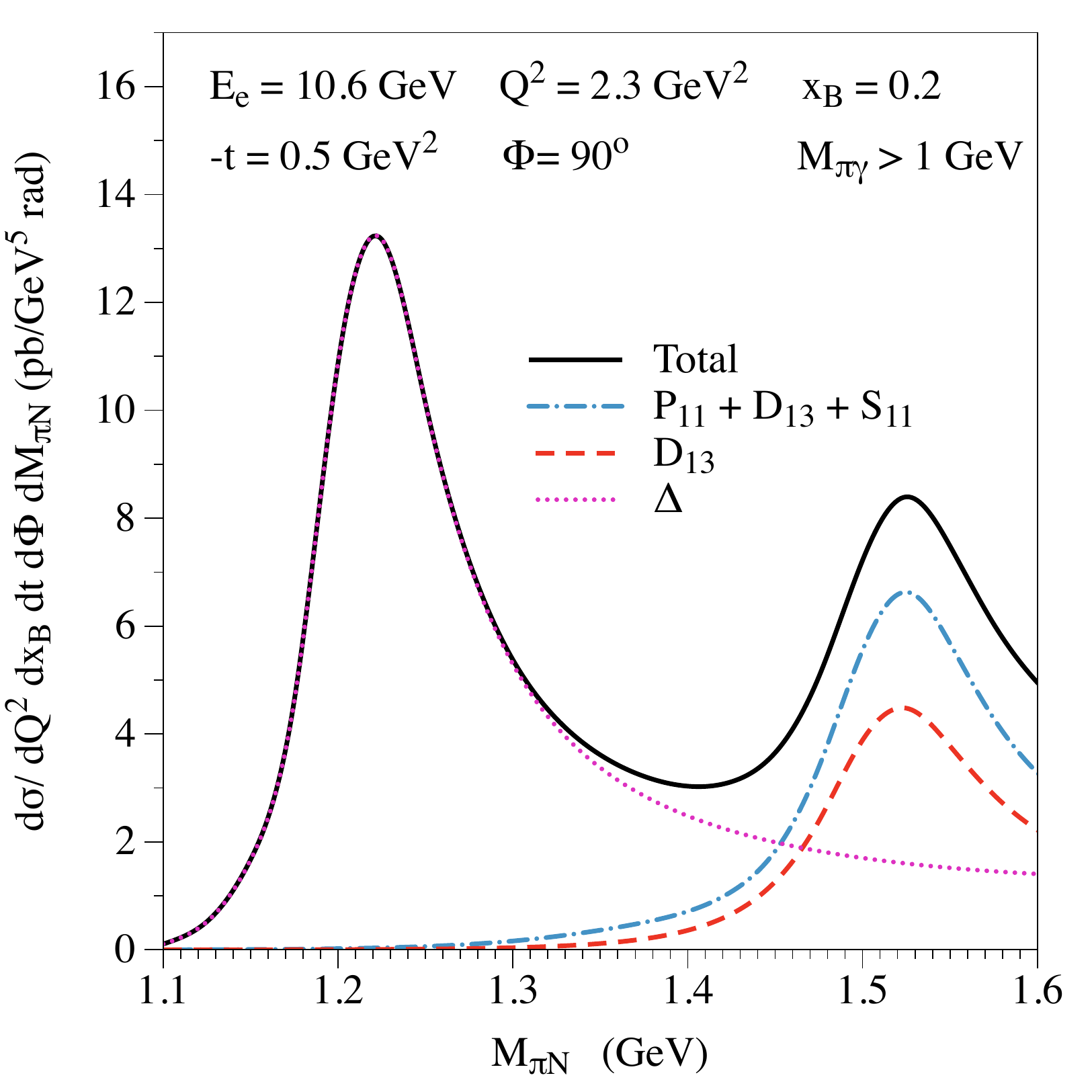}
\includegraphics[width=0.45\textwidth]{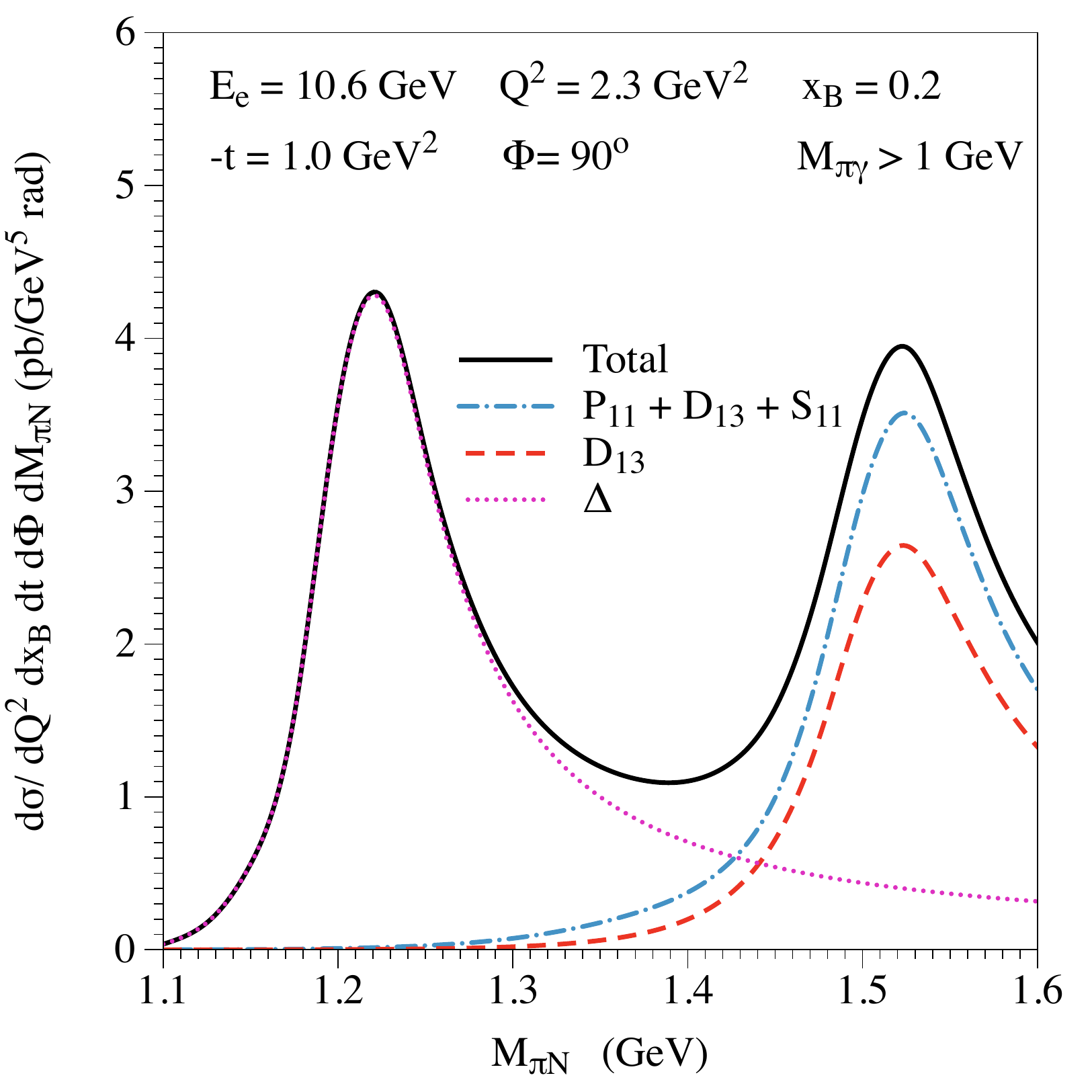}
\includegraphics[width=0.45\textwidth]{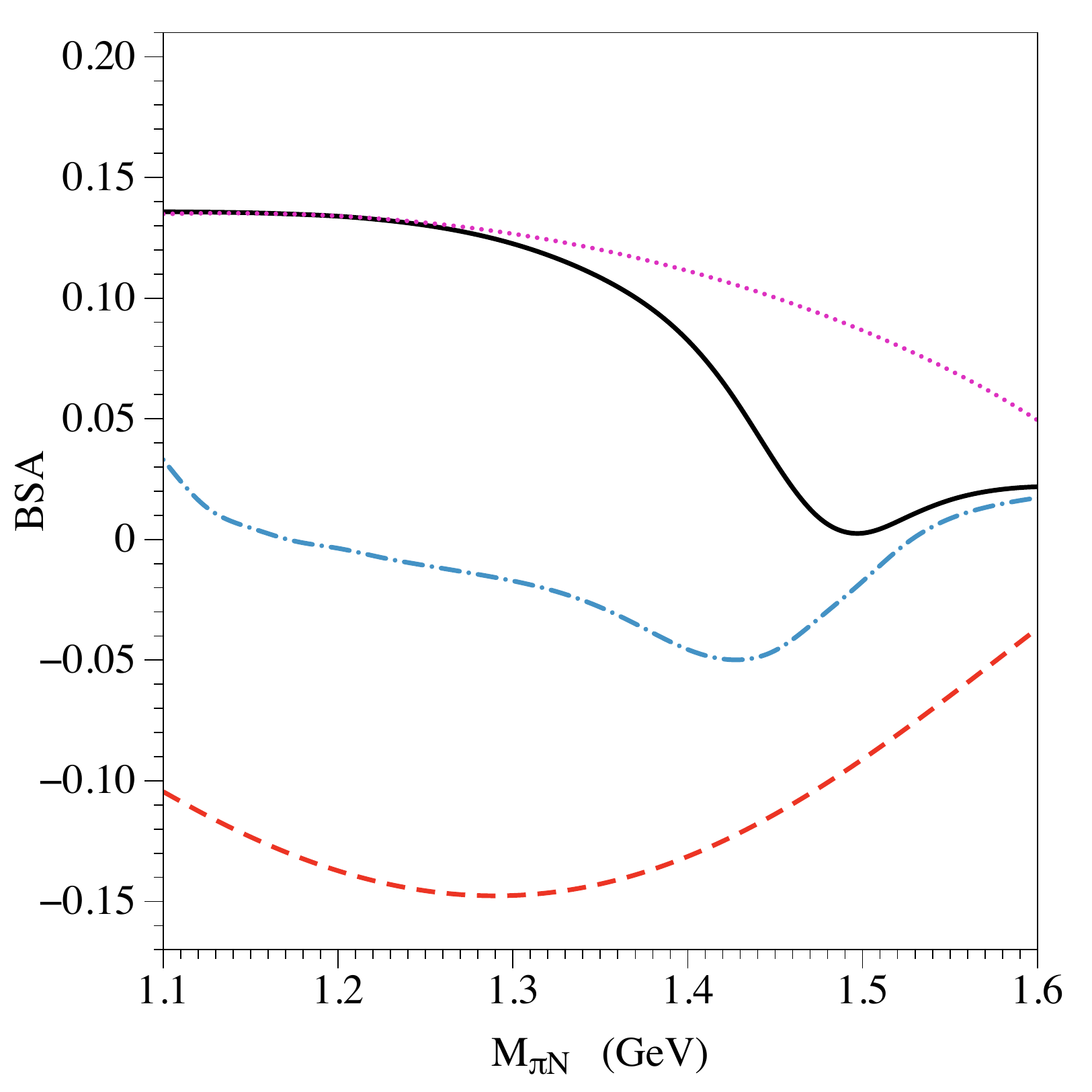}
\includegraphics[width=0.45\textwidth]{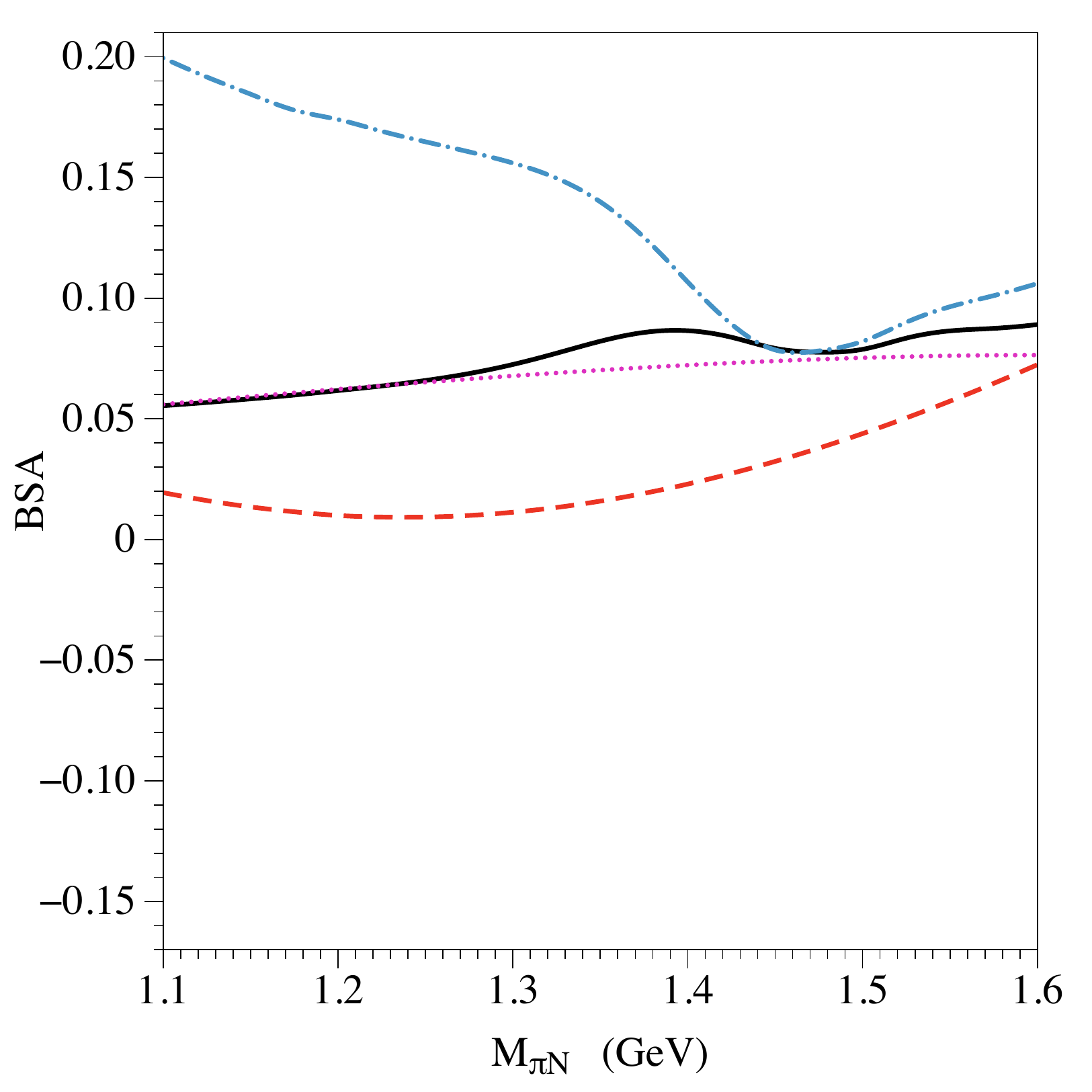}
\caption[]{Dependence on the invariant mass of the $\pi^+ n$ system ($M_{\pi N}$) 
of the $e^- p \to e^- \gamma R \to e^- \gamma \pi^+ n$ 
cross section (upper panels) and corresponding beam-spin asymmetry (lower panels), integrated over the decay pion solid angle, 
with the cut $M_{\pi \gamma} > 1$~GeV, for two values of $-t$.  
Magenta dotted curves: BH + DVCS process for $R = \Delta(1232)$; 
red dashed curves: BH + DVCS process for $R = D_{13}(1520)$; 
blue dashed-dotted curves: BH + DVCS process for $R = P_{11}(1440) + D_{13}(1520) + S_{11}(1535)$;
black solid curves: BH + DVCS process for $R = \Delta(1232) + P_{11}(1440) + D_{13}(1520) + S_{11}(1535)$. }
\label{fig:2ndres_5f}
\end{figure*}

We next show our results for the $N \to R$ DVCS process in the second nucleon resonance region, i.e. for $R = P_{11}(1440), D_{13}(1520), S_{11}(1535)$. We are choosing again kinematical settings close to those of forthcoming data for the $e^- p \to e^- \gamma R \to e^- \gamma \pi^+ n$ reaction from the CLAS12 experiment at JLab. 

In Fig.~\ref{fig:2ndres_5f}, we show the $M_{\pi N}$ invariant mass dependence in the first and second nucleon resonance region of the $e^- p \to e^- \gamma \pi^+ n$ cross section and corresponding BSA with cut $M_{\pi \gamma} > 1$~GeV. The latter is again chosen in order to minimize the possible contamination from the $\rho^+$ production channel, with subsequent decay $\rho^+ \to \gamma \pi^+$. We firstly notice from the cross section behavior that with increasing values of $-t$ the second nucleon resonance region becomes more important relative to the $\Delta(1232)$ resonance region. This can be understood because the $\gamma^\ast N \Delta$ transition form factors are known to drop faster with increasing $-t$ values in comparison with the corresponding ones for the $D_{13}(1520)$ and $S_{11}(1535)$ resonances. One furthermore notices that in the second nucleon resonance region, the $D_{13}(1520)$ excitation provides the largest contribution followed by the $S_{11}(1535)$ resonance. On the other hand, the contribution of the $P_{11}(1440)$ excitation to the unpolarized cross section is only very small. For the BSA one notices that at $-t = 0.5$~GeV$^2$ it reaches a value around 10 \% in the $\Delta$-resonance region, and shows a sharp drop in the second resonance region, mainly driven by the opposite sign of the BSA of the $D_{13}(1520)$ and $S_{11}(1535)$ resonance productions. 
Fig.~\ref{fig:2ndres_5f} also shows that, 
with increasing value of $-t$,  
the BSA for the $D_{13}(1520)$ displays a sign change, resulting in a positive BSA in the second nucleon resonance region at $-t = 1$~GeV$^2$.  

\begin{figure}[!]
\includegraphics[width=0.475\columnwidth]{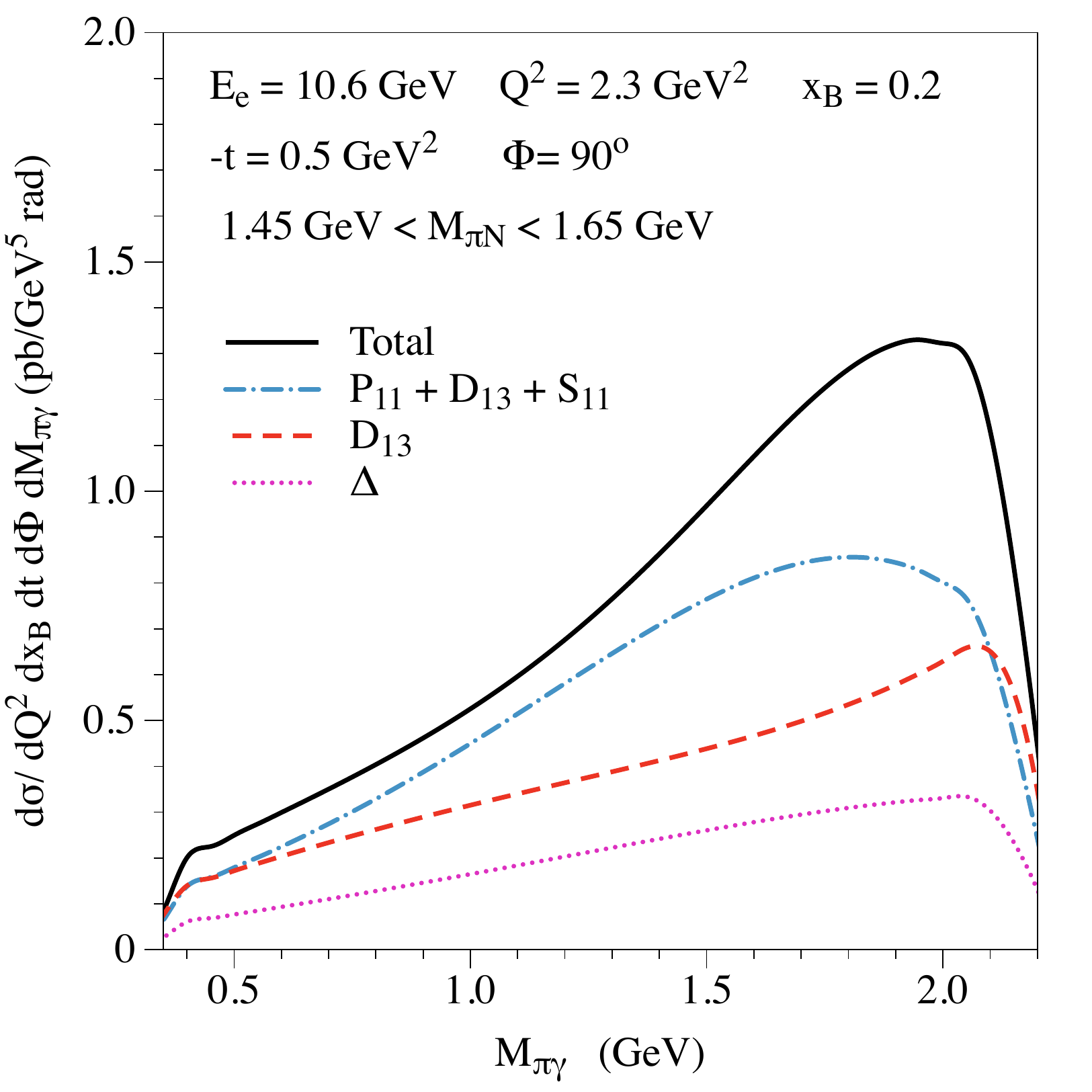}
\includegraphics[width=0.475\columnwidth]{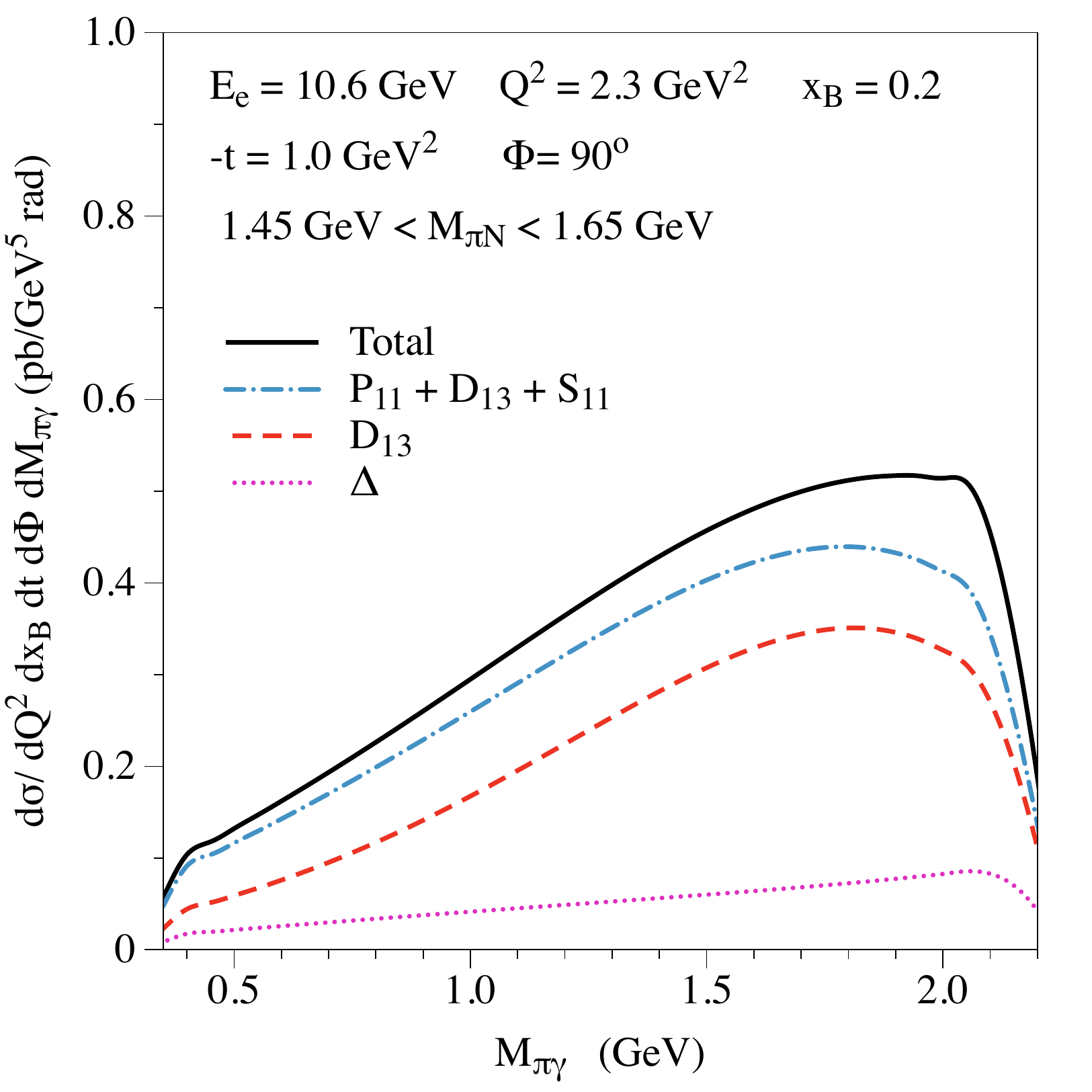}
\includegraphics[width=0.475\columnwidth]{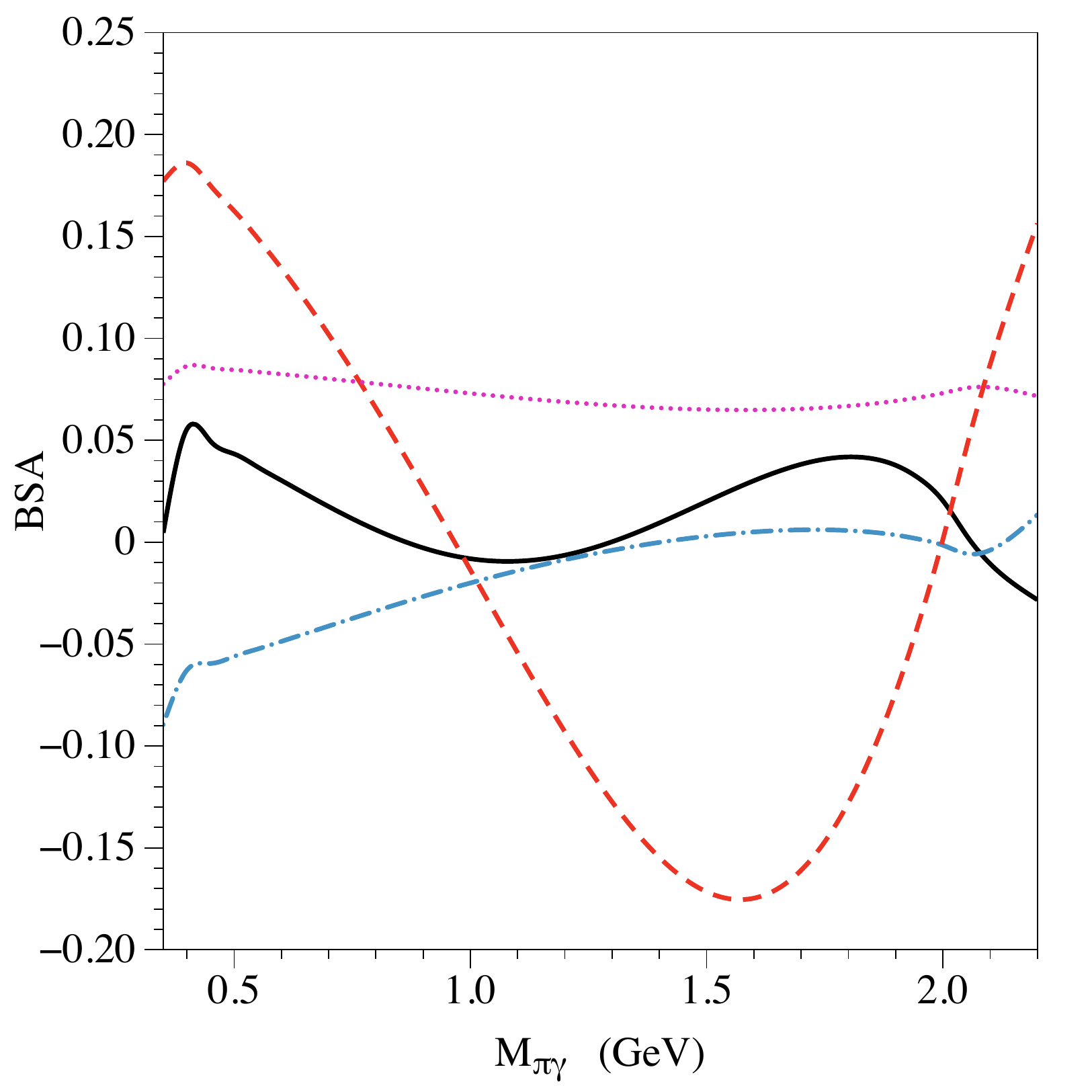}
\includegraphics[width=0.475\columnwidth]{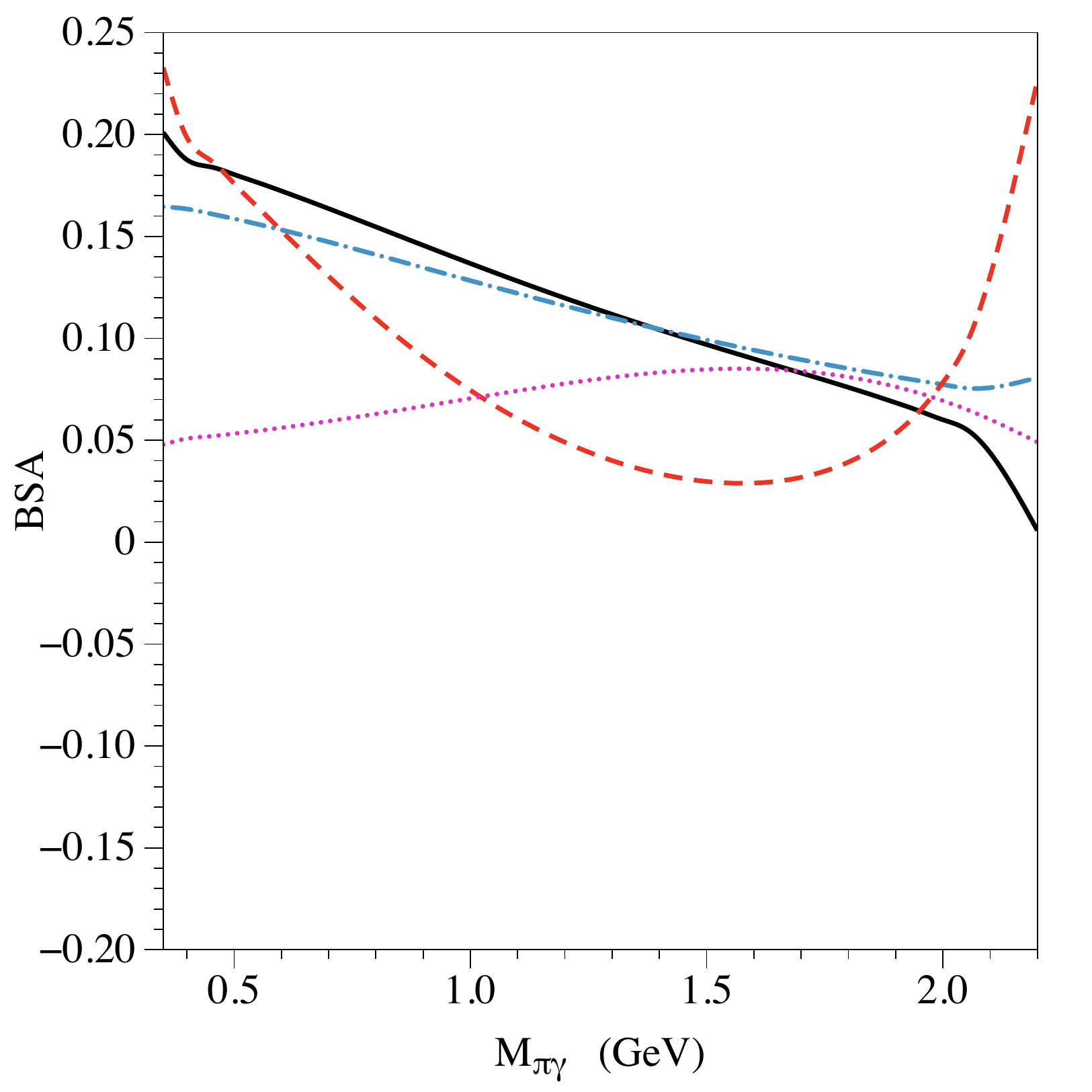}
\caption[]{Dependence on the invariant mass of the $\gamma \pi^+$ system ($M_{\pi \gamma}$) 
of the $e^- p \to e^- \gamma R \to e^- \gamma \pi^+ n$ 
cross section (upper panels) and 
corresponding beam-spin asymmetry (lower panels), integrated over the bin: 
1.45~GeV $\leq M_{\pi N} \leq$ 1.65~GeV, corresponding with the second nucleon resonance region, for two values of $-t$. 
Curve conventions as in Fig.~\ref{fig:2ndres_5f}. 
}
\label{fig:2ndres_mpiga}
\end{figure}

In Fig.~\ref{fig:2ndres_mpiga}, we show the 
$M_{\pi \gamma}$ invariant mass dependence of the 
$e^- p \to e^- \gamma R \to e^- \gamma \pi^+ n$ cross section contribution when integrating over the second nucleon resonance region, i.e. for 1.45~GeV $\leq M_{\pi N} \leq$ 1.65~GeV. As for the result in the first resonance region, which was shown in Fig.~\ref{fig:5fmpiga}, we notice that the process with nucleon resonance production shows a strong rise with increasing value of $M_{\pi \gamma}$. In the second nucleon resonance region, the relative contribution of the region $M_{\pi \gamma} > 1$~GeV as compared to the region $M_{\pi \gamma} < 1$~GeV is even larger than in the $\Delta$ region, facilitating the separation from the an expected $e^- p \to e^- \rho^+(770) n \to e^- \gamma \pi^+ n$ background process. 

\begin{figure}[!]
    \centering
\includegraphics[width=0.49\columnwidth]{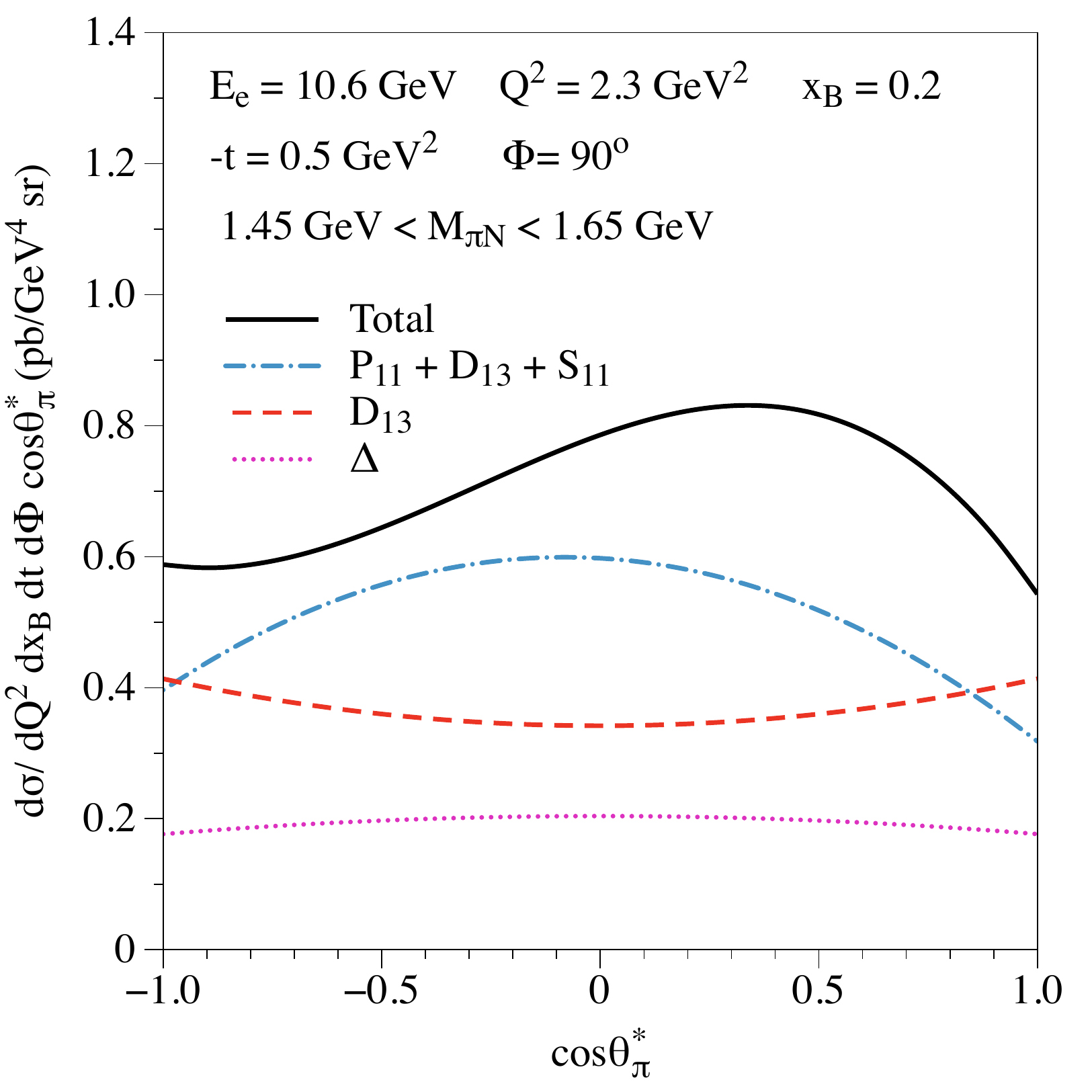}
\includegraphics[width=0.49\columnwidth]{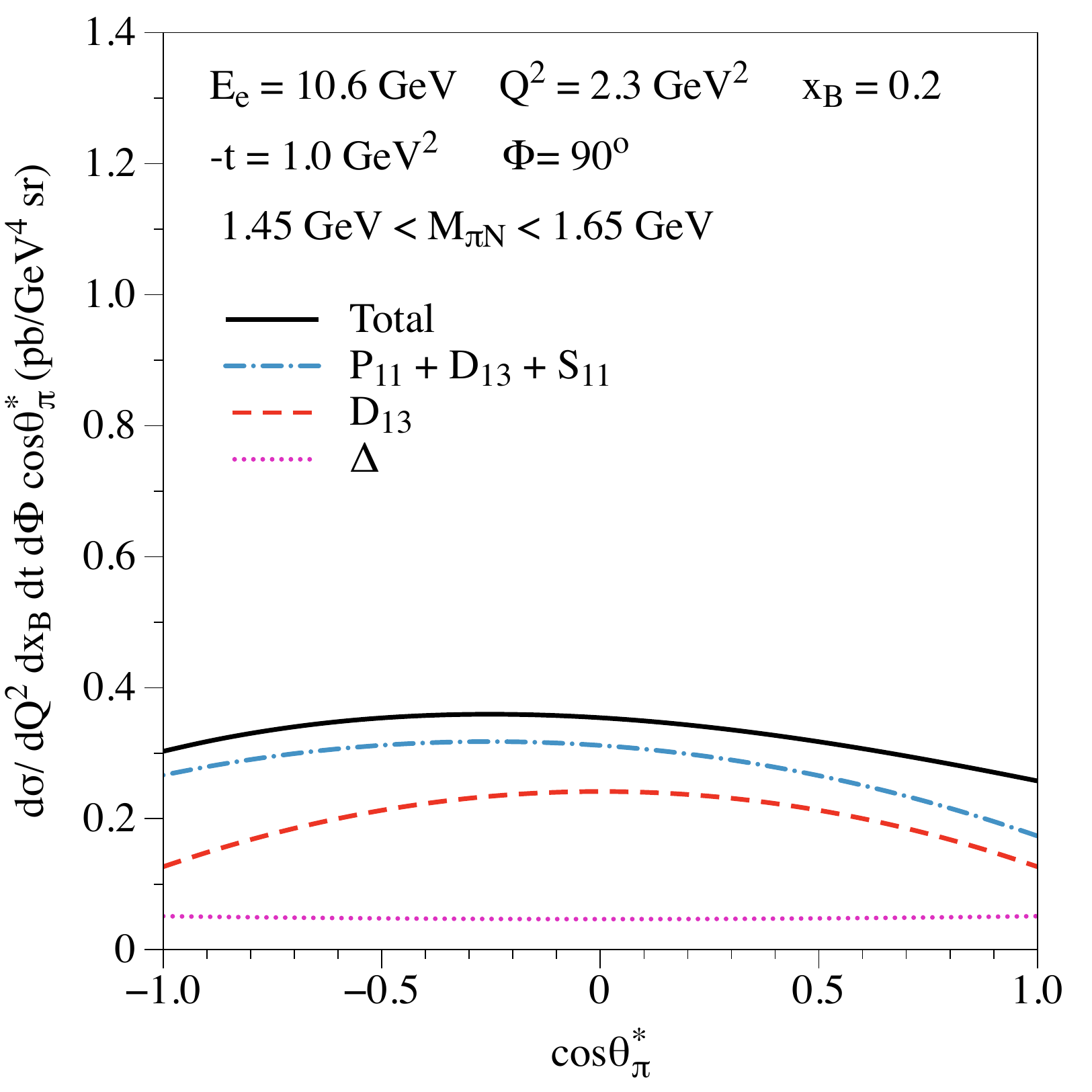}
\includegraphics[width=0.49\columnwidth]{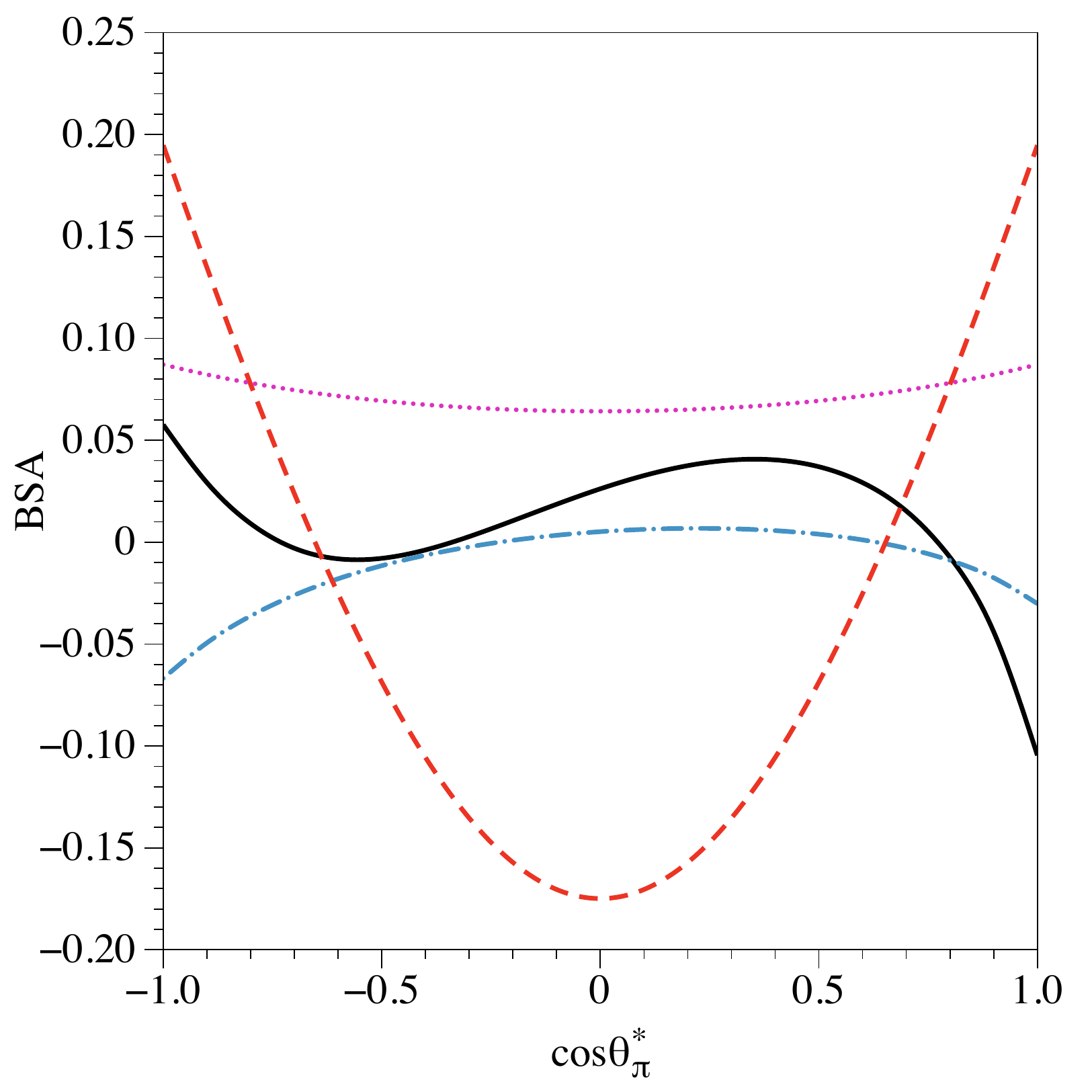}
\includegraphics[width=0.49\columnwidth]{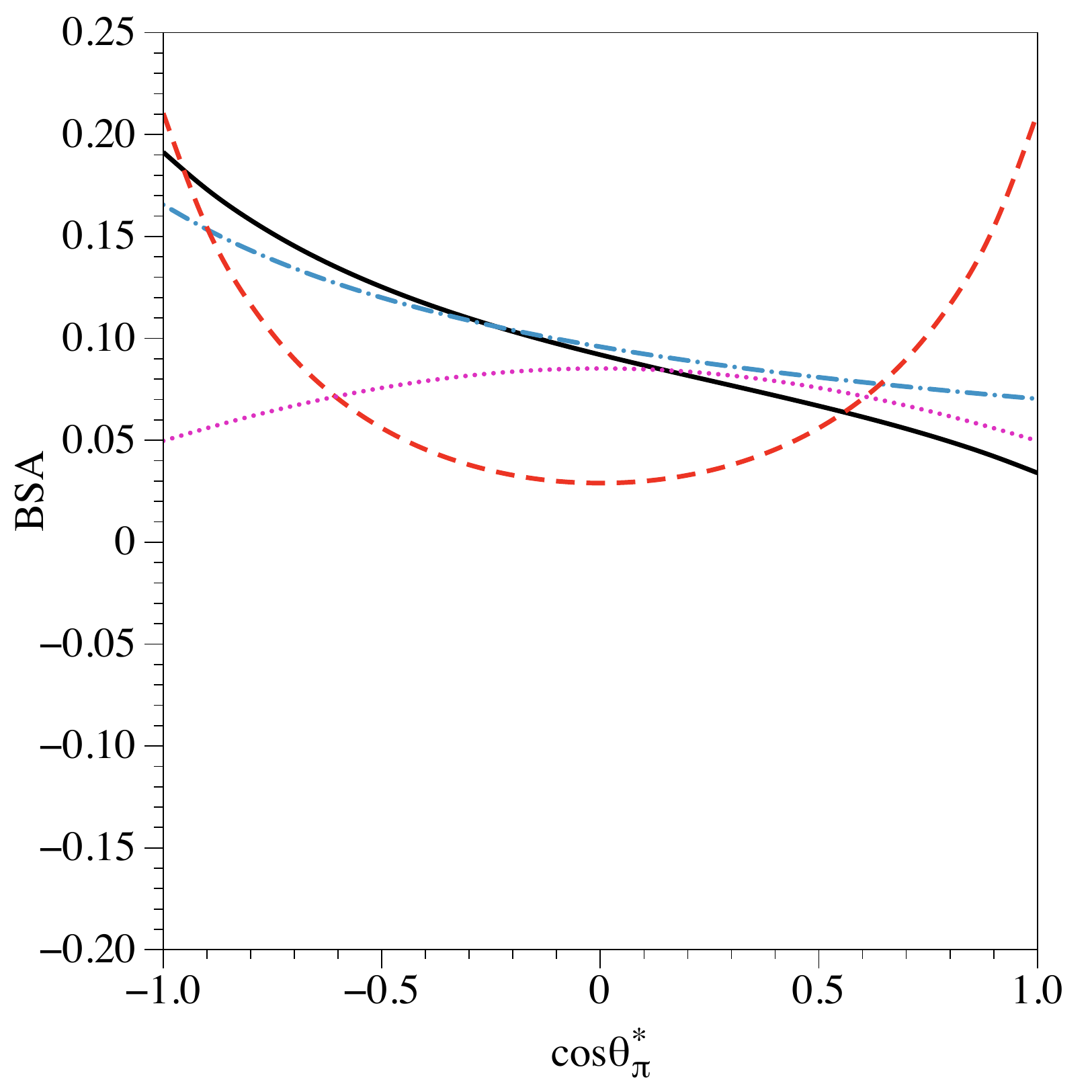}
\caption[]{Decay pion angular distribution (defined in the $\pi N$ rest frame) 
 of the $e^- p \to e^- \gamma R \to e^- \gamma \pi^+ n$ 
cross section (upper panels) and corresponding beam-spin asymmetry (lower panels), 
integrated over the bin: 1.45~GeV $\leq M_{\pi N} \leq$ 1.65~GeV, corresponding with the second nucleon resonance region, for two values of $-t$. 
Curve conventions as in Fig.~\ref{fig:2ndres_5f}. 
}
\label{fig:2ndres_piang}
\end{figure}

Fig.~\ref{fig:2ndres_piang} shows the decay pion angular distribution of the cross section and BSA integrated over the second nucleon resonance region, i.e. for 1.45~GeV $\leq M_{\pi N} \leq$ 1.65~GeV. We notice that while the angular distribution for a single partial wave, e.g. $D_{13}$, 
is symmetric in $\cos \theta_\pi^\ast$, the result for $P_{11} + D_{13} + S_{11}$ as well as the total result, corresponding with $P_{33} + P_{11} + D_{13} + S_{11}$, shows a forward-backward asymmetry in $\theta_\pi^\ast$, which is due to the interference between even and odd partial waves. It is seen that this is even more pronounced in the BSA. 
This opens up the prospect that a sufficiently  accurate angular coverage of such observables in future experiments will allow to perform a partial-wave analysis.

\section{Conclusions and Outlook}
\label{sec6}

Summarizing, in this work we have studied the deeply-virtual Compton process $e^- N \to e^- \gamma R \to e^- \gamma \pi N$ in the first and second $\pi N$-resonance region, within the framework of $N \to R$ generalized parton distributions (GPDs). We started our investigation by a study of the $e^- N \to e^- \gamma R$ amplitude in the Bjorken regime, and subsequently discussed the two main contributions. Firstly, the Bethe-Heitler (BH) process in which the photon is emitted from the lepton line, and secondly the deeply-virtual Compton scattering (DVCS) process which originates from the Compton subprocess on a quark line in the $N \to R$ transition. 

The BH process can be calculated exactly and was fully expressed in terms of the $N \to R$ electromagnetic transition form factors
(FFs). The latter have been measured in quite detail in the first and second nucleon resonance region by several experiments. In particular, the CLAS6 experiment at JLab has achieved the largest kinematic coverage so far in the measurement of these transition FFs. 

The DVCS amplitude was subsequently discussed for the most prominent $\Delta(1232)$-resonance and for the three main resonances in the second $\pi N$-resonance region: 
$P_{11}(1440)$, $D_{13}(1520)$, and $S_{11}(1535)$. 

For the $N \to \Delta$ transition, we extended earlier work and expressed the $N \to \Delta$ matrix element of the quark bi-local vector and axial-vector operators each in terms of $4$
tensor structures, introducing $8$
GPDs in total. As we aim to also describe the amplitudes away from the resonance position, we took care in defining all spin-3/2 tensor structures such that they satisfy the spin-3/2 consistency condition, i.e. don't mix with unphysical spin-1/2 degrees of freedom of the Rarita-Schwinger field when going away from the on-shell resonance position.  
The first moments of three vector GPDs are constrained by the three $N \to \Delta$ electromagnetic transition FFs, whereas the fourth structure has a vanishing first moment. The first moments of the four axial-vector GPDs are constrained by the four $N \to \Delta$ axial-vector FFs. In our estimates for the $N \to \Delta$ GPDs, we used the large-$N_c$ limit relations, allowing to relate the dominant magnetic dipole vector GPDs as well as the two dominant axial-vector GPDs to the corresponding nucleon GPDs. 

We next discussed the $N \to R$ transition in the second $\pi N$ resonance region, discussing the states: $R = P_{11}(1440), D_{13}(1520), S_{11}(1535)$, which have isospin $1/2$ and 
correspond with spin-parity quantum numbers 
$J^P = \frac{1}{2}^+, \frac{3}{2}^-, \frac{1}{2}^-$
respectively.  
For these different transitions, 
we have presented in this work the $N \to R$ DVCS twist-2 amplitude, and introduced the corresponding GPDs. For the vector GPDs, we incorporated the first moment constraints in terms of electromagnetic transition FFs, and for the axial-vector GPDs in terms of the axial-vector transition FFs. For the latter, we explicitly gave the expression for the pion-pole contribution, and used the generalized Goldberger-Treiman relations to constrain the dominant axial-vector FF at low-momentum transfer. In order to provide estimates for the $N \to R$ DVCS amplitude in the second nucleon resonance region, we provided  the simplest factorized parameterization for the corresponding $N \to R$ GPDs, consisting of a product of the transition FFs, and a valence quark parameterization for the quark longitudinal momentum fraction behavior. 

Having constructed the $e^- N \to e^- \gamma R$ amplitude, we then discussed the $R \to \pi N$ decay and calculated the full amplitude for the $e^- N \to e^- \gamma \pi N$ process. As the decay pion angular distribution depends on the quantum numbers of the resonance considered, we subsequently discussed this distribution for the four prominent resonances in the first and second $\pi N$ resonance regions. 

We have presented numerical estimates aimed at providing a theoretical guidance for ongoing analyses of CLAS12 data from JLab. We have shown results for the cross section and beam spin asymmetries for the 
$e^- p \to e^- \gamma R \to e^- \gamma \pi^+ n$ process in the Bjorken limit both in the first and second $\pi N$ resonance region. In the $\Delta(1232)$ resonance region, our estimates 
have shown that in the momentum transfer range $-t = 0.5 - 1$~GeV$^2$, the cross section changes behavior from being BH dominated at $-t = 0.5$~GeV$^2$ to being DVCS dominated at $-t = 1.0$~GeV$^2$. Such behavior is unlike the DVCS process on a nucleon in similar valence region kinematics, for which the BH dominates in both cases. The faster drop with increasing values of $-t$ of the 
$N \to \Delta$ BH amplitude is due to the faster drop of the $N \to \Delta$ magnetic dipole transition FF as compared to the nucleon Dirac FF. We have shown that the BSA for the $e^- p \to e^- \gamma \Delta^+ \to e^- \gamma \pi^+ n$ process is in the 5 - 10\% range. 

Our results have furthermore shown that with increasing values of $-t$ the cross section in the second nucleon resonance region becomes more important relative to the $\Delta(1232)$ resonance region. This was understood from the behavior of the $\gamma^\ast N \Delta$ transition FFs, which are known to drop faster with increasing $-t$ values in comparison with the corresponding ones for the $D_{13}(1520)$ and $S_{11}(1535)$ resonances. In the second nucleon resonance region, the $D_{13}(1520)$ excitation was found to provide the largest contribution followed by the $S_{11}(1535)$ resonance.
We have also shown results for the decay pion angular distribution of the cross section and BSA in first and second $\pi N$-resonance regions. While the angular distribution for a single partial wave, e.g. $P_{33}$ or $D_{13}$, 
is symmetric in $\cos \theta_\pi^\ast$, the result for $P_{33} + P_{11} + D_{13} + S_{11}$ shows a forward-backward asymmetry in $\theta_\pi^\ast$, which is due to the interference between even and odd partial waves. This effect was seen to be even more pronounced in the BSA. 
This opens up the prospect that a sufficiently  accurate angular coverage of such observables in future experiments will allow to perform a partial-wave analysis of the final $\pi N$ system. 

This work can be extended along several different avenues. 

\begin{itemize}
\item
A first extension is to provide model  calculations of $N \to R$ GPDs. In particular in the valence region, dynamical quark/di-quark approaches have proven to yield a good understanding of the $N \to R$ electromagnetic transition FFs~\cite{Aznauryan:2011qj,Burkert:2017djo}. They may therefore be very useful to also get estimates of the largely unmeasured $N \to R$ axial-vector FFs, which constrain the first moments of the axial-vector GPDs. Furthermore, they may also yield estimates of the valence quark longitudinal momentum distribution, and shed light on its correlation with the quark's transverse position distribution in the $N \to R$ transitions. 

\item
Another extension of the present work is to describe the non-diagonal DVCS process in terms of more general $N \to \pi N$ GPDs~\cite{Polyakov:2006dd}, as sketched in ~\cite{Goeke:2001tz}. In such extension, the results of the present work will serve as resonance contribution of four prominent partial-waves in the first and second $\pi N$-resonance region. In a more general study of the spin-independent sector, a useful intermediate step will be the study of the $e^- \pi \to e^- \gamma \pi \pi$ process in terms of non-diagonal $\pi \to \pi \pi$ GPDs. Such studies will set the stage for a full partial-wave analysis of the  
$e^- N \to e^- \gamma \pi N$ process. 

\item
For a detailed comparison with forthcoming CLAS12 data from JLab, it will also be important to provide an estimate for the $e^- p \to e^- \rho^+ n \to e^- \gamma \pi^+ n$ background process, in which a $\rho^+(770)$ is produced, followed by the decay $\rho^+ \to \pi^+ \gamma$. In the present work, we have presented several of our estimates using the cut $M_{\pi \gamma} > 1$~GeV on the $\pi^+ \gamma$ invariant mass in order to minimize such $\rho^+$ contribution. It has been known however at lower c.m. energies that in the extraction of the related $e^- p \to e^- \pi^- \Delta^{++}$ process from 
$e^- p \to e^- \pi^- \pi^+ p$ data~\cite{Wacker:1978at}, a knowledge of the $e^- p \to e^- \rho^0 p$ process, which yields the same final state, is needed.  

\item
A further development and extension of the present formalism is also timely in view of the prospects for future experiments 
at upgrades of the CEBAF accelerator at JLab~\cite{Arrington:2021alx}, 
as well as extensions to the small $x_B$ region 
at the Electron-Ion Collider~\cite{Burkert:2022hjz}.  

\end{itemize}

\section*{Acknowledgements}

The work of K.S. was supported by the National Research Foundation of Korea (NRF) under Grants No. NRF-2020R1A2C1007597 and No. NRF-2018R1A6A1A06024970 (Basic Science Research Program); and by the Foundation for the Advancement of Theoretical Physics and Mathematics ``BASIS''.

The work of M.V. was supported by the Deutsche Forschungsgemeinschaft (DFG, German Research Foundation), in part through the Cluster of Excellence [Precision Physics, Fundamental Interactions, and Structure of Matter] (PRISMA$^+$ EXC 2118/1) within the German Excellence Strategy (Project ID 39083149). 

The authors like to thank Stefan Diehl for many helpful discussions on the experimental analysis of the studied process. 
Furthermore, K.S. also acknowledges the enlightening discussions with Yongseok Oh, Bernard Pire and Lech Szymanowski.

\bibliography{biblio_NtoNstardvcs}

\begin{thebibliography}{49}
\expandafter\ifx\csname natexlab\endcsname\relax\def\natexlab#1{#1}\fi
\expandafter\ifx\csname bibnamefont\endcsname\relax
  \def\bibnamefont#1{#1}\fi
\expandafter\ifx\csname bibfnamefont\endcsname\relax
  \def\bibfnamefont#1{#1}\fi
\expandafter\ifx\csname citenamefont\endcsname\relax
  \def\citenamefont#1{#1}\fi
\expandafter\ifx\csname url\endcsname\relax
  \def\url#1{\texttt{#1}}\fi
\expandafter\ifx\csname urlprefix\endcsname\relax\def\urlprefix{URL }\fi
\providecommand{\bibinfo}[2]{#2}
\providecommand{\eprint}[2][]{\url{#2}}

\bibitem[{\citenamefont{M\"uller et~al.}(1994)\citenamefont{M\"uller,
  Robaschik, Geyer, Dittes, and Ho\v{r}ej\v{s}i}}]{Mueller:1998fv}
\bibinfo{author}{\bibfnamefont{D.}~\bibnamefont{M\"uller}},
  \bibinfo{author}{\bibfnamefont{D.}~\bibnamefont{Robaschik}},
  \bibinfo{author}{\bibfnamefont{B.}~\bibnamefont{Geyer}},
  \bibinfo{author}{\bibfnamefont{F.-M.} \bibnamefont{Dittes}},
  \bibnamefont{and}
  \bibinfo{author}{\bibfnamefont{J.}~\bibnamefont{Ho\v{r}ej\v{s}i}},
  \bibinfo{journal}{Fortsch. Phys.} \textbf{\bibinfo{volume}{42}},
  \bibinfo{pages}{101} (\bibinfo{year}{1994}).

\bibitem[{\citenamefont{Radyushkin}(1997)}]{Radyushkin:1997ki}
\bibinfo{author}{\bibfnamefont{A.~V.} \bibnamefont{Radyushkin}},
  \bibinfo{journal}{Phys. Rev. D} \textbf{\bibinfo{volume}{56}},
  \bibinfo{pages}{5524} (\bibinfo{year}{1997}).

\bibitem[{\citenamefont{Ji}(1997{\natexlab{a}})}]{Ji:1996nm}
\bibinfo{author}{\bibfnamefont{X.-D.} \bibnamefont{Ji}},
  \bibinfo{journal}{Phys. Rev. D} \textbf{\bibinfo{volume}{55}},
  \bibinfo{pages}{7114} (\bibinfo{year}{1997}{\natexlab{a}}).

\bibitem[{\citenamefont{Goeke et~al.}(2001)\citenamefont{Goeke, Polyakov, and
  Vanderhaeghen}}]{Goeke:2001tz}
\bibinfo{author}{\bibfnamefont{K.}~\bibnamefont{Goeke}},
  \bibinfo{author}{\bibfnamefont{M.~V.} \bibnamefont{Polyakov}},
  \bibnamefont{and}
  \bibinfo{author}{\bibfnamefont{M.}~\bibnamefont{Vanderhaeghen}},
  \bibinfo{journal}{Prog. Part. Nucl. Phys.} \textbf{\bibinfo{volume}{47}},
  \bibinfo{pages}{401} (\bibinfo{year}{2001}).

\bibitem[{\citenamefont{Diehl}(2003)}]{Diehl:2003ny}
\bibinfo{author}{\bibfnamefont{M.}~\bibnamefont{Diehl}},
  \bibinfo{journal}{Phys. Rept.} \textbf{\bibinfo{volume}{388}},
  \bibinfo{pages}{41} (\bibinfo{year}{2003}).

\bibitem[{\citenamefont{Belitsky and Radyushkin}(2005)}]{Belitsky:2005qn}
\bibinfo{author}{\bibfnamefont{A.}~\bibnamefont{Belitsky}} \bibnamefont{and}
  \bibinfo{author}{\bibfnamefont{A.}~\bibnamefont{Radyushkin}},
  \bibinfo{journal}{Phys. Rept.} \textbf{\bibinfo{volume}{418}},
  \bibinfo{pages}{1} (\bibinfo{year}{2005}).

\bibitem[{\citenamefont{Boffi and Pasquini}(2007)}]{Boffi:2007yc}
\bibinfo{author}{\bibfnamefont{S.}~\bibnamefont{Boffi}} \bibnamefont{and}
  \bibinfo{author}{\bibfnamefont{B.}~\bibnamefont{Pasquini}},
  \bibinfo{journal}{Riv. Nuovo Cim.} \textbf{\bibinfo{volume}{30}},
  \bibinfo{pages}{387} (\bibinfo{year}{2007}).

\bibitem[{\citenamefont{Ralston and Pire}(2002)}]{Ralston:2001xs}
\bibinfo{author}{\bibfnamefont{J.~P.} \bibnamefont{Ralston}} \bibnamefont{and}
  \bibinfo{author}{\bibfnamefont{B.}~\bibnamefont{Pire}},
  \bibinfo{journal}{Phys. Rev. D} \textbf{\bibinfo{volume}{66}},
  \bibinfo{pages}{111501} (\bibinfo{year}{2002}).

\bibitem[{\citenamefont{Dupr\'e
  et~al.}(2017{\natexlab{a}})\citenamefont{Dupr\'e, Guidal, and
  Vanderhaeghen}}]{Dupre:2016mai}
\bibinfo{author}{\bibfnamefont{R.}~\bibnamefont{Dupr\'e}},
  \bibinfo{author}{\bibfnamefont{M.}~\bibnamefont{Guidal}}, \bibnamefont{and}
  \bibinfo{author}{\bibfnamefont{M.}~\bibnamefont{Vanderhaeghen}},
  \bibinfo{journal}{Phys. Rev. D} \textbf{\bibinfo{volume}{95}},
  \bibinfo{pages}{011501} (\bibinfo{year}{2017}{\natexlab{a}}).

\bibitem[{\citenamefont{Dupr\'e
  et~al.}(2017{\natexlab{b}})\citenamefont{Dupr\'e, Guidal, Niccolai, and
  Vanderhaeghen}}]{Dupre:2017hfs}
\bibinfo{author}{\bibfnamefont{R.}~\bibnamefont{Dupr\'e}},
  \bibinfo{author}{\bibfnamefont{M.}~\bibnamefont{Guidal}},
  \bibinfo{author}{\bibfnamefont{S.}~\bibnamefont{Niccolai}}, \bibnamefont{and}
  \bibinfo{author}{\bibfnamefont{M.}~\bibnamefont{Vanderhaeghen}},
  \bibinfo{journal}{Eur. Phys. J. A} \textbf{\bibinfo{volume}{53}},
  \bibinfo{pages}{171} (\bibinfo{year}{2017}{\natexlab{b}}).

\bibitem[{\citenamefont{Burkardt}(2000)}]{Burkardt:2000za}
\bibinfo{author}{\bibfnamefont{M.}~\bibnamefont{Burkardt}},
  \bibinfo{journal}{Phys. Rev. D} \textbf{\bibinfo{volume}{62}},
  \bibinfo{pages}{071503} (\bibinfo{year}{2000}), \bibinfo{note}{[Erratum:
  Phys.Rev.D {\bf 66}, 119903 (2002)]}.

\bibitem[{\citenamefont{Ji}(1997{\natexlab{b}})}]{Ji:1996ek}
\bibinfo{author}{\bibfnamefont{X.-D.} \bibnamefont{Ji}},
  \bibinfo{journal}{Phys. Rev. Lett.} \textbf{\bibinfo{volume}{78}},
  \bibinfo{pages}{610} (\bibinfo{year}{1997}{\natexlab{b}}).

\bibitem[{\citenamefont{Polyakov}(2003)}]{Polyakov:2002yz}
\bibinfo{author}{\bibfnamefont{M.~V.} \bibnamefont{Polyakov}},
  \bibinfo{journal}{Phys. Lett. B} \textbf{\bibinfo{volume}{555}},
  \bibinfo{pages}{57} (\bibinfo{year}{2003}).

\bibitem[{\citenamefont{Polyakov and Schweitzer}(2018)}]{Polyakov:2018zvc}
\bibinfo{author}{\bibfnamefont{M.~V.} \bibnamefont{Polyakov}} \bibnamefont{and}
  \bibinfo{author}{\bibfnamefont{P.}~\bibnamefont{Schweitzer}},
  \bibinfo{journal}{Int. J. Mod. Phys. A} \textbf{\bibinfo{volume}{33}},
  \bibinfo{pages}{1830025} (\bibinfo{year}{2018}).

\bibitem[{\citenamefont{Burkert et~al.}(2018)\citenamefont{Burkert,
  Elouadrhiri, and Girod}}]{Burkert:2018bqq}
\bibinfo{author}{\bibfnamefont{V.~D.} \bibnamefont{Burkert}},
  \bibinfo{author}{\bibfnamefont{L.}~\bibnamefont{Elouadrhiri}},
  \bibnamefont{and} \bibinfo{author}{\bibfnamefont{F.~X.} \bibnamefont{Girod}},
  \bibinfo{journal}{Nature} \textbf{\bibinfo{volume}{557}},
  \bibinfo{pages}{396} (\bibinfo{year}{2018}).

\bibitem[{\citenamefont{Kumeri\v{c}ki}(2019)}]{Kumericki:2019ddg}
\bibinfo{author}{\bibfnamefont{K.}~\bibnamefont{Kumeri\v{c}ki}},
  \bibinfo{journal}{Nature} \textbf{\bibinfo{volume}{570}}, \bibinfo{pages}{E1}
  (\bibinfo{year}{2019}).

\bibitem[{\citenamefont{Dutrieux et~al.}(2021)\citenamefont{Dutrieux, Lorc\'e,
  Moutarde, Sznajder, Trawi\'nski, and Wagner}}]{Dutrieux:2021nlz}
\bibinfo{author}{\bibfnamefont{H.}~\bibnamefont{Dutrieux}},
  \bibinfo{author}{\bibfnamefont{C.}~\bibnamefont{Lorc\'e}},
  \bibinfo{author}{\bibfnamefont{H.}~\bibnamefont{Moutarde}},
  \bibinfo{author}{\bibfnamefont{P.}~\bibnamefont{Sznajder}},
  \bibinfo{author}{\bibfnamefont{A.}~\bibnamefont{Trawi\'nski}},
  \bibnamefont{and} \bibinfo{author}{\bibfnamefont{J.}~\bibnamefont{Wagner}},
  \bibinfo{journal}{Eur. Phys. J. C} \textbf{\bibinfo{volume}{81}},
  \bibinfo{pages}{300} (\bibinfo{year}{2021}).

\bibitem[{\citenamefont{Burkert et~al.}(2021)\citenamefont{Burkert,
  Elouadrhiri, and Girod}}]{Burkert:2021ith}
\bibinfo{author}{\bibfnamefont{V.~D.} \bibnamefont{Burkert}},
  \bibinfo{author}{\bibfnamefont{L.}~\bibnamefont{Elouadrhiri}},
  \bibnamefont{and} \bibinfo{author}{\bibfnamefont{F.~X.} \bibnamefont{Girod}}
  (\bibinfo{year}{2021}), \eprint{2104.02031}.

\bibitem[{\citenamefont{Polyakov}(1998)}]{Polyakov:1998sz}
\bibinfo{author}{\bibfnamefont{M.~V.} \bibnamefont{Polyakov}}, in
  \emph{\bibinfo{booktitle}{{8th International Conference on the Structure of
  Baryons}}} (\bibinfo{year}{1998}), pp. \bibinfo{pages}{765--769}.

\bibitem[{\citenamefont{Frankfurt et~al.}(1998)\citenamefont{Frankfurt,
  Polyakov, and Strikman}}]{Frankfurt:1998jq}
\bibinfo{author}{\bibfnamefont{L.~L.} \bibnamefont{Frankfurt}},
  \bibinfo{author}{\bibfnamefont{M.~V.} \bibnamefont{Polyakov}},
  \bibnamefont{and} \bibinfo{author}{\bibfnamefont{M.}~\bibnamefont{Strikman}},
  in \emph{\bibinfo{booktitle}{{Workshop on Jefferson Lab Physics and
  Instrumentation with 6-12-GeV Beams and Beyond}}} (\bibinfo{year}{1998}),
  \eprint{hep-ph/9808449}.

\bibitem[{\citenamefont{Frankfurt et~al.}(2000)\citenamefont{Frankfurt,
  Polyakov, Strikman, and Vanderhaeghen}}]{Frankfurt:1999xe}
\bibinfo{author}{\bibfnamefont{L.~L.} \bibnamefont{Frankfurt}},
  \bibinfo{author}{\bibfnamefont{M.~V.} \bibnamefont{Polyakov}},
  \bibinfo{author}{\bibfnamefont{M.}~\bibnamefont{Strikman}}, \bibnamefont{and}
  \bibinfo{author}{\bibfnamefont{M.}~\bibnamefont{Vanderhaeghen}},
  \bibinfo{journal}{Phys. Rev. Lett.} \textbf{\bibinfo{volume}{84}},
  \bibinfo{pages}{2589} (\bibinfo{year}{2000}).

\bibitem[{\citenamefont{Polyakov and Stratmann}(2006)}]{Polyakov:2006dd}
\bibinfo{author}{\bibfnamefont{M.~V.} \bibnamefont{Polyakov}} \bibnamefont{and}
  \bibinfo{author}{\bibfnamefont{S.}~\bibnamefont{Stratmann}}
  (\bibinfo{year}{2006}), \eprint{hep-ph/0609045}.

\bibitem[{\citenamefont{Guichon et~al.}(2003)\citenamefont{Guichon, Moss\'e,
  and Vanderhaeghen}}]{Guichon:2003ah}
\bibinfo{author}{\bibfnamefont{P.~A.~M.} \bibnamefont{Guichon}},
  \bibinfo{author}{\bibfnamefont{L.}~\bibnamefont{Moss\'e}}, \bibnamefont{and}
  \bibinfo{author}{\bibfnamefont{M.}~\bibnamefont{Vanderhaeghen}},
  \bibinfo{journal}{Phys. Rev. D} \textbf{\bibinfo{volume}{68}},
  \bibinfo{pages}{034018} (\bibinfo{year}{2003}).

\bibitem[{\citenamefont{\"Ozdem and Azizi}(2020)}]{Ozdem:2019pkg}
\bibinfo{author}{\bibfnamefont{U.}~\bibnamefont{\"Ozdem}} \bibnamefont{and}
  \bibinfo{author}{\bibfnamefont{K.}~\bibnamefont{Azizi}},
  \bibinfo{journal}{Phys. Rev. D} \textbf{\bibinfo{volume}{101}},
  \bibinfo{pages}{054031} (\bibinfo{year}{2020}).

\bibitem[{\citenamefont{Polyakov and Tandogan}(2020)}]{Polyakov:2020rzq}
\bibinfo{author}{\bibfnamefont{M.~V.} \bibnamefont{Polyakov}} \bibnamefont{and}
  \bibinfo{author}{\bibfnamefont{A.}~\bibnamefont{Tandogan}},
  \bibinfo{journal}{Phys. Rev. D} \textbf{\bibinfo{volume}{101}},
  \bibinfo{pages}{118501} (\bibinfo{year}{2020}).

\bibitem[{\citenamefont{Azizi and \"Ozdem}(2021)}]{Azizi:2020jog}
\bibinfo{author}{\bibfnamefont{K.}~\bibnamefont{Azizi}} \bibnamefont{and}
  \bibinfo{author}{\bibfnamefont{U.}~\bibnamefont{\"Ozdem}},
  \bibinfo{journal}{Nucl. Phys. A} \textbf{\bibinfo{volume}{1015}},
  \bibinfo{pages}{122296} (\bibinfo{year}{2021}).

\bibitem[{\citenamefont{\"Ozdem and Azizi}(2022)}]{Ozdem:2022zig}
\bibinfo{author}{\bibfnamefont{U.}~\bibnamefont{\"Ozdem}} \bibnamefont{and}
  \bibinfo{author}{\bibfnamefont{K.}~\bibnamefont{Azizi}}
  (\bibinfo{year}{2022}), \eprint{2212.07290}.

\bibitem[{\citenamefont{Pascalutsa et~al.}(2007)\citenamefont{Pascalutsa,
  Vanderhaeghen, and Yang}}]{Pascalutsa:2006up}
\bibinfo{author}{\bibfnamefont{V.}~\bibnamefont{Pascalutsa}},
  \bibinfo{author}{\bibfnamefont{M.}~\bibnamefont{Vanderhaeghen}},
  \bibnamefont{and} \bibinfo{author}{\bibfnamefont{S.~N.} \bibnamefont{Yang}},
  \bibinfo{journal}{Phys. Rept.} \textbf{\bibinfo{volume}{437}},
  \bibinfo{pages}{125} (\bibinfo{year}{2007}).

\bibitem[{\citenamefont{Aznauryan and Burkert}(2012)}]{Aznauryan:2011qj}
\bibinfo{author}{\bibfnamefont{I.~G.} \bibnamefont{Aznauryan}}
  \bibnamefont{and} \bibinfo{author}{\bibfnamefont{V.~D.}
  \bibnamefont{Burkert}}, \bibinfo{journal}{Prog. Part. Nucl. Phys.}
  \textbf{\bibinfo{volume}{67}}, \bibinfo{pages}{1} (\bibinfo{year}{2012}).

\bibitem[{\citenamefont{Burkert et~al.}(2022)}]{Burkert:2022hjz}
\bibinfo{author}{\bibfnamefont{V.}~\bibnamefont{Burkert}} \bibnamefont{et~al.}
  (\bibinfo{year}{2022}), \eprint{2211.15746}.

\bibitem[{\citenamefont{Arndt et~al.}(2006)\citenamefont{Arndt, Briscoe,
  Strakovsky, and Workman}}]{Arndt:2006bf}
\bibinfo{author}{\bibfnamefont{R.~A.} \bibnamefont{Arndt}},
  \bibinfo{author}{\bibfnamefont{W.~J.} \bibnamefont{Briscoe}},
  \bibinfo{author}{\bibfnamefont{I.~I.} \bibnamefont{Strakovsky}},
  \bibnamefont{and} \bibinfo{author}{\bibfnamefont{R.~L.}
  \bibnamefont{Workman}}, \bibinfo{journal}{Phys. Rev. C}
  \textbf{\bibinfo{volume}{74}}, \bibinfo{pages}{045205}
  (\bibinfo{year}{2006}).

\bibitem[{\citenamefont{Kobzarev and Okun}(1962)}]{Kobzarev:1962wt}
\bibinfo{author}{\bibfnamefont{I.~Y.} \bibnamefont{Kobzarev}} \bibnamefont{and}
  \bibinfo{author}{\bibfnamefont{L.~B.} \bibnamefont{Okun}},
  \bibinfo{journal}{Zh. Eksp. Teor. Fiz.} \textbf{\bibinfo{volume}{43}},
  \bibinfo{pages}{1904} (\bibinfo{year}{1962}).

\bibitem[{\citenamefont{Guidal et~al.}(2003)\citenamefont{Guidal, Bouchigny,
  Didelez, Hadjidakis, Hourany, and Vanderhaeghen}}]{Guidal:2003ji}
\bibinfo{author}{\bibfnamefont{M.}~\bibnamefont{Guidal}},
  \bibinfo{author}{\bibfnamefont{S.}~\bibnamefont{Bouchigny}},
  \bibinfo{author}{\bibfnamefont{J.~P.} \bibnamefont{Didelez}},
  \bibinfo{author}{\bibfnamefont{C.}~\bibnamefont{Hadjidakis}},
  \bibinfo{author}{\bibfnamefont{E.}~\bibnamefont{Hourany}}, \bibnamefont{and}
  \bibinfo{author}{\bibfnamefont{M.}~\bibnamefont{Vanderhaeghen}},
  \bibinfo{journal}{Nucl. Phys. A} \textbf{\bibinfo{volume}{721}},
  \bibinfo{pages}{327} (\bibinfo{year}{2003}).

\bibitem[{\citenamefont{Moreno}(2009)}]{Moreno:2009oga}
\bibinfo{author}{\bibfnamefont{B.}~\bibnamefont{Moreno}}, \bibinfo{type}{Phd
  thesis}, \bibinfo{school}{Orsay, IPN} (\bibinfo{year}{2009}).

\bibitem[{\citenamefont{{Diehl, S.}}(2022)}]{SDiehlTrento}
\bibinfo{author}{\bibnamefont{{Diehl, S.}}} (\bibinfo{year}{2022}),
  \bibinfo{note}{talk presented at ``Opportunities with JLab Energy and
  Luminosity Upgrade'', ECT$^*$ Trento, Italy, 26-30 September 2022;
  \url{https://indico.ectstar.eu/event/152/contributions/3151/attachments/2027/2644/Diehl_ECT_2022_tranistionGPDs.pdf}}.

\bibitem[{\citenamefont{Kroll and Passek-Kumeri\v{c}ki}(2022)}]{Kroll:2022roq}
\bibinfo{author}{\bibfnamefont{P.}~\bibnamefont{Kroll}} \bibnamefont{and}
  \bibinfo{author}{\bibfnamefont{K.}~\bibnamefont{Passek-Kumeri\v{c}ki}}
  (\bibinfo{year}{2022}), \eprint{2211.09474}.

\bibitem[{\citenamefont{Jones and Scadron}(1973)}]{Jones:1972ky}
\bibinfo{author}{\bibfnamefont{H.~F.} \bibnamefont{Jones}} \bibnamefont{and}
  \bibinfo{author}{\bibfnamefont{M.~D.} \bibnamefont{Scadron}},
  \bibinfo{journal}{Annals Phys.} \textbf{\bibinfo{volume}{81}},
  \bibinfo{pages}{1} (\bibinfo{year}{1973}).

\bibitem[{\citenamefont{Drechsel et~al.}(2007)\citenamefont{Drechsel, Kamalov,
  and Tiator}}]{Drechsel:2007if}
\bibinfo{author}{\bibfnamefont{D.}~\bibnamefont{Drechsel}},
  \bibinfo{author}{\bibfnamefont{S.}~\bibnamefont{Kamalov}}, \bibnamefont{and}
  \bibinfo{author}{\bibfnamefont{L.}~\bibnamefont{Tiator}},
  \bibinfo{journal}{Eur. Phys. J. A} \textbf{\bibinfo{volume}{34}},
  \bibinfo{pages}{69} (\bibinfo{year}{2007}).

\bibitem[{\citenamefont{Tiator et~al.}(2011)\citenamefont{Tiator, Drechsel,
  Kamalov, and Vanderhaeghen}}]{Tiator:2011pw}
\bibinfo{author}{\bibfnamefont{L.}~\bibnamefont{Tiator}},
  \bibinfo{author}{\bibfnamefont{D.}~\bibnamefont{Drechsel}},
  \bibinfo{author}{\bibfnamefont{S.}~\bibnamefont{Kamalov}}, \bibnamefont{and}
  \bibinfo{author}{\bibfnamefont{M.}~\bibnamefont{Vanderhaeghen}},
  \bibinfo{journal}{Eur. Phys. J. ST} \textbf{\bibinfo{volume}{198}},
  \bibinfo{pages}{141} (\bibinfo{year}{2011}).

\bibitem[{\citenamefont{Tiator and Vanderhaeghen}(2009)}]{Tiator:2008kd}
\bibinfo{author}{\bibfnamefont{L.}~\bibnamefont{Tiator}} \bibnamefont{and}
  \bibinfo{author}{\bibfnamefont{M.}~\bibnamefont{Vanderhaeghen}},
  \bibinfo{journal}{Phys. Lett. B} \textbf{\bibinfo{volume}{672}},
  \bibinfo{pages}{344} (\bibinfo{year}{2009}).

\bibitem[{\citenamefont{Adler}(1968)}]{Adler:1968tw}
\bibinfo{author}{\bibfnamefont{S.~L.} \bibnamefont{Adler}},
  \bibinfo{journal}{Annals Phys.} \textbf{\bibinfo{volume}{50}},
  \bibinfo{pages}{189} (\bibinfo{year}{1968}).

\bibitem[{\citenamefont{Adler et~al.}(1975)\citenamefont{Adler, Dashen, Healy,
  Karliner, Lieberman, Ng, and Tsao}}]{Adler:1975tm}
\bibinfo{author}{\bibfnamefont{S.~L.} \bibnamefont{Adler}},
  \bibinfo{author}{\bibfnamefont{R.~F.} \bibnamefont{Dashen}},
  \bibinfo{author}{\bibfnamefont{J.~B.} \bibnamefont{Healy}},
  \bibinfo{author}{\bibfnamefont{I.}~\bibnamefont{Karliner}},
  \bibinfo{author}{\bibfnamefont{J.}~\bibnamefont{Lieberman}},
  \bibinfo{author}{\bibfnamefont{Y.~J.} \bibnamefont{Ng}}, \bibnamefont{and}
  \bibinfo{author}{\bibfnamefont{H.-S.} \bibnamefont{Tsao}},
  \bibinfo{journal}{Phys. Rev. D} \textbf{\bibinfo{volume}{12}},
  \bibinfo{pages}{3522} (\bibinfo{year}{1975}).

\bibitem[{\citenamefont{Kitagaki et~al.}(1990)}]{Kitagaki:1990vs}
\bibinfo{author}{\bibfnamefont{T.}~\bibnamefont{Kitagaki}}
  \bibnamefont{et~al.}, \bibinfo{journal}{Phys. Rev. D}
  \textbf{\bibinfo{volume}{42}}, \bibinfo{pages}{1331} (\bibinfo{year}{1990}).

\bibitem[{\citenamefont{Burkert and Roberts}(2019)}]{Burkert:2017djo}
\bibinfo{author}{\bibfnamefont{V.~D.} \bibnamefont{Burkert}} \bibnamefont{and}
  \bibinfo{author}{\bibfnamefont{C.~D.} \bibnamefont{Roberts}},
  \bibinfo{journal}{Rev. Mod. Phys.} \textbf{\bibinfo{volume}{91}},
  \bibinfo{pages}{011003} (\bibinfo{year}{2019}).

\bibitem[{\citenamefont{Workman and Others}(2022)}]{Workman:2022ynf}
\bibinfo{author}{\bibfnamefont{R.~L.} \bibnamefont{Workman}} \bibnamefont{and}
  \bibinfo{author}{\bibnamefont{Others}} (\bibinfo{collaboration}{Particle Data
  Group}), \bibinfo{journal}{PTEP} \textbf{\bibinfo{volume}{2022}},
  \bibinfo{pages}{083C01} (\bibinfo{year}{2022}).

\bibitem[{\citenamefont{Drechsel et~al.}(1999)\citenamefont{Drechsel, Hanstein,
  Kamalov, and Tiator}}]{Drechsel:1998hk}
\bibinfo{author}{\bibfnamefont{D.}~\bibnamefont{Drechsel}},
  \bibinfo{author}{\bibfnamefont{O.}~\bibnamefont{Hanstein}},
  \bibinfo{author}{\bibfnamefont{S.~S.} \bibnamefont{Kamalov}},
  \bibnamefont{and} \bibinfo{author}{\bibfnamefont{L.}~\bibnamefont{Tiator}},
  \bibinfo{journal}{Nucl. Phys. A} \textbf{\bibinfo{volume}{645}},
  \bibinfo{pages}{145} (\bibinfo{year}{1999}).

\bibitem[{\citenamefont{Chiang et~al.}(2002)\citenamefont{Chiang, Yang, Tiator,
  and Drechsel}}]{Chiang:2001as}
\bibinfo{author}{\bibfnamefont{W.-T.} \bibnamefont{Chiang}},
  \bibinfo{author}{\bibfnamefont{S.-N.} \bibnamefont{Yang}},
  \bibinfo{author}{\bibfnamefont{L.}~\bibnamefont{Tiator}}, \bibnamefont{and}
  \bibinfo{author}{\bibfnamefont{D.}~\bibnamefont{Drechsel}},
  \bibinfo{journal}{Nucl. Phys. A} \textbf{\bibinfo{volume}{700}},
  \bibinfo{pages}{429} (\bibinfo{year}{2002}).

\bibitem[{\citenamefont{Wacker et~al.}(1978)}]{Wacker:1978at}
\bibinfo{author}{\bibfnamefont{K.}~\bibnamefont{Wacker}} \bibnamefont{et~al.},
  \bibinfo{journal}{Nucl. Phys. B} \textbf{\bibinfo{volume}{144}},
  \bibinfo{pages}{269} (\bibinfo{year}{1978}).

\bibitem[{\citenamefont{Arrington et~al.}(2022)}]{Arrington:2021alx}
\bibinfo{author}{\bibfnamefont{J.}~\bibnamefont{Arrington}}
  \bibnamefont{et~al.}, \bibinfo{journal}{Prog. Part. Nucl. Phys.}
  \textbf{\bibinfo{volume}{127}}, \bibinfo{pages}{103985}
  (\bibinfo{year}{2022}).

\end{thebibliography}

\end{document}